\documentclass[superscriptaddress,amsmath,amssymb, reprint]{revtex4-1}

\usepackage{graphicx}
\usepackage{dcolumn}
\usepackage{bm}

\usepackage[utf8]{inputenc}
\usepackage[T1]{fontenc}
\usepackage{mathptmx}
\usepackage{etoolbox}

\makeatother
\begin{document}

\title[Non-adiabatic electronic relaxation of tetracene]{Non-adiabatic electronic relaxation of tetracene from its brightest singlet excited state}
\author{A. Scognamiglio}
 \altaffiliation{These authors contributed equally to this work.}

\author{K. S. Thalmann}

\altaffiliation[]{These authors contributed equally to this work.}
\author{S. Hartweg}
\email{sebastian.hartweg@physik.uni-freiburg.de}

\author{$^{, *}$ N. Rendler}
 \affiliation{University of Freiburg, Institute of Physics, Hermann-Herder-Str.\,3, Freiburg, Germany.}

\author{L. Bruder}
 \affiliation{University of Freiburg, Institute of Physics, Hermann-Herder-Str.\,3, Freiburg, Germany.}

\author{P. B. Coto}
 \affiliation{Materials Physics Center (CFM), CSIC and Donostia International Physics Center (DIPC), Paseo Manuel de Lardizabal 5, 20018 Donostia - San Sebastián, Spain }

\author{M. Thoss}
 \affiliation{University of Freiburg, Institute of Physics, Hermann-Herder-Str.\,3, Freiburg, Germany.}

 \author{F. Stienkemeier}
 \affiliation{University of Freiburg, Institute of Physics, Hermann-Herder-Str.\,3, Freiburg, Germany.}

\date{\today}

\begin{abstract}
The ultrafast relaxation dynamics of tetracene following UV excitation to the bright singlet state S$_6$ has been studied with time-resolved photoelectron spectroscopy. With the help of high-level \textit{ab-initio} multireference perturbation theory calculations, we assign photoelectron signals to intermediate dark electronic states S$_3$, S$_4$ and S$_5$ as well as a to a low-lying electronic state S$_2$. The energetic structure of these dark states has not been determined experimentally previously. The time-dependent photoelectron yields assigned to the states S$_6$, S$_5$ and S$_4$ have been analyzed and reveal the depopulation of S$_6$ within 60\,fs, while S$_5$ and S$_4$ are populated with delays of about 50 and 80\,fs. The dynamics of the lower-lying states S$_3$ and S$_2$ seem to agree with a delayed population coinciding with the depopulation of the higher-lying states S$_4$-S$_6$, but could not be elucidated in full detail due to the low signal levels of the corresponding two-photon ionization probe processes.
\end{abstract}

\maketitle

\section{Introduction}
Polyacenes, a class of polyaromatic hydrocarbons (PAH) consisting of linearly fused benzene rings, currently receive scientific attention from two different fields. While on the one hand they are promising candidates for organic semiconductor and optoelectronic applications\cite{Anthony2008,Clarke2010, Xue2012}, they are on the other hand, like other PAH also relevant for astrochemistry and astrophysics\cite{Tielens_25_2011, Joblin_PAHs_2011, Omont2016}. Research on polyacene semiconductor applications, including transitors\cite{Li2012}, flexible displays\cite{Rogers2001} and photovoltaics\cite{Xia2017} focuses on their bulk or thin film properties, like charge carrier mobility and ability to undergo singlet fission, i.e. the ability to produce two excited triplet states after the initial excitation of a singlet exciton. 
The progress made with these applications may hide the fact that our understanding of the fundamental molecular properties remains incomplete. Fundamental studies of these properties, and how intermolecular interactions affect them \cite{Izadnia2017,Hoche2021,Bohlen2022,Bogomolov2023} can provide deep physical insights into relevant mechanisms for application-guided designs.
In astrophysics and astrochemistry the interest in PAH stems from their role as carriers of diffuse interstellar bands in the IR region\cite{Tielens_25_2011}, as well as from their importance as reservoir and source of carbon atoms in the interstellar space. Due to these aspects research focuses on IR spectroscopic studies\cite{Colangeli1992,Oomens_lab_2003, Lemmens_farIR_2020} and interactions of PAH with radiation in the UV to XUV range present in the interstellar medium and contributing to ionization and dissociation of these species \cite{ Marciniak2015}. 

From an experimental point of view, most previous spectroscopic studies of tetracene have focused on the bright transition to the lowest electronically excited state and its pronounced vibrational progression\cite{amirav_1979,amirav_1981}. 
Few studies extend far into the UV region of the electromagnetic spectrum, such that the much brighter optical transitions lying in this region have only been studied in crystals, solutions\cite{Tanaka2006} and hot molecular vapors\cite{Morris_1965}, offering only limited spectral resolution.
Recently, a spectrum of parts of the intense absorption feature recorded with a picosecond laser system has been published\cite{Hoche2021}, but without providing spectral assignments.
The electronic structure, composed of a few bright states and many dark states has however been studied theoretically\cite{Canuto1991, Heinze2000,Grimme2003,dorando_targeted_2007,hachmann_radical_2007,sony_large-scale_2007,Marian2008,Knippenberg2010,lopata_excited-state_2011,pelzer_strong_2011,sauri_multiconfigurational_2011,zimmerman_mechanism_2011,casanova_electronic_2014,kurashige_theoretical_2014,yanai_density_2015,casanova_quantifying_2016,yang_nature_2016,bettanin_singlet_2017,bohl_low-lying_2017,plasser_detailed_2017,c_a_valente_excitonic_2021,manian_dominant_2022,pomogaeva_manifestations_2022,dai_impact_2023,dreuw_influence_2023}.
Detailed benchmarking of these calculations against experiments was so far not possible, since most of the calculated states could not be observed experimentally.
Also the application of high levels of theory for such large molecules remains challenging.
Concerning single-photon XUV photoelectron spectroscopy, spectra of tetracene and other polyacenes have been measured\cite{Clark1972,Brogli1972,Schmidt1977,Coropceanu} obtaining the vibronic structure of the tetracene cation.

Only a handful of studies are dedicated to the investigation of non-adiabatic relaxations in neutral PAH excited to high-lying singlet states. 
Noble \textit{et al.}\cite{Noble2019} have investigated the non-adiabatic relaxation of pyrene upon excitation to the third singlet excited state $\mathrm{S}_3$ accessible with a UV photon, revealing consecutive relaxation processes involving several electronic states and various timescales.
Radloff \textit{et al.}\cite{Radloff1997} have studied the ultrafast internal conversion in benzene and its dimer from the excited $\mathrm{S}_2$ state to the $\mathrm{S}_1$ and $\mathrm{S}_0$ states. 
Blanchet \textit{et al.}\cite{Blanchet2008} report a two-color pump-probe experiment in azulene, where a monoexponential decay of several tens of picoseconds is attributed to the relaxation from the $\mathrm{S}_4$ state to the $\mathrm{S}_2$ and to the ground state via highly excited vibrational states. 
Further femtosecond photoionization studies of PAH including acenes\cite{Kjellberg2010,Johansson2013,Johansson2014} have revealed peculiarities in the ionization mechanisms, including the emission of thermal electrons, caused by the rapid heating of the electronic subsystem of such molecules, with much slower equilibration with nuclear degrees of freedom.
Coronene\cite{bohl_low-lying_2017} as well as azulene and naphtalene\cite{Kuthirummal2003} have been studied with femtosecond photoelectron spectroscopy, revealing the energetic structure of Rydberg states and their sensitivity to the molecular species. Internal conversion between highly excited valence states, Rydberg states and super excited states seem ubiquitous in aromatic molecules\cite{Schick2001,Kuthirummal2003,Blanchet2008,Spesyvtsev2012} and the mixing between states with valence and Rydberg character influences the non-adiabatic dynamics in various systems \cite{Reisler2009}. 
The conversion of highly excited states of polyacenes to low-lying ones has been recently modelled with a computational approach combining density functional theory and trajectory surface hopping calculations \cite{Posenitskiy_2019, Posenitskiy_2020}. 

In this article we present an experimental time-resolved photoelectron spectroscopy study of the relaxation of the tetracene molecule excited to a bright state in the UV, supported by high-level \textit{ab-initio} multireference perturbation theory calculations.
The electronic relaxation of highly excited electronic states of isolated PAH molecules is a potential pathway for the production of vibrationally hot ground state molecules that are interesting for their potential importance as interstellar emitters of infrared radiation. 
Furthermore, the general understanding of the electronic structure and dynamics of isolated PAH molecules, including the involvement and characterization of dark states can be a prerequisite for the comprehension of collective molecular effects, like those involved in singlet fission processes. 
The quantitative comparison of experimental and theoretical methods provides a valuable check for their respective performance on these complex molecular species, and paves the way for future studies.

\section{Methods}
\subsection{Experimental}

Most aspects of the experimental setup for femtosecond time-resolved photoelectron velocity map imaging (VMI) used in this study have been described previously \cite{rendler2021}. 
An effusive molecular beam of tetracene was produced by a radiatively heated oven at 135\textdegree C.
The tetracene molecules were excited and ionized by consecutive femtosecond laser pulses intersecting the effusive beam at a right angle in the ionization region of a VMI spectrometer. 
The femtosecond laser pulses, centered around 400\,nm (VIS, 2.6\,nm FWHM) and 267\,nm (UV, 1.7\,nm FWHM) were created by collinear second and third harmonic (SH, TH) generation of 1.7\,W of the output of a regenerative titanium-sapphire amplifier (Coherent Legend, 5\,kHz, 2.5\,W, 125\,fs). The second and third harmonic were generated in BBO crystals of 0.5\,mm and 0.2\,mm thickness. In between the two crystals the group delay between the SH and the fundamental was compensated by a calcite plate and the polarization of the fundamental was rotated by 90° using a dual waveplate.
The VIS and UV beams were separated from each other and the remaining fundamental, before the TH pulse was delayed on a variable delay stage.
The cross correlation between the VIS and UV pulses was determined by measuring the time-resolved two-color photoelectron spectrum of effusive sodium atoms via the 6p state. The step-like signal increase was fitted with an error function, obtaining a cross correlation of 83$\pm$5 fs ($\sigma$), corresponding to a FWHM of 195$\pm$12 fs, see details in the supplementary material.
The position of the zero delay was determined directly from the obtained tetracene signals (see below).
The power of each color was adjusted with a $\lambda$/2 plate and polarizer, with the polarizer setting the horizontal polarization, i.e.\ parallel to the molecular beam axis and perpendicular to the extraction axis of the VMI spectrometer.
Both colors were focused into the interaction region of the VMI spectrometer with individual lenses (focal length 75\,cm) and overlapped collinearly using a dichroic mirror.
The beam profile at the focal point was measured with a camera 
to be Gaussian with a beam diameter of 40\,$\mu$m for the TH and 55\,$\mu$m for the SH. With these values we obtain intensities between 2.5 and $3\cdot 10^{11}\,\mathrm{W}\,\mathrm{cm}^{-2}$, for a UV power of 2.5\,mW, and between 3 and $3.5\cdot10^{11}\,\mathrm{W}\,\mathrm{cm}^{-2}$, for a VIS power of 5\,mW. 
The obtained photoelectron images were reconstructed following published procedures to obtain photoelectron spectra\cite{Dick_2014}. 
The energy calibration of the VMI spectrometer was verified using two-photon ionization signals of an effusive beam of sodium atoms, and the energy resolution was determined to be about 2\% at roughly 75\% of the detector radius, see supplementary material.
For the time-resolved two-color photoelectron signals (267\,nm pump, 400\,nm probe), the single-color photoelectron spectra of the VIS and UV pulses were subtracted. These single-color signals corresponded to 14\% and 23\%, respectively, of the two-color electron yield for zero delay between the two pulses. Also, we subtract an additional weak two-color signal arising from excitation by the 400\,nm pulse and subsequent ionization with the 267\,nm pulse that could also be observed in the data. The latter feature was determined at long negative delays (VIS before UV) and assumed to show a transient signal level described by an error function with a width determined by an estimated cross-correlation of 195\,fs (FWHM, see supplementary material for details on the background subtraction).

\subsection{Computational}
For the characterization of the electronic spectra of tetracene and its radical cation first we have obtained the equilibrium structures enforcing D$_{\textnormal{2h}}$ symmetry, using density functional theory (DFT) with the B3LYP exchange-correlation functional\cite{vosko_accurate_1980,becke_density-functional_1988,lee_development_1988,becke_density-functional_1992,becke_density-functional_1993}, and a def2-TZVP basis set\cite{weigend_balanced_2005}. Dispersion interaction corrections were applied following the DFT-D3 method of Grimme \textit{et al.}\cite{grimme_consistent_2010} with Becke-Johnson damping\cite{grimme_effect_2011}. A frequency analysis was carried out to ensure that the equilibrium structure is a minimum on the potential energy surface. The calculations described above were carried out with the Turbomole V7.5 package\cite{noauthor_turbomole_nodate,balasubramani_turbomole_2020}.

The vertical excitation energies of the neutral tetracene and its radical cation were calculated employing the CASPT2/CASSCF level of theory\cite{andersson_multiconfigurational_1993,helgaker_molecular_2000}. 
The reference wave functions in D$_{\textnormal{2h}}$ symmetry were obtained using 15 roots equal weights state-average (SA) CASSCF calculations with an active space of 16 electrons and 16 $\pi$-like orbitals  employing an ANO-L-VTZP basis set\cite{widmark_density_1990,widmark_density_1991,pou-amerigo_density_1995,pierloot_density_1995} (selected after calibration, see supplementary material for further details). The CASPT2 calculations were carried out using an imaginary shift of 0.05\,au\cite{forsberg_multiconfiguration_1997} to mitigate the impact of intruder states. An IPEA shift of 0.25\,au was used for the zero order Hamiltonian\cite{ghigo_modified_2004}. The oscillator strengths were calculated using the CASSCF transition dipole moments and the CASPT2 energies.

In addition, the Dyson orbitals were calculated, as described in ref.~\cite{tenorio_photoionization_2022}, using the overlap equation:
\begin{equation}
\label{eq:overlap}
	\phi_{IF}^d(x_1) = \sqrt{N}\int{\Psi_F^{N-1}(x_2,\cdots,x_N)\Psi_I^N(x_1,x_2,\cdots,x_N)dx_2\cdots dx_N}
\end{equation}
with the initial and final wavefunctions $\Psi_I^N$ and $\Psi_F^{N-1}$ of the $N$-electron system and its cation, respectively. These calculations were carried out using MOLCAS 8.4\cite{aquilante_span_2016} and OpenMOLCAS~v20.10\cite{fdez_galvan_openmolcas_2019}.

\section{Results and discussion}

\subsection{Single-color and two-color photoelectron spectra}
Figure 1 a) depicts a single-color photoelectron spectrum recorded with a UV femtosecond laser pulse centered around 267\,nm. The spectrum shows a structured plateau region between 1.5 and 2.5\,eV electron kinetic energy and a broad unstructured band below electron energies of 1\,eV.
The feature at the highest electron kinetic energy of 2.3\,eV (label A) is in agreement with direct two-photon ionization (2 $\times$ 4.64\,eV) to the cationic ground state (IP 6.97\,eV\cite{Schmidt1977}) while the onset of the low energy feature at around 0.9\,eV is in agreement with two-photon ionization to the first two electronically excited cationic states, both located at 8.41\,eV\cite{Brogli1972,Clark1972,Schmidt1977,Szczepanski}.

\begin{figure}
\centering
  \includegraphics[width=0.5\textwidth]{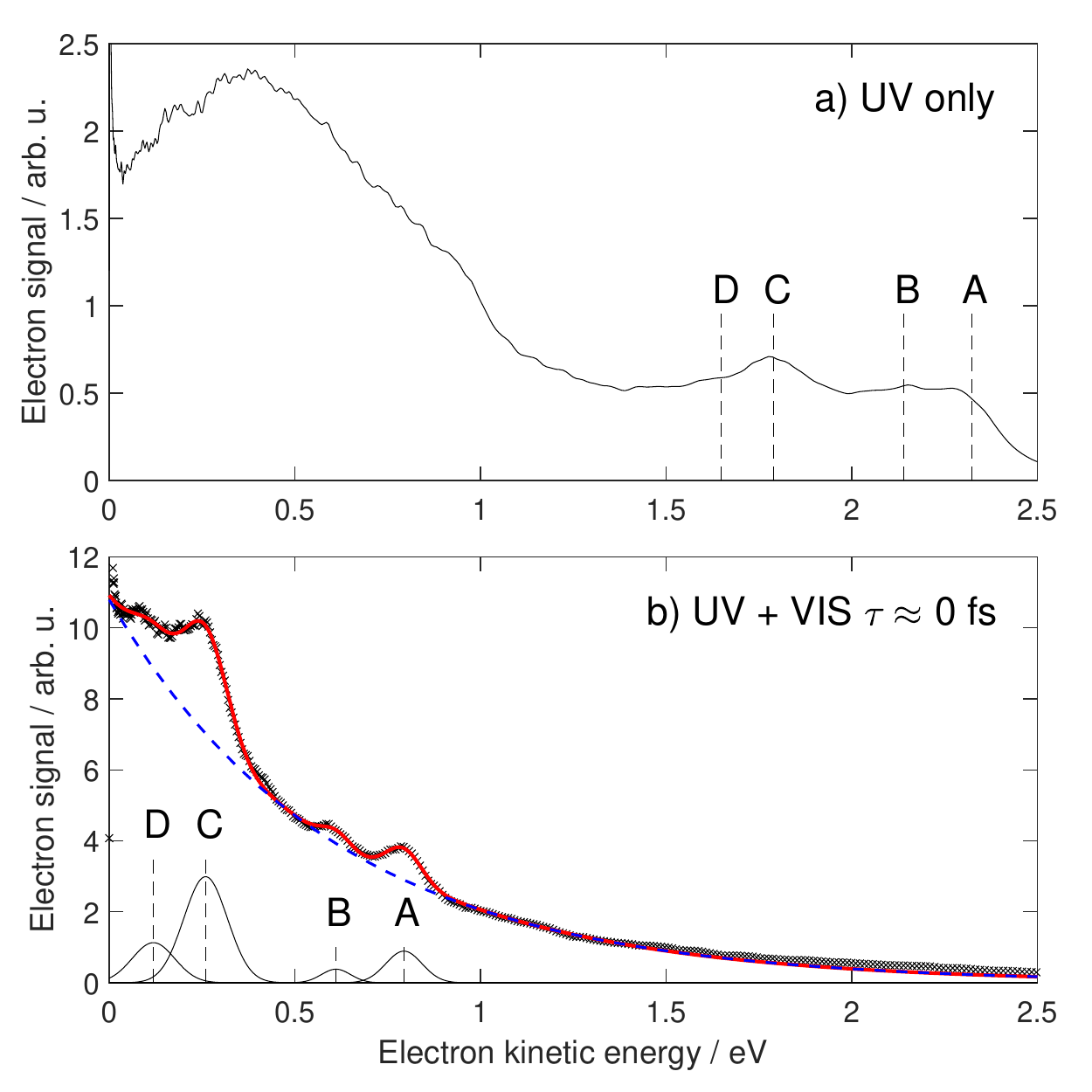}
  \caption{a) One-color two-photon photoelectron spectrum of tetracene, recorded with a single fs laser pulse centered around 267\,nm (4.64\,eV). b) Two-color two-photon photoelectron spectrum recorded for temporally overlapping ($\tau\approx0$\,fs) laser pulses centered at 267 and 400\,nm (4.64 and 3.1\,eV). The red full line in b) shows a least squares fit to the experimental data (symbols) with a model consisting of Gaussian peaks (thin black lines) and an exponential background (blue dashed line).}
  \label{fig:UV_pp_2_delays}
\end{figure}
The photoelectron spectra obtained by overlapping both laser pulses (UV and VIS) in the ionization region are depicted in figure 1 b) for a delay $\tau\approx0$\,fs.
The same features that are observed on the structured plateau of figure~1~a) (labels A-D) are visible in the two-color photoelectron spectra, shifted according to the lower ionizing photon energy.
The photoelectron spectra in panel b) shows a prominent exponential background, which is caused by thermal electron emission previously observed for other PAH interacting with intense fs laser pulses\cite{Kjellberg2010,Johansson2013,Johansson2014,bohl_low-lying_2017}. This autoionization process occurs after the absorption of several photons creates a hot electronic subsystem that cannot equilibrate with the nuclear degrees of freedom on this ultrafast timescale.
Close inspection of the photoelectron spectrum in figure 1 a) shows that the thermal electron emission is also present in the UV only spectrum, albeit significantly less pronounced.
To extract reliable positions of the observed photoelectron features as well as delay-dependent intensities in the presence of the pronounced exponential background, we have performed a least squares fit of the two-color spectra employing a model consisting of an exponential background and a number of Gaussian bands:
    \begin{equation}
        \label{eq:thermal}
        f_{PE}=A\,e^{-\frac{\epsilon}{k_BT}} + \sum_{n_{peaks}=5} a_n\,e^{\left(\frac{-(\epsilon-\epsilon_n)}{\sqrt{2}\sigma_n}\right)^2}
    \end{equation}
\begin{figure}
\centering
  \includegraphics[width=0.5\textwidth]{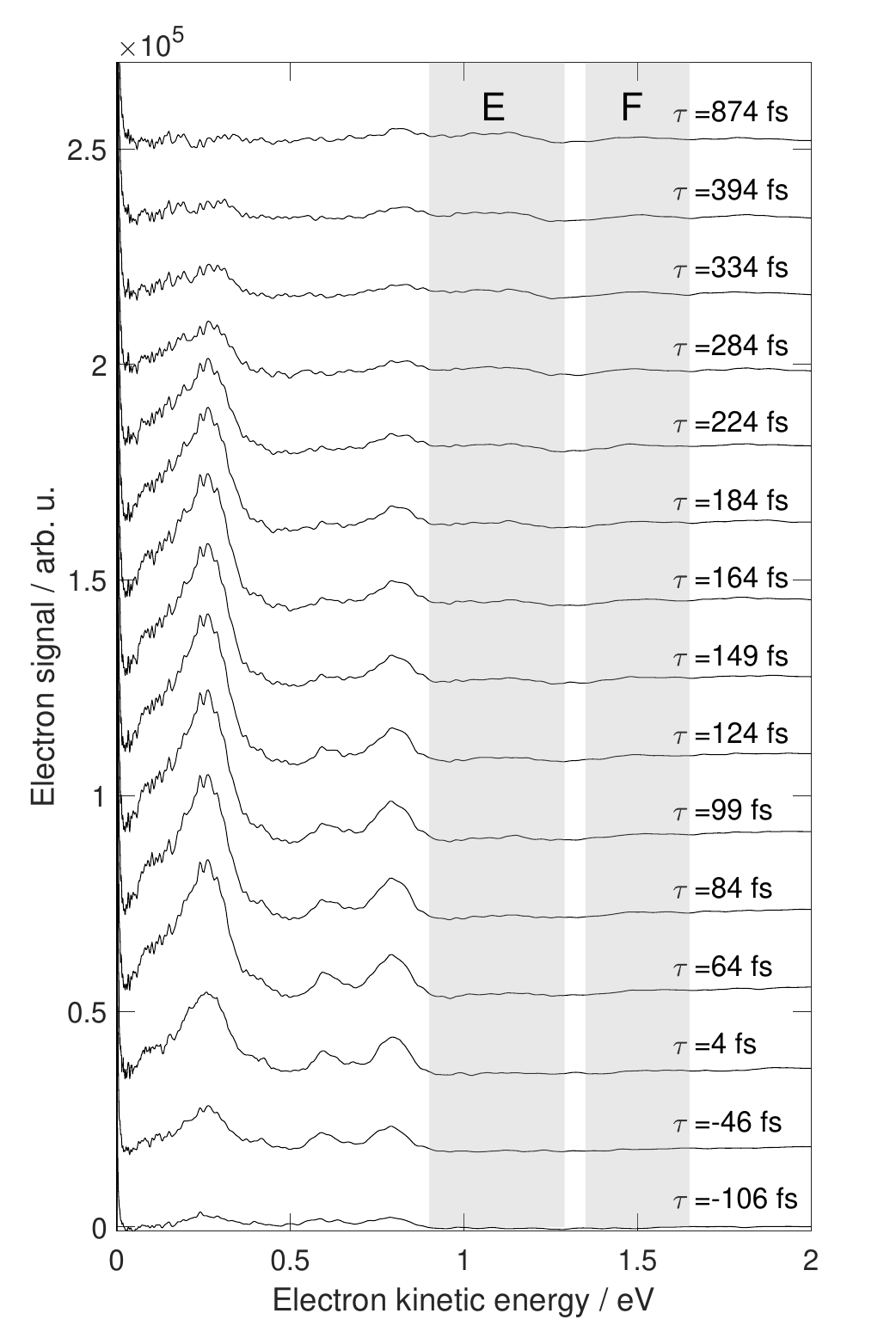}
  \caption{Photoelectron spectra recorded for selected time delays between the UV and VIS pulses, after subtraction of exponential signal contributions obtained from the least-squares fit. Gray-shaded areas labeled E and F indicate the position of two weak features arising at positive delay times.}
  \label{fig:pp-exp}
\end{figure}
Below 0.03 eV there is a signal of zero kinetic energy electrons and above 1.4 eV there are weak features in addition to the exponential background (see discussion below). Since the fit model does not include these low and high kinetic energy features, we only fit data points between 0.03 and 1.4\,eV. The fit is performed in a global fashion, such that for all pump-probe delays only one global set of five Gaussian band positions and band widths is fitted, while the amplitudes of the Gaussians and of the exponential background are fitted for each delay individually. 
The resulting fit for zero time delay is included in figure 1 b) as a solid red line with the exponential and Gaussian components shown separately as a blue dashed and a black solid line, respectively. Vertical dashed lines indicate the positions of the Gaussian bands. The obtained fits for photoelectron spectra at other pump-probe delays are displayed in the supplementary material.
The fitting procedure retrieves band positions $E_\mathrm{PE}^{\mathrm{A/B/C/D}}$ of 0.79\,eV, 0.61\,eV, 0.26\,eV and 0.12\,eV for the photoelectron bands A-D. We estimate the uncertainties of these values to be 15 meV, including the accuracy of the fitting procedure, the energy calibration as well as the spectrometer resolution.
The fit obtains a position of 1.04 eV for the fifth Gaussian band, which shows a non-zero intensity only at larger positive time delays (see supplementary material). Due to the large uncertainties in the parameters of this last Gaussian band we consider its values less reliable than those of the other four bands and exclude this band from the quantitative discussion.

Feature A ($E_\mathrm{PE}^\mathrm{A}$=0.79\,eV) is in agreement with a state of excitation energy 4.66 eV ($E_\mathrm{ex}^\mathrm{A}=\mathrm{IP}+E_\mathrm{PE}-\nu_\mathrm{probe}$) in good agreement with the used UV photon energy.
A possible explanation of the other bands is the population of several lower-lying neutral excited states of tetracene, occurring via non-adiabatic relaxation of the molecule within the duration of the fs laser pulses. 
While absorption studies did not report electronic states between the brightest state in the UV ($\approx$ 4.82 eV)\cite{Morris_1965} and the lowest-lying bright state S$_1$ (2.77 eV)\cite{Morris_1965,amirav_1979,amirav_1981}, different theoretical works\cite{Grimme2003,Marian2008,Knippenberg2010,kurashige_theoretical_2014} predict a number of dark states in that region as well, supporting this explanation.

Figure \ref{fig:pp-exp} shows the two-color photoelectron spectra for different delays between the excitation (UV) and ionization (VIS) pulses. For better visibility of the individual bands the exponential signal component obtained from the fitting procedure has been removed and the spectra are separated by vertical offsets. The bands labeled A-D above show a pronounced time-dependence in agreement with the transient population and depopulation of lower-lying states by non-adiabatic dynamics. 
In addition to these bands, signals rising toward later delay times seem to appear in the two gray-shaded areas labeled E and F. 
Due to their weak intensities, we do not try to describe these bands quantitatively using the fit model.
Neutral states lying higher in energy than the initially excited state (4.66\,eV) cannot be populated over time. However, the features E and F located at $E_\mathrm{PE}^{\mathrm{E}}$=1.1\,eV and $E_\mathrm{PE}^{\mathrm{F}}$=1.5\,eV, respectively, can be explained by lower lying excited states ionized by two probe photons from the VIS pulse. 

\subsection{Comparison with calculations}
\begin{figure}
\centering
  \includegraphics[width=0.5\textwidth]{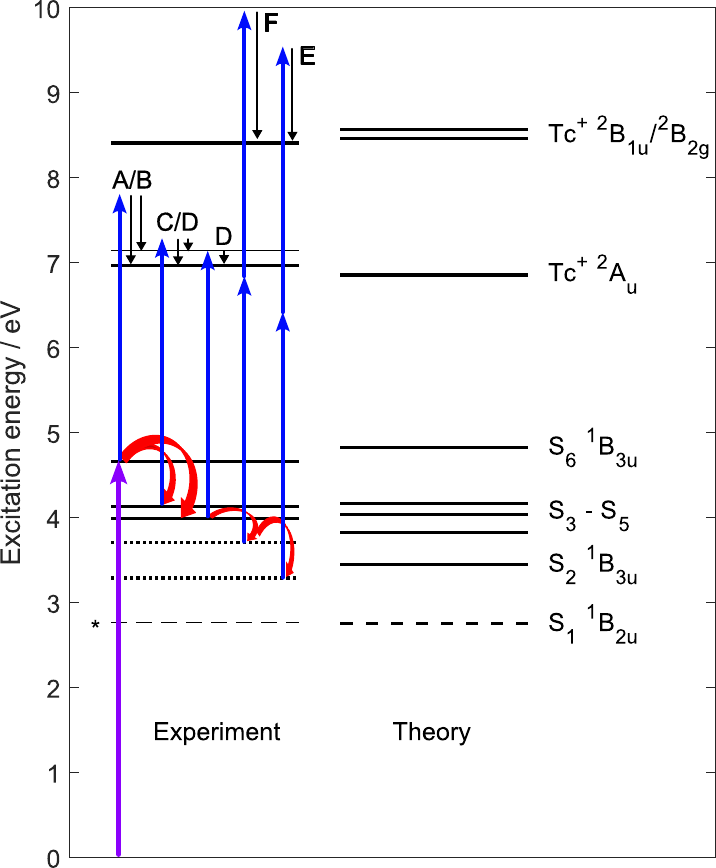}
\caption{Energy level diagram obtained from our photoelectron data and CASPT2 calculations. The additional line close to the Tc$^+$ $^2$A$_\textrm{u}$ state indicates the first vibrationally excited state of the cation. Dotted lines indicate states deduced from weak features tentatively assigned to two-photon ionization to excited cationic states, while dashed lines indicate data for the S$_1$ state. Experimental excitation, relaxation and ionization pathways are indicated schematically. $\ast$ Value taken from Ref.~\cite{amirav_1979}.}
  \label{fig:assignment}
\end{figure}

The calculated vertical state energies of neutral tetracene and their respective oscillator strengths $f$ to the ground state S$_0$ are shown in table \ref{tab:comp}. Furthermore, table~\ref{tab:comp_cat} provides the energies of the three lowest-lying doublet states of the radical cation, calculated at the equilibrium structure of the tetracene cation, and the Dyson intensities between the respective neutral and cationic states. For comparison, we also give the experimental values according to our assignments. These experimental values have been obtained from the fitted electron kinetic energy band position (E$_\mathrm{PE}^\mathrm{A/C/D}$) of features A, C, and D, the known ionization potential of (IP = 6.97 eV\cite{Schmidt1977}) and the probe photon energy ($\nu_\mathrm{probe}=3.1$ eV) as $E_\mathrm{ex}^\mathrm{A/C/D}=\mathrm{IP}+E_\mathrm{PE}-\nu_\mathrm{probe}$.
For features E and F the corresponding expression takes into account the ionization energy to the two lowest-lying excited cationic states  (IP$_2$ = 8.41 eV\cite{Schmidt1977,Szczepanski}), and twice the probe photon energy to account for the two-photon ionization process, $E_\mathrm{ex}^\mathrm{F}=\mathrm{IP_2}+E_\mathrm{PE}-2\nu_\mathrm{probe}$. Note, that assuming two-photon ionization to the cationic ground state would yield unphysical negative excitation energies for the features E and F. 
See also figure \ref{fig:assignment} for a sketch of the ionization scheme, and energy level diagram.
\begin{table}
    \caption{\label{tab:comp}Comparison of experimentally observed photoelectron band positions E$_\mathrm{PE}$, excitation energies to which these states correspond E$_\mathrm{ex}$ and calculated excitation energies E$_\mathrm{theo}$, oscillator strengths \textit{f} and symmetries of the six lowest-lying singlet excited electronic states of tetracene.}
\begin{ruledtabular}
    \begin{tabular}{c|ccc|cc}
    Left\footnote{a: value taken from Amirav \textit{et al.} \cite{amirav_1981}}
    & \multicolumn{3}{c}{experiment}&\multicolumn{2}{c}{theory}\\
    state &  feature&    E$_\mathrm{PE}$ / eV&    E$_\mathrm{ex}$ / eV& E$_\mathrm{theo}$ / eV ($f$)&   symmetry\\\hline
    S$_1$   &      &    &    2.77$^a$ & 2.75 (0.081) &     B$_\mathrm{2u}$ \\
    S$_2$   &E     & 1.1   &   3.3 & 3.45 (0.002) &     B$_\mathrm{3u}$  \\
    S$_3$   &F    & 1.5  &  3.7 & 3.83 (0.000) &     A$_\mathrm{g}$ \\
    S$_4$   &D     & 0.12   &   3.99 & 4.04 (0.000) &     B$_\mathrm{1g}$ \\
    S$_5$   &C &  0.26 &  4.13 & 4.17 (0.000) &     B$_\mathrm{1g}$\\
    S$_6$   &A/B    & 0.79   &   4.66 & 4.83 (2.732) &     B$_\mathrm{3u}$ \\
    \end{tabular}
    \end{ruledtabular}

\end{table}

The largest oscillator strength was calculated for the S$_6$ state, which indicates that it is the high-lying bright state. Moreover, the energy of 4.83 eV calculated for the S$_6$ state agrees well with the maximum of the absorption spectrum observed for tetracene vapor ($\approx$ 4.82\,eV)\cite{Morris_1965}. Our experimental value of 4.64 eV, determined by the choice of our UV photon energy, is located in the rising flank of this strong absorption band, consisting of an unresolved broad vibrational progression. Thus, it is clear that our UV excitation populates a set of vibrational states of the S$_6$ state, below its vertical excitation. This excitation gives rise to the feature A in the photoelectron spectrum upon ionization with the VIS probe pulse.
The experimentally observed feature B would correspond to an excitation energy of 4.48\,eV which is not in agreement with any of the calculated neutral excited states. 
Feature B is located 0.18\,eV below feature A, and can thus be explained as an ionizing transition from the same initial state as feature A toward the first vibrationally excited state of the cationic ground state. 
The vibrational splitting observed in XUV photoelectron spectra of ground state tetracene was reported to be 0.17\,eV \cite{Clark1972,Schmidt1977,Coropceanu}, which is in good agreement with our observed splitting of 0.18\,eV.

A similar splitting of 0.14\,eV is also observed between the states deduced from features C and D. However, in this case the calculations also show two dark states S$_5$ and S$_4$ of B$_\mathrm{1g}$ symmetry (4.17\,eV and 4.04\,eV) in excellent agreement with the experimental values of 4.13\,eV and 3.99\,eV. The separation between the features C and D, which deviates slightly from the expected vibrational splitting, favors an assignment as two separate electronic states S$_5$ and S$_4$. 
If the bands C and D are nevertheless caused by two ionizing transitions of the same intermediate state, this state could either be the S$_5$ or the S$_4$ state. The small difference in the calculated excitation energy does not allow an unambiguous assignment. 
The unresolved nature of the photoelectron spectra (figure \ref{fig:pp-exp}) does not allow a final conclusion, and would even support the combination of both possible explanations.
The absence of an observable ionizing transition toward the vibrationally excited cation would indicate a similarity between the equilibrium geometries of states S$_4$ and S$_5$ to the cationic equilibrium geometry. As a consequence, vibrational energy that is created in the neutral relaxation process will be transferred to the cationic ground state without being significantly affected by the ionization process.
The features will be further discussed below in the analysis of the dynamical photoelectron yields.

The excited states deduced from the bands E and F located at excitation energies of 3.3\,eV and 3.7\,eV agree well with the calculated values for the S$_2$ (B$_\mathrm{3u}$ symmetry) and S$_3$ (A$_\mathrm{g}$ symmetry) states. 
To the best of our knowledge none of the states S$_2$-S$_5$ have previously been observed and assigned experimentally. While the transitions from the ground state to the states S$_3$-S$_5$ are forbidden by symmetry and expected to be dark, the transition to the S$_2$ state is expected to be weakly allowed, which agrees with its calculated oscillator strength.
For the S$_1$ state the reported experimental value of 2.77 eV \cite{amirav_1979,amirav_1981} is in excellent agreement with our theoretical result of 2.75 eV. Furthermore, the vertical excitation energies of S$_1$ and S$_2$ agree very well with the reported energies in previous theoretical works~\cite{Marian2008,Grimme2003,bettanin_singlet_2017}.
The lowest-lying adiabatic ionization potentials of the radical cation agree well with the previously measured experimental values (further information about the vertical excitation energies is given in the supplementary material).
The calculation of the Dyson intensities gives insight about likely transitions between the neutral and cationic states of the molecule~\cite{arbelo-gonzalez_steady_2016}. The intensities of D$_\mathrm{S_1}$, D$_\mathrm{S_2}$, D$_\mathrm{S_4}$ and D$_\mathrm{S_6}$ to the cationic Tc$^{+}$ $^2$A$_\mathrm{u}$ state are much larger than D$_\mathrm{S_3}$ and D$_\mathrm{S_5}$. 
The values of D$_\mathrm{S_4}$, D$_\mathrm{S_5}$ and D$_\mathrm{S_6}$ as well as the probe photon energy of 3.1 eV emphasize the one-photon ionization process from S$_6$, S$_5$ and S$_4$ governing feature A/B and C/D, respectively. 
The Dyson intensities for the states S$_2$ and S$_3$ show that the ionization toward an excited cationic state is more likely than ionization to the cationic ground state. This supports our assignment of the features E and F as two-photon ionization signals of these states to the excited cationic states. The very low value of D$_\mathrm{S_3}$ for the cationic ground state Tc$^+$~$^2$A$_\mathrm{u}$ rationalizes that a corresponding two-photon ionization signal toward the cationic ground state cannot be observed in the photoelectron spectra. For the S$_2$ state the absence of such a photoelectron band (around 2.5\,eV) is less clear from the values of D$_\mathrm{S_2}$ and has to be ascribed to the very low signal levels.

\begin{table}[ht]

 	\caption{\label{tab:comp_cat}Comparison of the experimentally observed ionization potentials E$_\mathrm{IP}$\cite{Schmidt1977} to the calculated energies at the equilibrium structure of the radical cation E$_\mathrm{IP}^\mathrm{theo}$ as well as the Dyson intensities $\mathrm{D}$ from different neutral states to the respective cationic state of tetracene.}
 \begin{ruledtabular}
	\begin{tabular}{c|ccc}
		 &     Tc$^{+}$ $^2$A$_\mathrm{u}$& Tc$^{+}$ $^2$B$_\mathrm{1u}$ & Tc$^{+}$ $^2$B$_\mathrm{2g}$  \\\hline
		E$_\mathrm{IP}$ / eV & 6.97 & 8.41 & 8.41 \\
		E$_\mathrm{IP}^\mathrm{theo}$ / eV & 6.857 & 8.468 & 8.565 \\\hline
		D$_\mathrm{S_1}$ & 0.427 & 0.014 & 0.341 \\
		D$_\mathrm{S_2}$ & 0.200 & 0.212 & 0.005 \\
		D$_\mathrm{S_3}$ & 0.008 & 0.013 & 0.519 \\
		D$_\mathrm{S_4}$ & 0.240 & 0.007 & 0.026 \\
		D$_\mathrm{S_5}$ & 0.088 & 0.001 & 0.061 \\
		D$_\mathrm{S_6}$ & 0.230 & 0.221 & 0.004 \\
	\end{tabular}
  \end{ruledtabular}

\end{table}

\subsection{Relaxation dynamics}

\begin{figure}
\centering
  \includegraphics[width=0.5\textwidth]{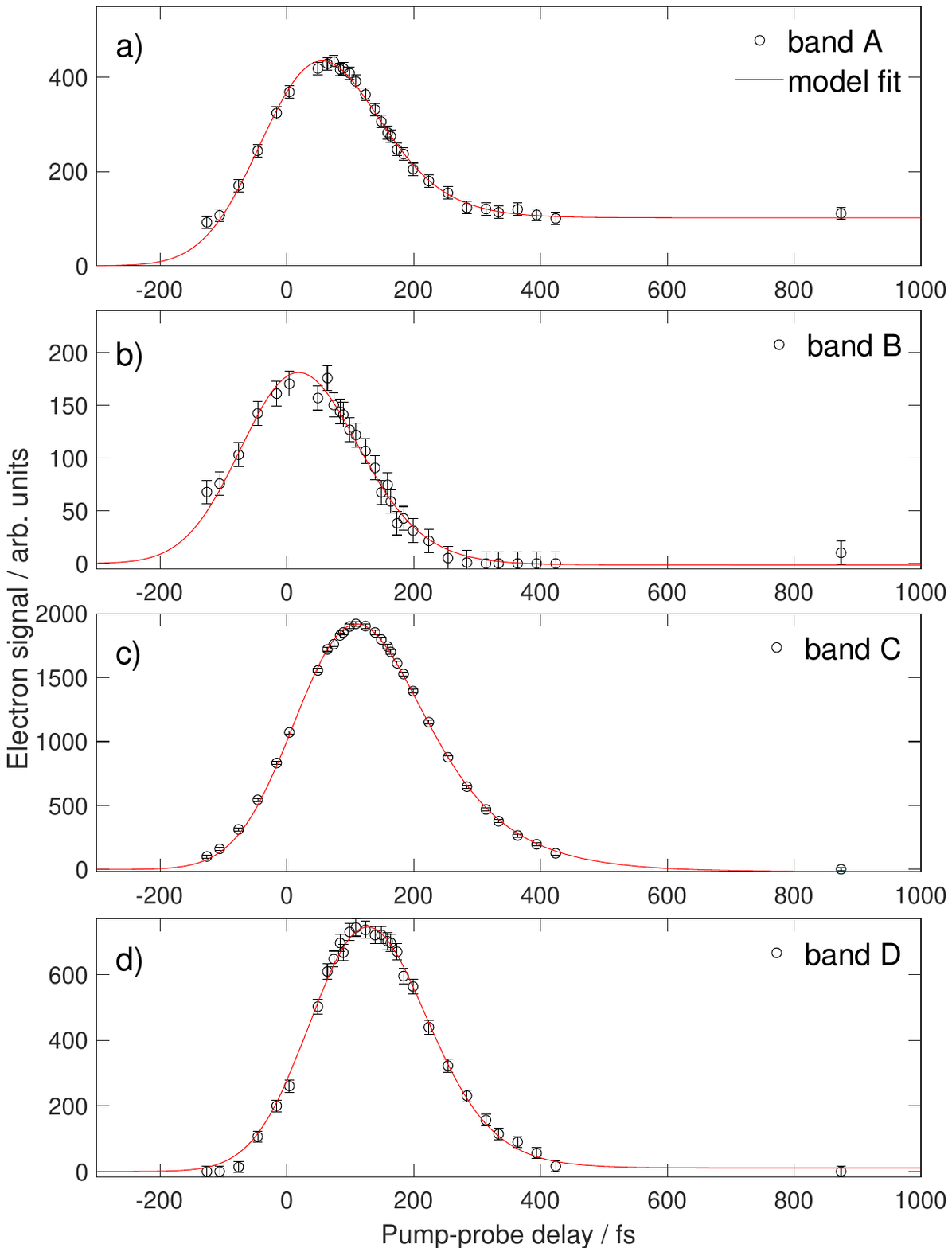}
  \caption{\label{fig:dynamics}Time-dependent photoelectron yields of the features A-D. The red solid lines are model fits obtained using a step function with an exponential decay, broadened by the experimental cross-correlation between the two laser pulses, see equation \ref{eq:fit}.}
\end{figure}

The fitting procedure described above was used to retrieve individual time-dependent photoelectron yields for four different Gaussian bands (A-D) observed in the two-color photoelectron spectra. For the two weaker contributions (E and F) an unambiguous retrieval of amplitudes was not possible, and we therefore refrain from discussing the details of the dynamics of these states further.
The resulting photoelectron intensities are displayed as a function of the delay between pump and probe pulses in figure \ref{fig:dynamics}.

The yield of band A is maximal close to a nominal pump-probe delay of zero, and shows an approximate Gaussian shape with a plateau for positive pump-probe delays $\tau>$300\,fs. The almost Gaussian appearance is an indication that the shape is dominated by the experimental cross-correlation of the laser pulses of about 195\,fs (FWHM).
The yield of band B shows a similar Gaussian shape but its maximum seems slightly shifted to earlier pump-probe delays and there is no plateau at positive pump-probe delays. 
The overall data quality for band B is lower, as shown by the errorbars ($\pm\sigma$, estimated from the covariance matrix of the model fit), due to the lower signal levels of this band.

Band C and D show maximal photoelectron yield at slightly more positive pump-probe delays. 
Also both time-dependent photoelectron yields deviate from a Gaussian shape by a slight asymmetry toward positive delay times, in agreement with a lifetime comparable to or longer than the cross-correlation of the laser pulses. The curve of the transient photoelectron yield of band D appears slightly more narrow than the one for band C, which indicates a shorter lifetime. 
To asses the observed differences more quantitatively, we fit all four transient photoelectron yields with a fit model according to 
\begin{multline}
\label{eq:fit}  
f(t)=Y_1\cdot e^{-\frac{t-\Delta t}{\tau}}\cdot\left(1+\mathrm{erf}\left({\frac{t-\Delta t}{\sqrt{2}\tau_{cc}}}-\frac{\tau_{cc}}{\sqrt{2}\tau}\right)\right)\\+Y_2\cdot\left(1+\mathrm{erf}\left({\frac{t-\Delta t}{\sqrt{2}\tau_{cc}}}\right)\right).
\end{multline}
$Y_i$ describe amplitudes, $\tau$ is the decay time of a state and $\tau_{cc}$ is the cross-correlation time. An additional time shift $\Delta t$ is explicitly included to account for and quantify the apparent shifts of the individual signals arising from the delayed population of lower-lying states. Note, that the nominal delay axis has been shifted such that the fitting procedure for band A obtains a value of $\Delta t=0$. The second error function term is included to account for possible plateaus, see figure \ref{fig:dynamics} a).
The cross-correlation time $\tau_{cc}$ was fixed in the fitting procedure to the experimentally determined value of 83$\pm$5 fs. 
To estimate the uncertainties of the fitted parameters we varied the fixed value of the cross correlation constant $\tau_{cc}$ within its experimental uncertainties. The resulting error bars exceed the errors estimated from the fitting routine. 

The results of the fitting procedure for the bands A-D are summarized in table \ref{tab:param}.
\begin{table}
\caption{\label{tab:param}Obtained fit parameters for the transient photoelectron yields fitted individually. The cross-correlation was fixed at 83 fs (FWHM= $2\sqrt{2\mathrm{ln}2} \tau_{cc}$ = 195$\pm$12 fs).}
\begin{ruledtabular}
\begin{tabular}{l|llll}
band&$\tau$ / fs& $\tau_{cc}$ / fs& $\Delta t$ \\\hline
A&  58$\pm$15& 83$\pm$5& 0$\pm$10\\
B&  54$\pm$12&83$\pm$5& -24$\pm$8\\
C&96$\pm$13& 83$\pm$5 & 47$\pm$6\\
D&55$\pm$17& 83$\pm$5 & 83$\pm$12 \\
\end{tabular}
\end{ruledtabular}

\end{table}
The fitted parameters quantitatively describe what could already be observed in figure \ref{fig:dynamics}. Within errorbars, band A and B have the same decay times $\tau$, shorter than $\tau_{cc}$. Band B shows a negative $\Delta t$, however the difference of $\Delta t$ between bands A and B is almost contained in the combined uncertainties of the two values.
Band C has a $\Delta t$ of 47 fs and a decay time $\tau =$96\,fs, slightly longer than the cross-correlation time $\tau_{cc}$, and band D is with $\Delta t=$83\,fs even further shifted to positive delay times and has a decay time comparable to those of band A and B.

\begin{figure}
\centering
  \includegraphics[width=0.5\textwidth]{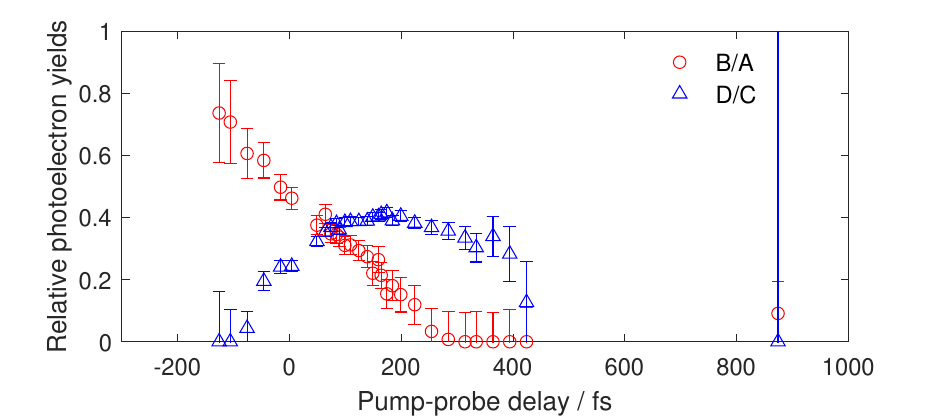}
  \caption{Time-dependent relative photoelectron yields of bands B/A and D/C.}
  \label{fig:FC}
\end{figure}

Bands A and B have above been assigned to the same initially excited electronic state, differing only by the vibrational excitation of the cationic ground state produced upon ionization. As such, we would expect the corresponding photoelectron yields to show an identical time-dependence. There is, however, a slight difference in the observed $\Delta t$ as well as the presence of a plateau in the yield of band A at large delay times, which is not present for band B.
The plateau observed for band A at large delays can arise for example from a part of the population becoming trapped in the state. However, in this case we would expect the plateau to be present for band B as well, unless the trapped part of the population would assume a nuclear geometry very close to the cationic equilibrium structure. In this case the Franck-Condon factor, governing the relative intensity of band B could become very small.
Figure \ref{fig:FC} shows the signal of band B relative to the signal of band A, decreasing approximately linearly with the delay time. This behavior is in principle in agreement with a nuclear relaxation process of the tetracene molecule in its initially excited state toward the cationic equilibrium geometry. It seems likely, that the equilibrium geometry of the $S_6$ state lies close to the equilibrium geometry of the cationic ground state, thus explaining this behavior.

Alternatively, the plateau in the yield of A may be explained by the internal conversion of a part of the initial population to a Rydberg state. An s-type Rydberg state has been computed to lie at an excitation energy of 4.92\,eV\cite{bohl_low-lying_2017} in the vicinity of the initially excited S$_6$ state. This Rydberg state could have a nuclear geometry closer to the geometry of the cationic ground state. Identical to the argumentation above this would avoid significant Franck-Condon factors for the ionization to the vibrationally excited cationic state, and thus explain the persistence of band A in the absence of band B at long delays. However, the almost isotropic photoelectron angular distributions (see supplementary material) show no change with pump-probe delay in the region around 0.79\,eV, which provides a strong argument against the involvement of an s-type Rydberg state.

The differences between the yields of bands C and D are, according to the fitted results, more significant than the differences between bands A and B. These differences include the shorter decay time $\tau$ as well as a larger value of $\Delta t$ of band D.
These differences between the two bands corroborate two different electronic states as the reason for the two photoelectron bands. The positive value of $\Delta t$ points to the delayed population of these states as the initially excited state is depopulated.  
An alternative explanation including two vibrational states of the cationic ground states like for bands A and B seems unlikely.
This is also reflected in the relative Franck-Condon factors depicted in figure \ref{fig:FC}. 
The ratio of the yields of bands C and D shows a complex behavior, which cannot easily explained by the motion of a nuclear wavepacket on a potential energy surface.
In combination with the band positions, we thus conclude that an assignment as two individual electronic states S$_4$ and S$_5$ seems more plausible than the interpretation as two ionizing transitions from the same electronic state. 
Nevertheless, it is not possible to fully rule out the latter explanation or a combination of the two.  

In summary, our data is in agreement with an initial excitation of the high-lying bright state S$_6$, which subsequently decays non-adiabatically via three dark states S$_5$, S$_4$ and S$_3$, into the state S$_2$. Most likely, the tetracene molecules decay in a final non-radiative step into the lowest excited state S$_1$ before decaying by fluorescence back to the ground state. 
While we can deduce a lifetime of about 58 fs of the S$_6$ state, our current fitting procedure treating each state individually, cannot give fully quantitative results. This task would require a detailed global fitting procedure based on rate equations taking into account all possible channels between the observed states. 
Currently, in part also limited by the cross-correlation of our laser pulses of 83 fs (195 fs FWHM), we cannot fully elucidate if the observed states are all populated consecutively or if there exist parallel decay channels.
Nevertheless, the differences in the $\Delta t$ values obtained from bands A, C and D agree with the states S$_5$ and S$_4$ being populated on slightly different timescales.

The non-adiabatic dynamics following the excitation to the brightest singlet state of tetracene and other PAH have previously been calculated by Posenitskiy \textit{et al.} \cite{Posenitskiy_2019,Posenitskiy_2020} based on a surface hopping algorithm using time-dependent density functional tight binding surfaces.
In these works an initial population of the brightest excited state is predicted to decay within 65 fs \cite{Posenitskiy_2020} toward a set of lower lying excited states. At the end of the 300 fs simulation window, there remains 8\% of the initial population in the initially excited brightest state.
The results from these theoretical studies agree qualitatively with our experimental observations, including the fast timescale of the depopulation of the initially excited state and the transient population of intermediate states.
A detailed comparison seems however infeasible, since the number of states predicted between the initially excited state and the ground state differs.

\section{Conclusion}
We have studied the ultra-fast dynamics of gas phase tetracene excited in the UV range with time-resolved photoelectron spectroscopy. Despite various background signals and many competing processes that occur in the interaction of PAH with femtosecond laser pulses, we could extract the signatures of the valence states that are transiently populated in the relaxation process.  
Using high-level \textit{ab-initio} calculations, we have characterized the low-lying singlet excited states of neutral tetracene and low-lying ionization potentials, calculated at the equilibrium structure of its radical cation. Comparing the experimental results and the calculations, we were able to assign the involved intermediate states to a series of dark states S$_3$-S$_5$ of  B$_\mathrm{1g}$ and A$_\mathrm{g}$ symmetry and to the S$_2$ state of B$_\mathrm{3u}$ symmetry. None of these states have to the best of our knowledge previously been measured or assigned experimentally, which highlights the power of time-resolved photoelectron spectroscopy in tracking excited state dynamics without restrictions due to dark states.
The initial population in the S$_6$ state decays within $\approx58$ fs. If the lower-lying states are all populated consecutively or partially in parallel, cannot be conclusively decided from our data, due to limitations of the achievable time-resolution. More reliable detection of the two low-lying states S$_2$ and S$_3$ could be achieved with higher probe photon energies, allowing single-photon ionization of these states. 
The future extension of these studies to higher probe energies supported by high-level quantum chemical calculations on polyacenes will provide deeper insights into the relaxation dynamics of this class of molecules.

\section{Supplementary material}
See the supplementary material for additional experimental details, details on data evaluation and computational details.

\section*{Conflicts of interest}
There are no conflicts to declare.

\section*{Acknowledgements}
We gratefully acknowledge funding from the Deutsche Forschungsgemeinschaft (grant numbers STI 125/25-1, RTG 2717) and the COST Action CA21101 “Confined Molecular Systems: From a New Generation of Materials to the Stars (COSY)”. Furthermore, the authors acknowledge support by the state of Baden-Württemberg through bwHPC and the German Research Foundation (DFG) through grant no INST 40/575-1 FUGG (JUSTUS 2 cluster).
The authors gratefully acknowledge the funding of this project by computing time provided by the Paderborn Center for Parallel Computing (PC2).
We thank Prof. Henrik Stapelfeldt for sharing and discussing his unpublished results with us.
A. S. acknowledges fruitful discussions with and support provided by Evgeny Posenitskyi, Mathias Rapacioli, Frank Lépine and Bernd v. Issendorff. P.B.C. acknowledges financial support under grant PID2021-124080OB-I00 funded by MCIN/AEI/10.13039/501100011033 and ERDF A way of Making Europe.

\section*{Data Availability Statement}
All data not provided in the supplementary material are available from the corresponding author upon reasonable request.

\nocite{*}
\bibliography{Main}
\bibliographystyle{SI} 
\end{document}


\title{Supplementary Material for:\\ Non-adiabatic electronic relaxation of tetracene from its brightest singlet state} 

\author{Audrey Scognamiglio,\textit{$^{a\ddag}$} Karin S. Thalmann,\textit{$^{a\ddag}$} Sebastian Hartweg,$^{\ast}$\textit{$^{a\ddag}$} \and Nicolas Rendler\textit{$^{a}$}, Lukas Bruder,\textit{$^{a}$} Pedro B. Coto,\textit{$^{b}$} Michael Thoss, \textit{$^{a}$} \and Frank Stienkemeier \textit{$^{a}$}} 
\date{}

\maketitle  
\noindent
$^\ast$ e-mail: sebastian.hartweg@physik.uni-freiburg.de\\
$^a$ University of Freiburg, Institute of Physics, Hermann-Herder-Str.\,3, Freiburg, Germany\\
$^b$ Materials Physics Center (CFM), CSIC and Donostia International Physics Center (DIPC), Paseo Manuel de Lardizabal 5, 20018 Donostia - San Sebastián, Spain

\section{Cross-correlation}
\begin{figure}
\centering
\includegraphics[width=0.55\textwidth]{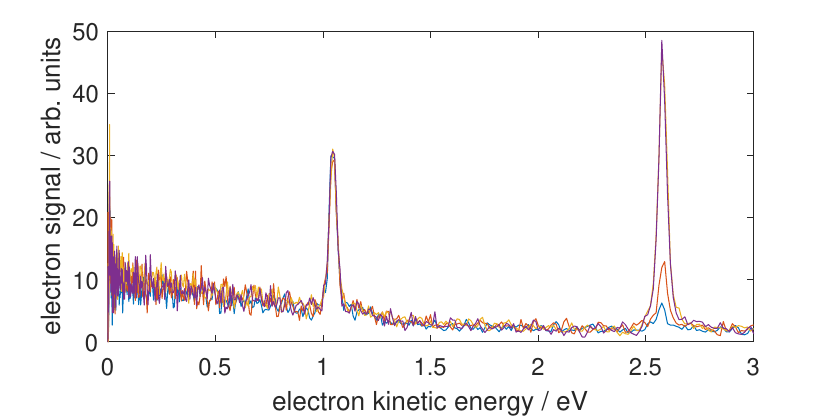}
\includegraphics[width=0.55\textwidth]{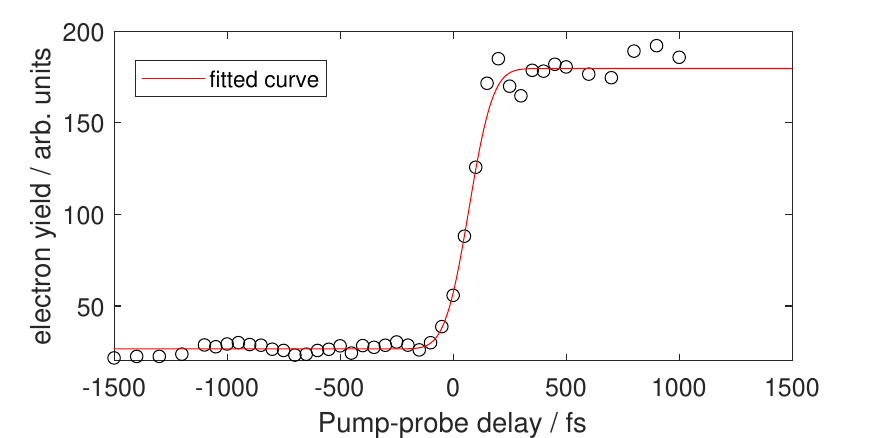}
\caption{\label{fig:x-co}Top panel: two-color photoelectron spectra of effusive sodium atoms at four different pump-probe delays. Bottom panel: Integrated photoelectron yields for the ionization from the Na 6p orbital by the 400 nm VIS pulse.}
\end{figure}
Figure \ref{fig:x-co} shows the measured cross correlation data. The 267\,nm UV pulse is used to excited the sodium 3s electron to the 6p orbital. The subsequent VIS pulse at 400 nm is sufficient to ionize the excited sodium atoms, giving rise to a photoelectron band at 2.575\,eV. The other band, located at 1.041\,eV is created by a combination of resonant two-photon ionization by the UV pulse and non-resonant three-photon ionization by the VIS pulse.
The integrated photoelectron yield of the band at 2.575 eV is plotted in the bottom panel of figure \ref{fig:x-co}. The red solid line shows the fitted curve according to 
\begin{equation}
f(t)=B+A\cdot\left(1+\textrm{erf}\left(\frac{t-\Delta t}{\sqrt2\tau_{cc}}\right)\right)
\end{equation}
used to deduce a cross correlation of $\tau_{cc}$=83$\pm$5, corresponding to a FWHM of 195$\pm$12 fs. $A$ is the fitted amplitude, $\Delta t$ is the position of the zero delay in the cross correlation measurement and $B$ accounts for a constant offset in the integrated photolectron yield.
The two sodium bands in the photoelectron spectra are used to verify the calibration of the VMI photoelectron spectrometer, which we assume to be accurate within 5 meV. The resolution can be determined to be slightly below 2\% for the band at 2.575\,eV.

\section{Details on background subtraction}
\begin{figure}
\centering
\includegraphics[width=0.55\textwidth]{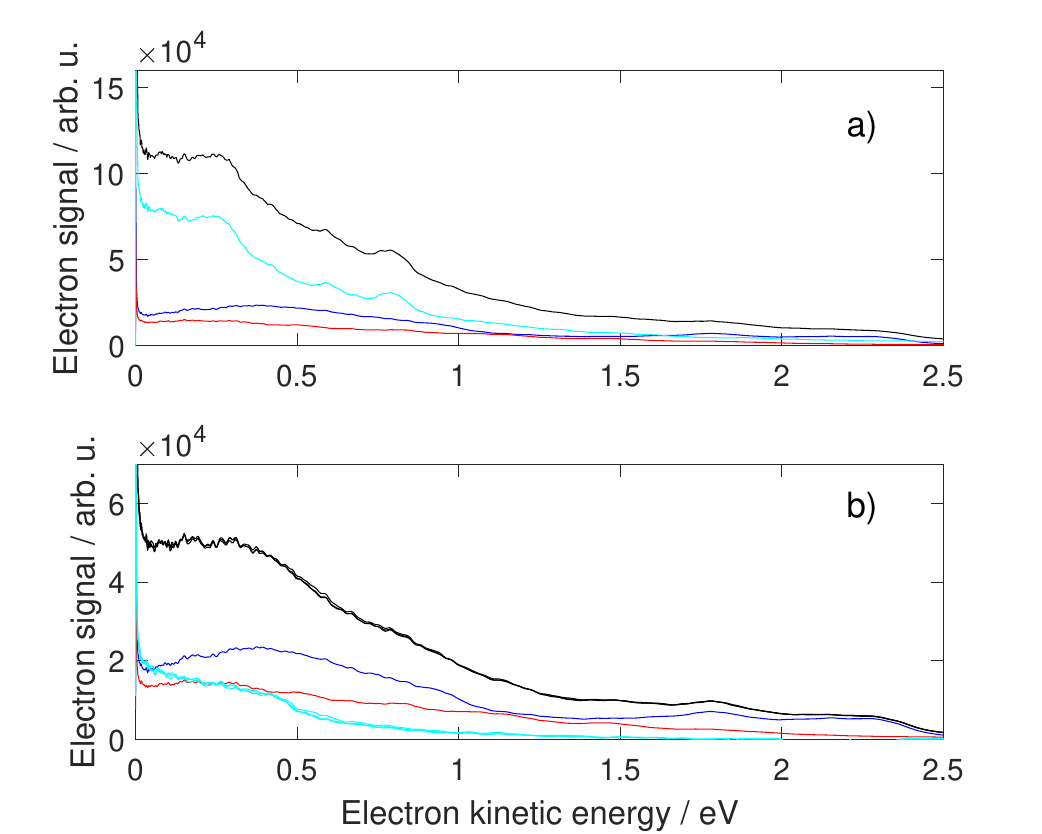}
\label{fig:bg}
\caption{Single-color background subtraction from time-resolved two-color photoelectron spectra at a) $\tau\approx0$~fs and b) $\tau=$-1430, -680 and -430~fs (lines overlapping). Black solid lines show data prior to background subtraction, blue shows the UV only background, red the VIS only background and cyan lines show data after background subtraction.}
\end{figure}
Figure \ref{fig:bg} shows the two-color photoelectron spectra obtained by the UV and VIS pulses prior to background subtraction (black solid lines) at a delay of $\tau\approx0$~fs a) and $\tau=$-1430, -680 and -430~fs b). The latter curves overlap almost perfectly, and are thus not visible as individual curves. The single-color backgrounds obtained with only the VIS (red lines) and only the UV pulses (blue lines) are shown in both panels, together with the background subtracted spectra (cyan lines).
The background subtracted spectra at negative delays show an exponentially decaying feature, with a single edge-like band on top, located at  $\approx0.43$~eV. This feature agrees well with the S$_1$ state (excitation energy 2.77~eV\cite{Amirav_1979}), ionized by the UV laser pulse. Note that the VIS pulse with a photon energy of 3.1 eV is in resonance with a vibrationally excited S$_1$ state, thus explaining the possible population of the S$_1$ state at negative pump-probe delays (i.e. the VIS pulse arriving prior to the UV pulse).

To remove the two-color pump-probe effect of exciting the tetracene molecules with the VIS pulse and probing with the UV pulse, we averaged the spectra obtained at the three negative delays given above (i.e. the three overlapping  cyan lines in figure \ref{fig:bg} b)). Given the fact that the S$_1$ state in the tetracene molecule has a lifetime of several nanoseconds\cite{Amirav_1979}, we can deduce the time-dependence of this pump-probe signal. The S$_1$ state should show a relative population increasing in a step-like fashion, broadened by the cross correlation of the laser pulses, given by 
\begin{equation}\label{eq:population}
p_{\mathrm{S_1}}(t)=\frac{1}{2}\cdot\left(1+\mathrm{erf}\left(-\frac{t}{\sqrt{2}\tau_{cc}}\right)\right),
\end{equation}
where $t$ is the time delay between the UV and VIS pulses and $\tau_{cc}$ is the cross correlation width. We multiply the averaged signal described above with the corresponding population according to equation \ref{eq:population} and subtract it from the time-resolved photoelectron spectra at the corresponding delay time.
\begin{figure}
\centering
\includegraphics[width=.45\textwidth]{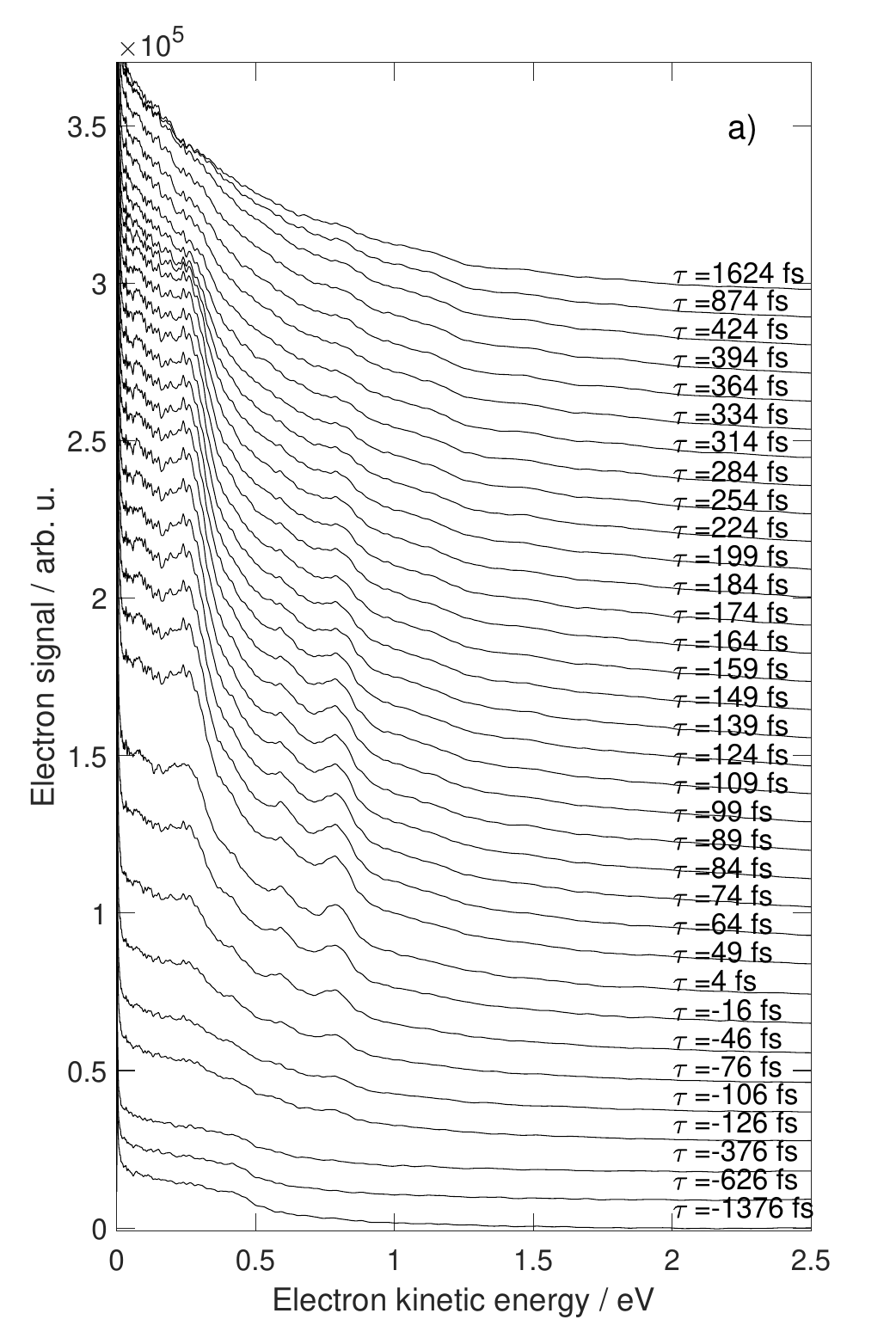}
\includegraphics[width=.45\textwidth]{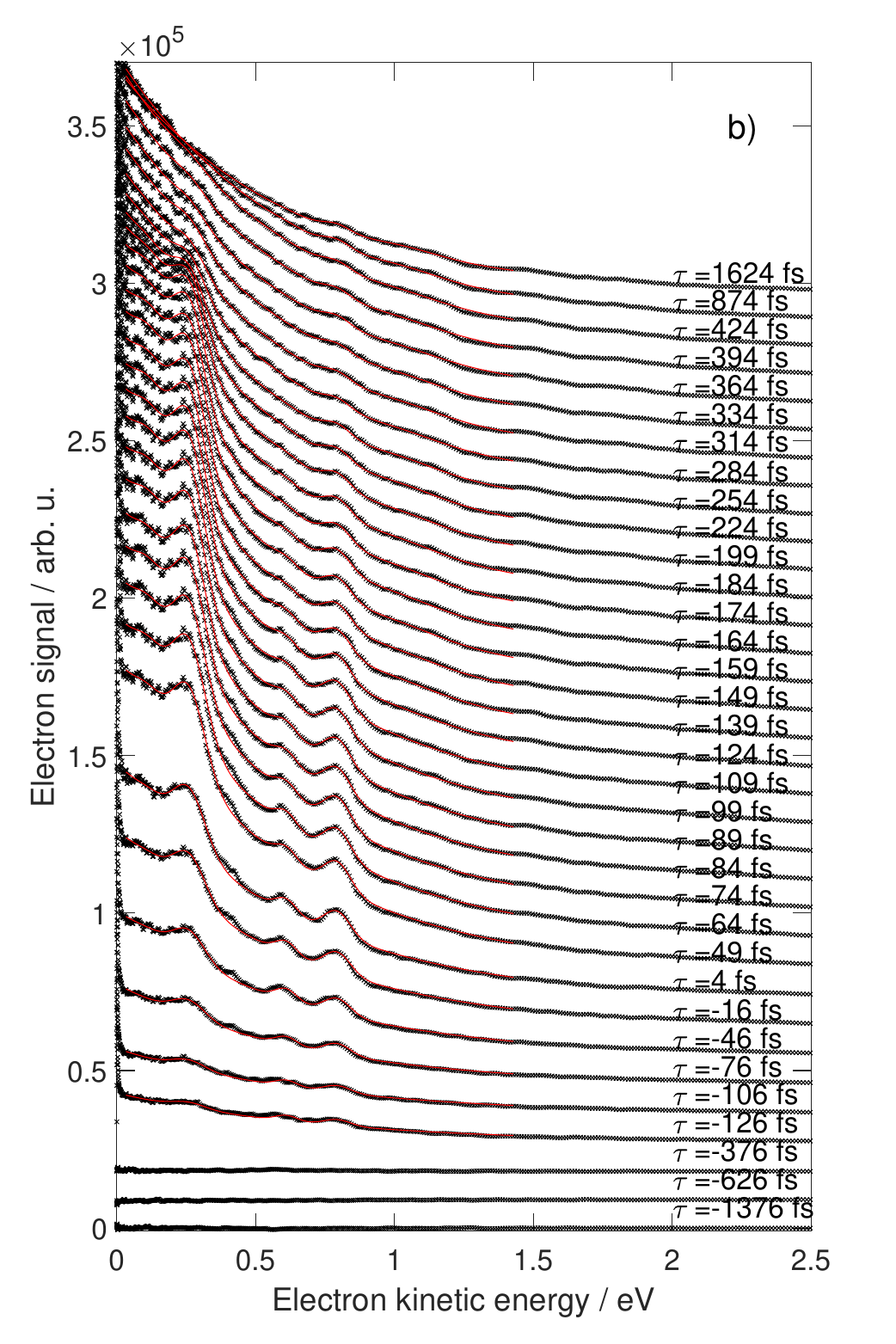}
\label{fig:negdelay}
\caption{Time-resolved photoelectron spectra without (panel a)) and with (panel b)) subtraction of the VIS pump-UV probe signals. Delays between the laser pulses are indicated for every spectrum. In panel b) we additionally include the results obtained from the least squares fit.}
\end{figure}
The effect of this additional background subtraction is depicted in figure \ref{fig:negdelay}, where panel a) depicts data without the subtraction of the VIS pump- UV probe signal, and the data including this subtraction is shown in panel b). 
\newpage

\section{Photoelectron images and angular distributions}
\begin{figure}
\centering
\includegraphics[width=0.45\textwidth]{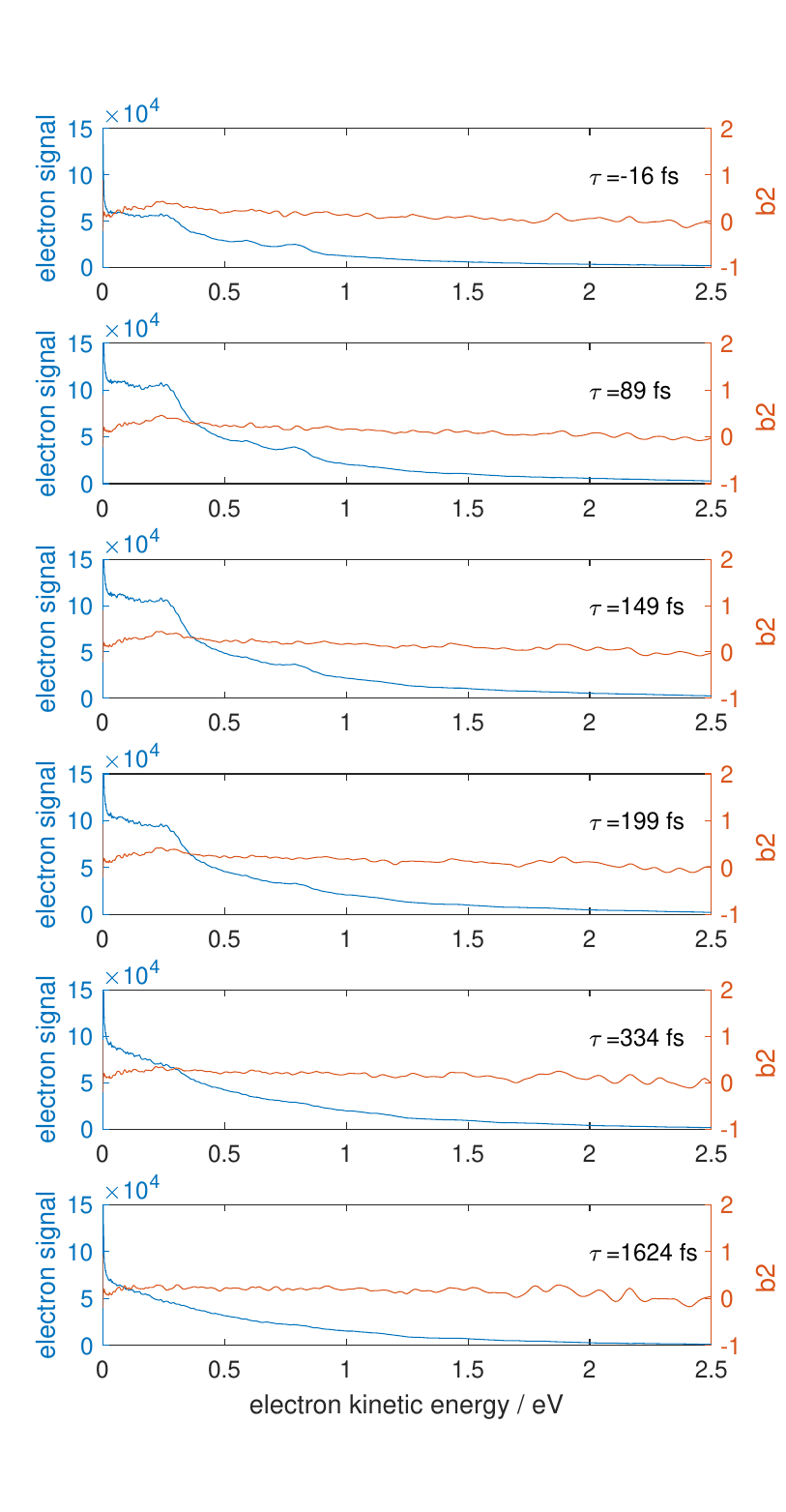}
\caption{\label{fig:b2}Photoelectron $b_2$ parameters for selected pump-probe delays.}
\end{figure}

\begin{figure}
\centering
\includegraphics[width=0.65\textwidth]{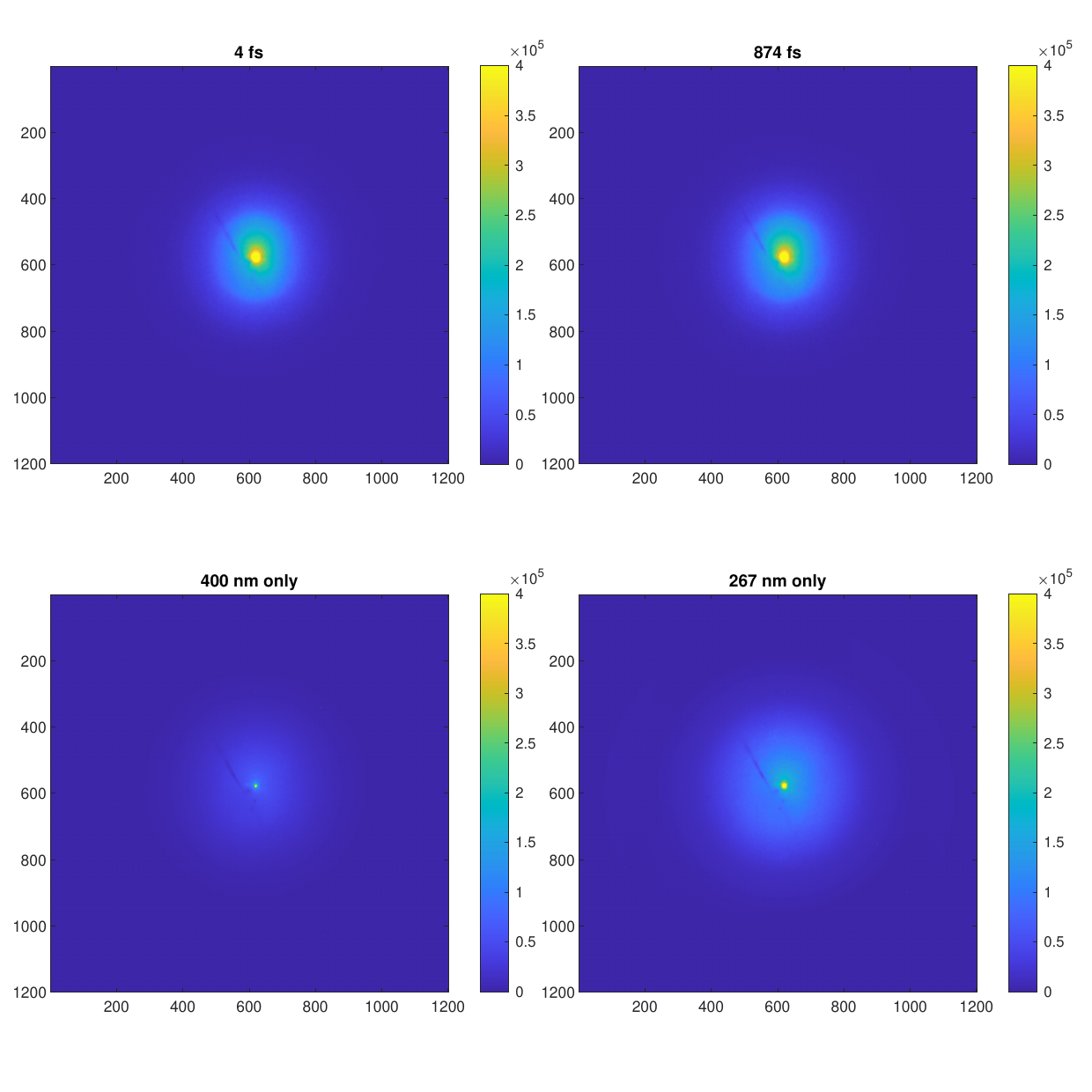}
\caption{\label{fig:image}Photoelectron images obtained at two different delays between the UV and VIS pulses, and for the UV and VIS pulses alone.}
\end{figure}

Photoelectron angular distributions (PAD) for selected pump probe delays are shown in the form of $b_2$ parameters in figure \ref{fig:b2}. $b_2$ specifies thereby the expansion coefficient of the second order Legendre polynomial, according to 
\begin{equation}
I(\theta)=I_0\cdot \left(1+\frac{b_2}{2}\left(3\cos^2(\theta)-1\right)\right).
\end{equation}
$I(\theta)$ is the angular dependence of the photoelectron distribution at any given photoelectron kinetic energy, $I_0$ is the angle integrated intensity at that kinetic energy and $b_2$ can typically take values between -1 and 2. 
The data shows overall close to isotropic photoemission with $b_2\approx0$. Only the band at 0.26\,eV shows a slightly positive $b_2$ value, indicating a slightly enhanced photoemission probability along the light polarization direction. Note that for a full quantification of the angular distributions for the individual electron bands a subtraction of the exponential signal contribution would be necessary. This would additionally require to make assumptions on the shape of the angular distribution of the exponential thermal electron signal.
Exemplary photoelectron images obtained with both laser pulses, as well as with the single laser pulses are displayed in figure \ref{fig:image}. The dominating exponential signal contribution makes it difficult to distinguish any structure in the images prior to reconstruction. 

\section{Details on least squares fit of the spectra}
\begin{figure}
\centering
\includegraphics[width=.6\textwidth]{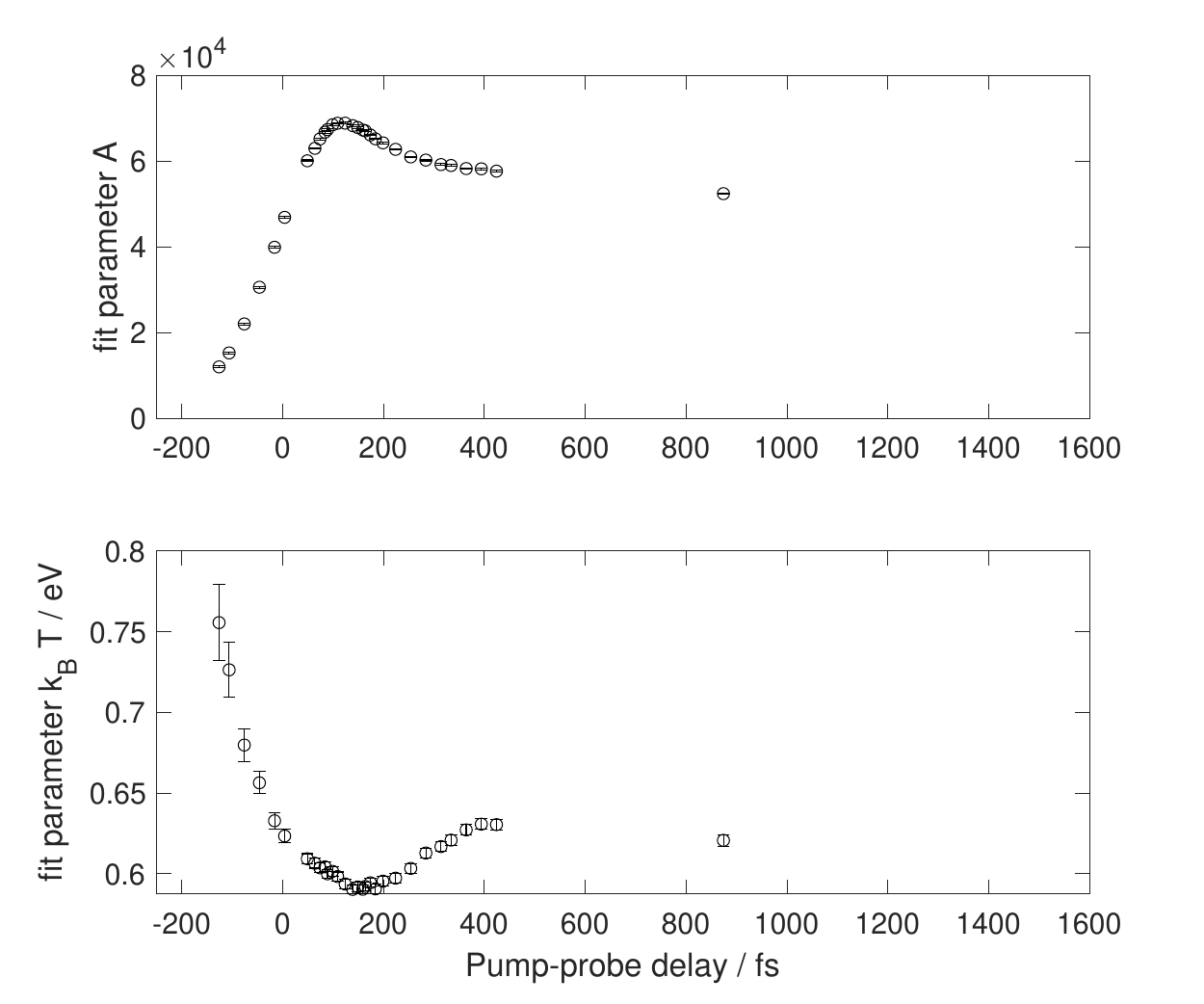}
\caption{\label{fig:exp}Obtained fit parameters for the exponential signal components of the photoelectron spectra at different pump-probe delays. Top panel: Exponential amplitude. Bottom panel: Decay constant.}
\end{figure}

 The obtained fits for the spectra are displayed in figure \ref{fig:negdelay} b). The time-dependent amplitude and decay constant of the exponential signal component are displayed in figure \ref{fig:exp}. The amplitude of the exponential signal component rises sharply from negative time delays and is maximal for slightly positive time delays before slowly decreasing for longer positive delays. It is evident that the UV excitation prior to the VIS pulse increases the yield of the exponential thermal electron signal, while the prior excitation with the VIS pulse does not create a similar enhancement. This is consistent with the information that the absorption cross section of tetracene is much higher at 267\,nm than at 400\,nm. The fact that even at long delays of 1.6 ps the thermal electron yield remains enhanced shows that the effect of the excitation persist. The lifetime of the lowest electronically excited singlet state S$_1$ is known to be in the nanosecond regime\cite{Amirav_1979}. Since our study could not resolve the dynamics of the relaxation from the low-lying electronic states S$_2$ and S$_3$ to the S$_1$ state, it remains unclear if the enhancement of the thermal electron yield is due a remaining electronic excitation, or due to the vibrational excitation created in the course of the electronic relaxation from the high-lying S$_6$ state. 
 The fitted exponential decay constant (k$_\textrm{B}$T) shows a behavior roughly inverted from the thermal electron yield. At negative delays, the decay constant decreases, indicating a trend toward a steeper exponential decay. The decay reaches a minimal decay constant (maximal steepness) close to its maximum yield, before the decay constant increases again slightly. Toward very long delays there seems to again be a trend of decreasing decay constant. 
 The behavior in the thermal electron yield as well as in the decay constant seems to indicate a complex dependence of the thermal electron emission on the vibronic state of the molecule. While there have been a significant number of studies on the emission of thermal electrons from fullerenes and other polycyclic aromatic hydrocarbons (e.g. references \cite{Johansson2013,Johansson2014} and references therein), there have to the best of our knowledge not been systematic studies on how the emission of thermal electrons depends on the initial electronic or vibrational states of molecules.
 
The time-dependent amplitudes of four of the five fitted Gaussian bands are shown and discussed in the main text. 
The obtained Gaussian band positions are 0.12, 0.26, 0.61, 0.79, and 1.04~eV, with widths ($\sigma$) of 0.057, 0.059, 0.039, 0.049, and 0.124~eV. The errors estimated from the covariance matrix of the fitting procedure are on the order of 1 meV. The value is likely reduced by the global nature of the fitting of the band positions, and appears too low. We estimate realistic uncertainties for the positions of the four bands discussed in the main text to be around 15 meV, including the uncertainty of the energy calibration and the spectrometer resolution.
\begin{figure}
\centering
\includegraphics[width=0.5\textwidth]{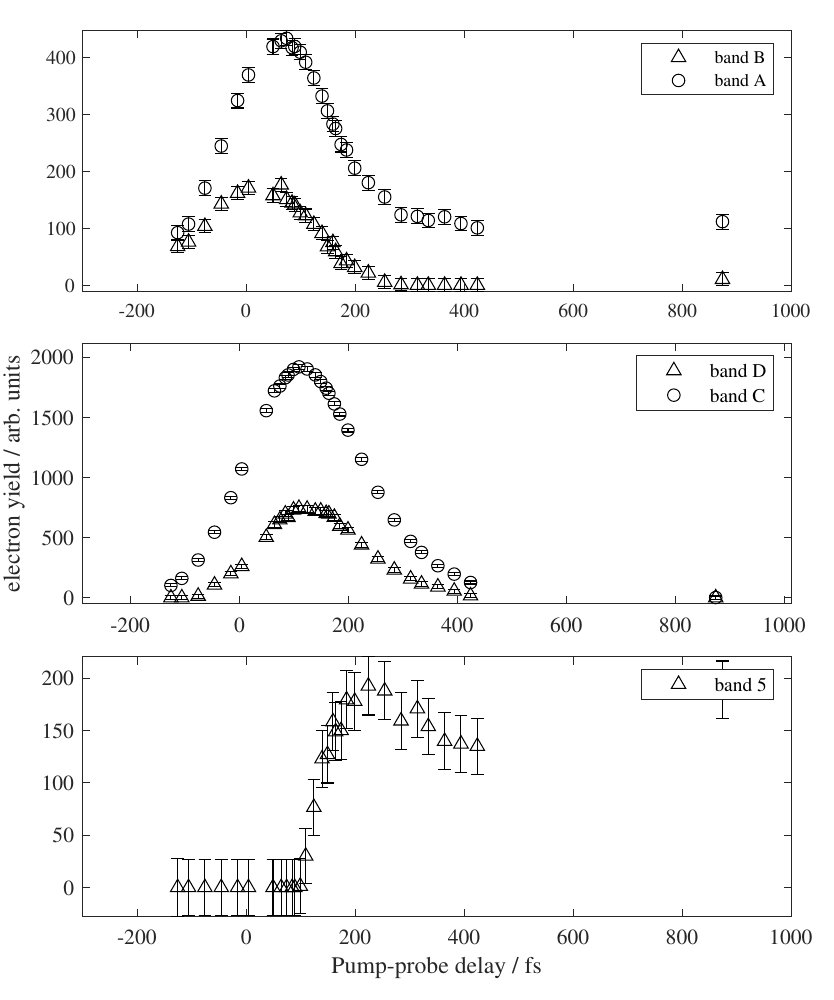}
\caption{\label{fig:band5}Fitted time-dependent amplitudes for all five photoelectron bands.}
\end{figure}
The amplitudes of all bands are again displayed in figure \ref{fig:band5}, including the amplitudes for the fifth band. While the first four bands are in detail discussed in the main text, we refrain from quantitatively discussing the amplitudes of the fifth band. While the overall time-dependence of the amplitude of the fifth photoelectron band is in agreement with the delayed population of a low-lying electronic state, ionized by two probe photons, we consider its values less reliable than those of the other four bands. We note, that including the fifth band in the fitting procedure improves the stability of the fitting procedure. 
 
\newpage
\section{Computational details}

In this section, additional information on the computational results for the excitation energies of tetracene in its neutral and cation ground state equilibrium structure is provided and the calibration of the calculations is discussed.

\subsection{Details on the electronic structure calculation of neutral tetracene}

\begin{figure}[h]
	\centering
	\begin{tabular}{c|cccc}
		sym. &A$_\text{u}$ & B$_\text{1u}$ & B$_\text{2g}$ & B$_\text{3g}$ \\
		\hline
		&\includegraphics[width=0.2\linewidth]{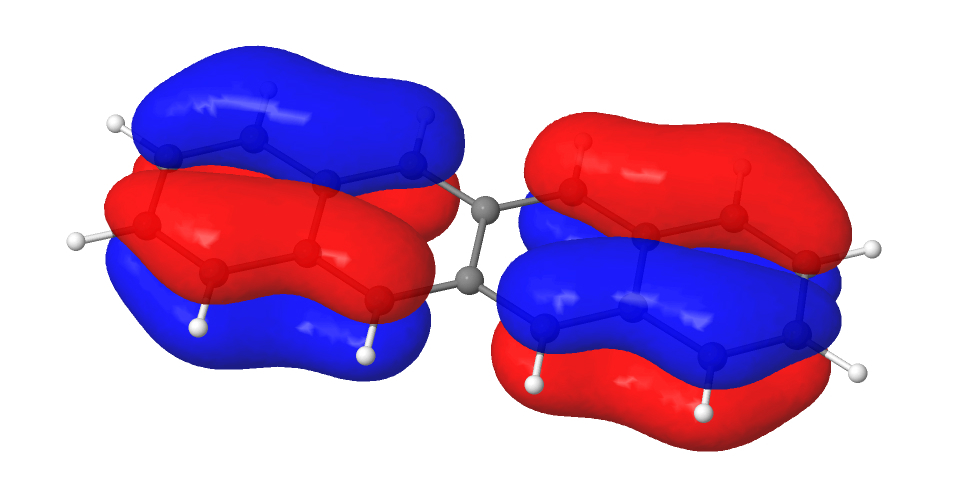} & \includegraphics[width=0.2\linewidth]{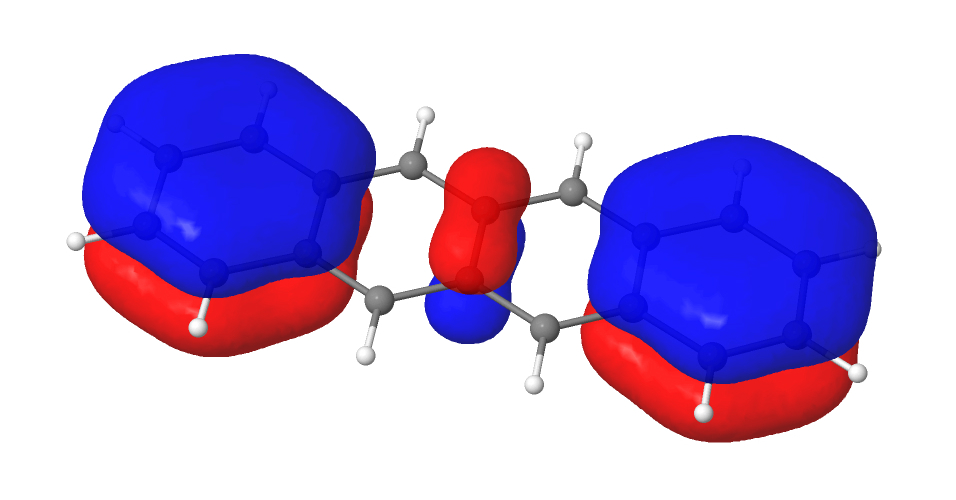} & \includegraphics[width=0.2\linewidth]{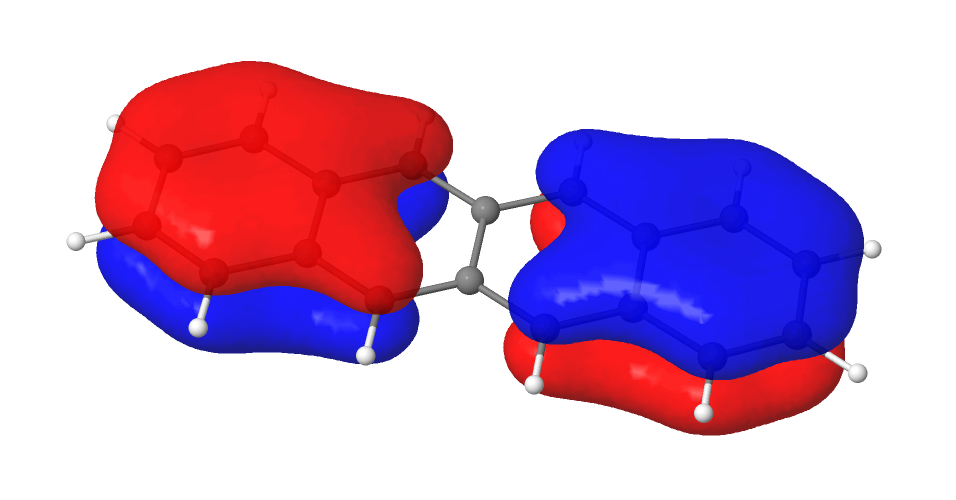} & \includegraphics[width=0.2\linewidth]{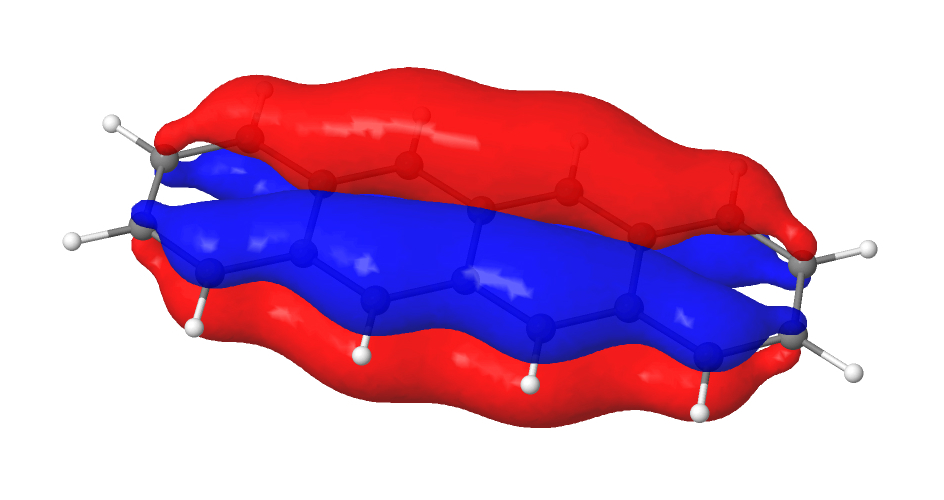} \\
		&\includegraphics[width=0.2\linewidth]{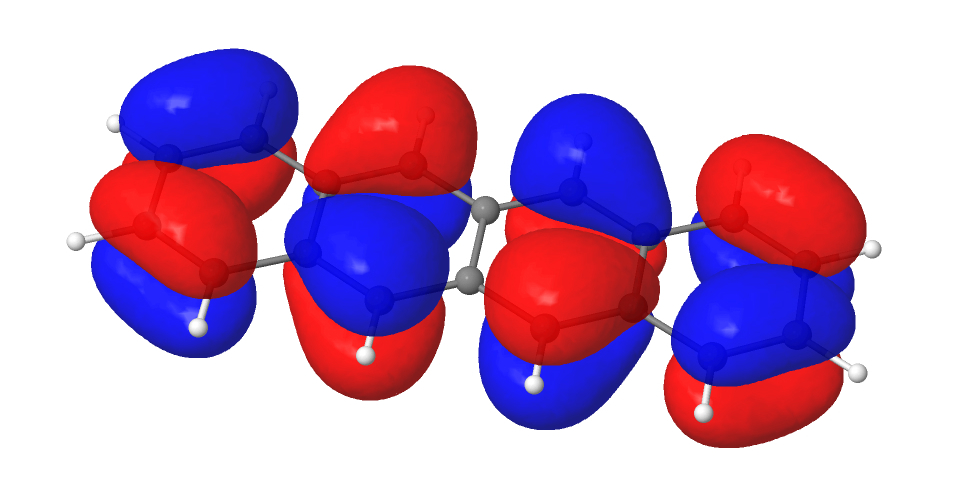} & \includegraphics[width=0.2\linewidth]{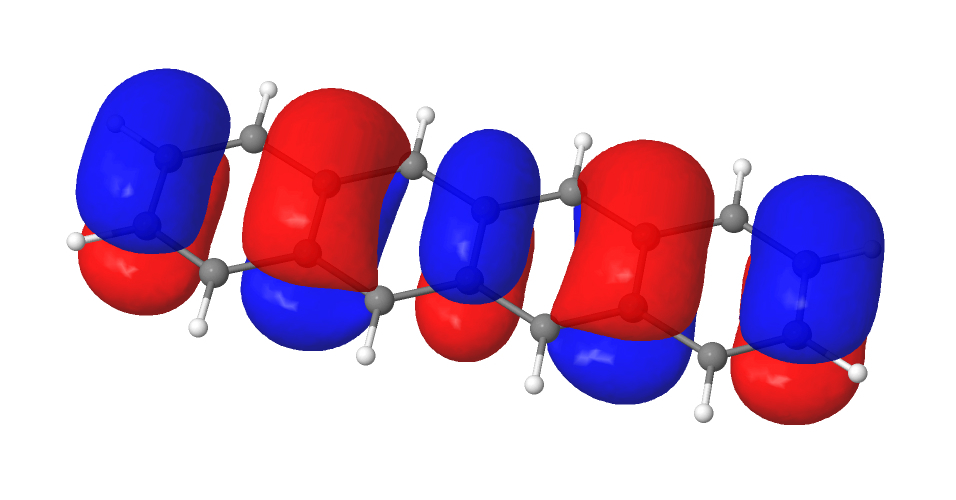} & \includegraphics[width=0.2\linewidth]{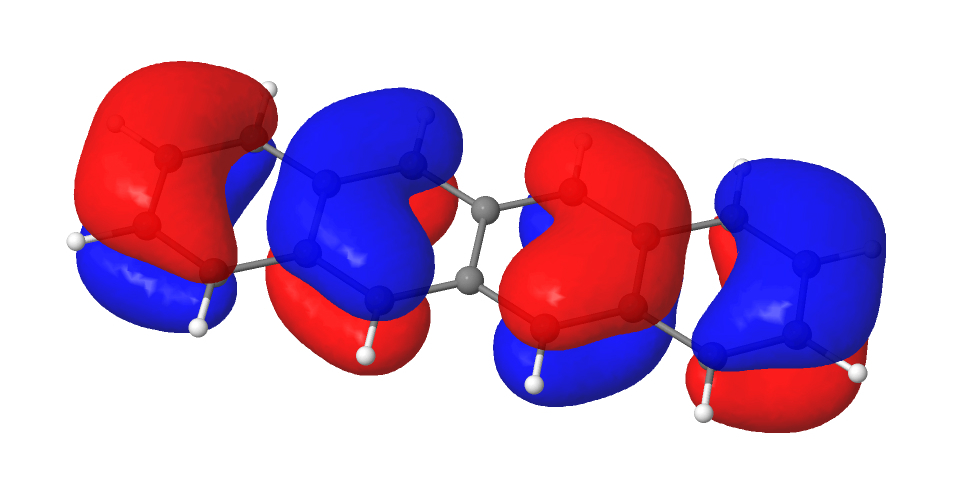} & \includegraphics[width=0.2\linewidth]{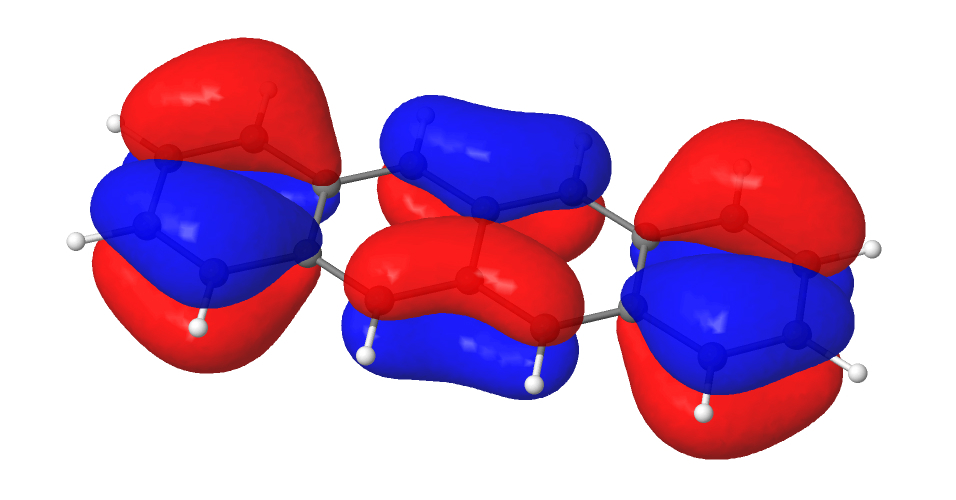} \\
		& \includegraphics[width=0.2\linewidth]{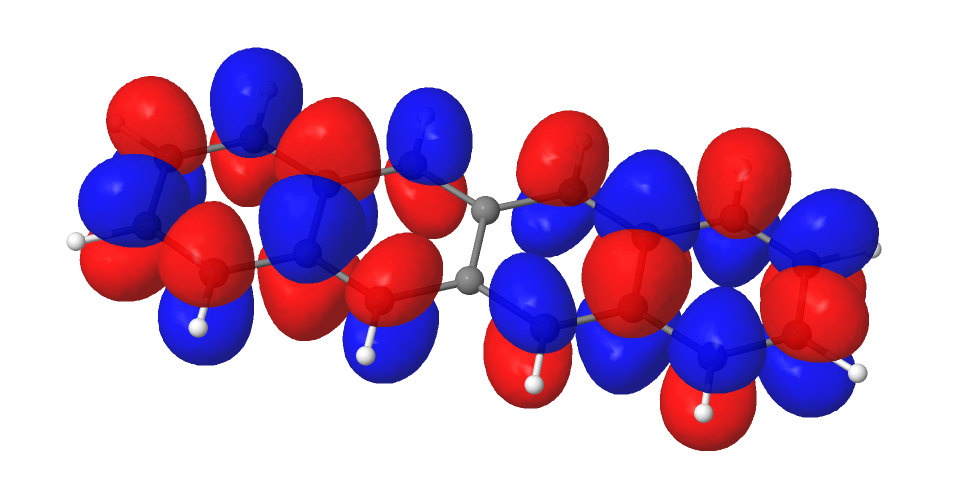} & \includegraphics[width=0.2\linewidth]{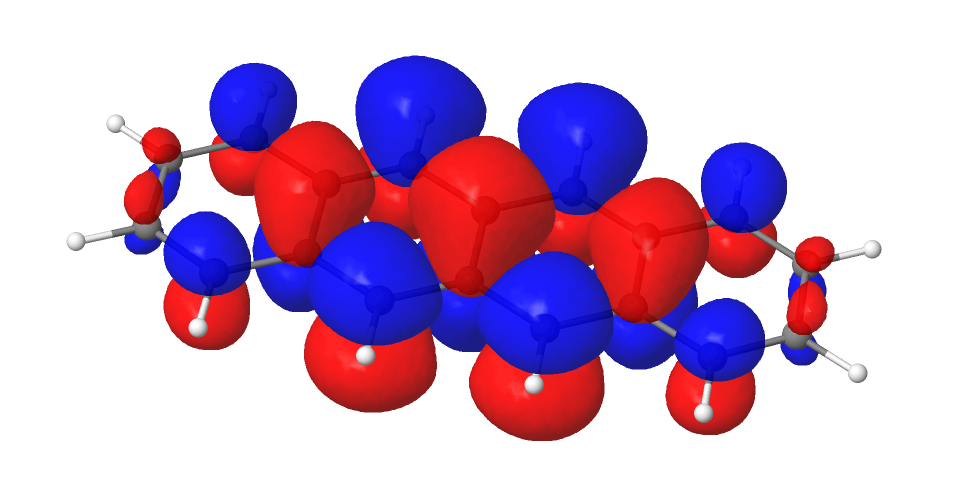} & \includegraphics[width=0.2\linewidth]{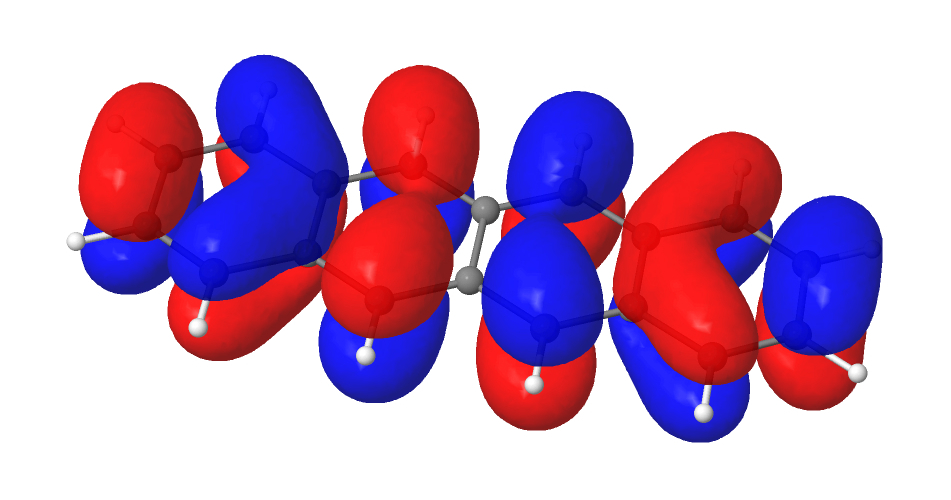} & \includegraphics[width=0.2\linewidth]{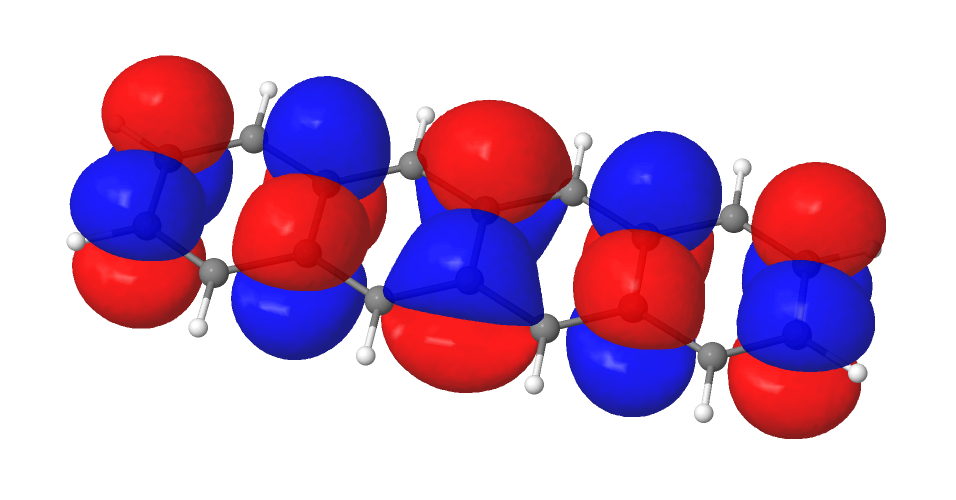} \\
		& \includegraphics[width=0.2\linewidth]{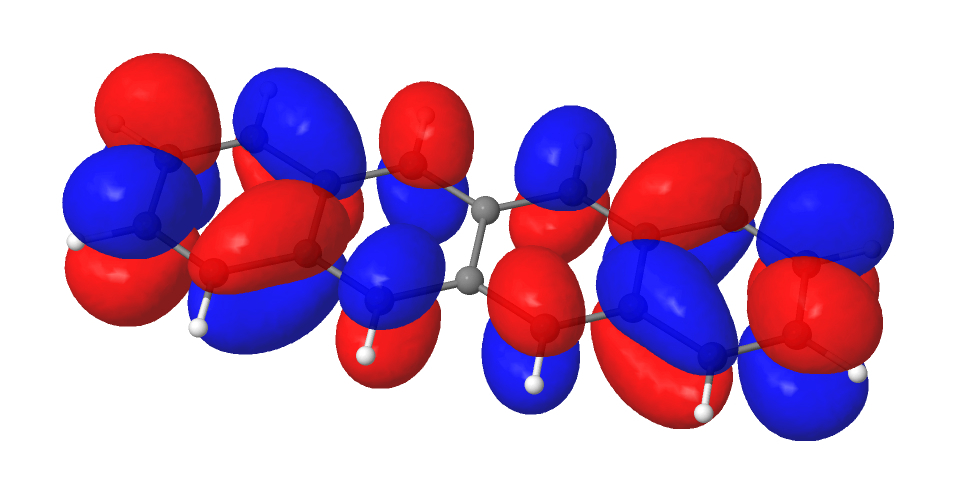} & \includegraphics[width=0.2\linewidth]{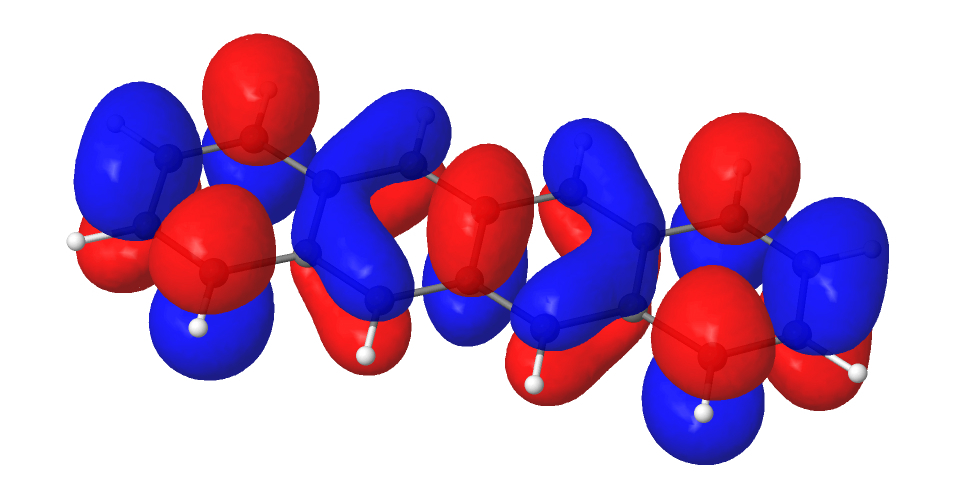} & \includegraphics[width=0.2\linewidth]{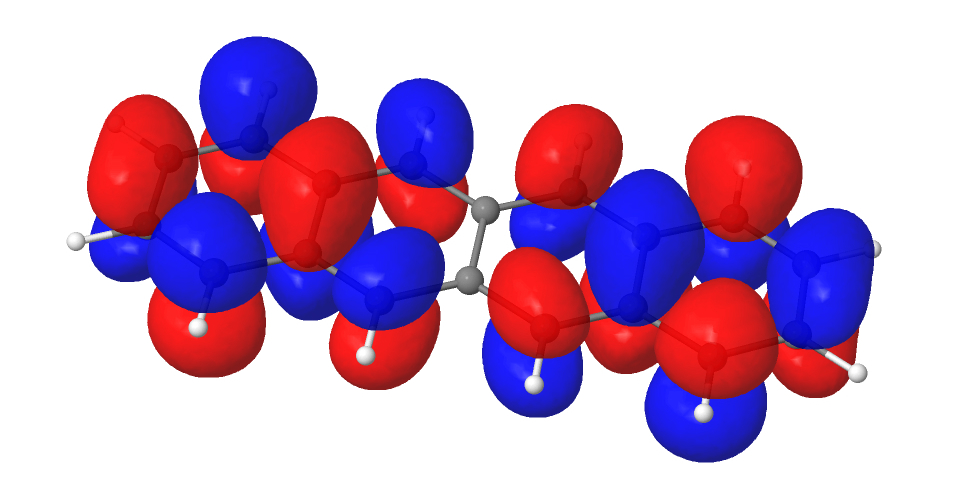} &  \includegraphics[width=0.2\linewidth]{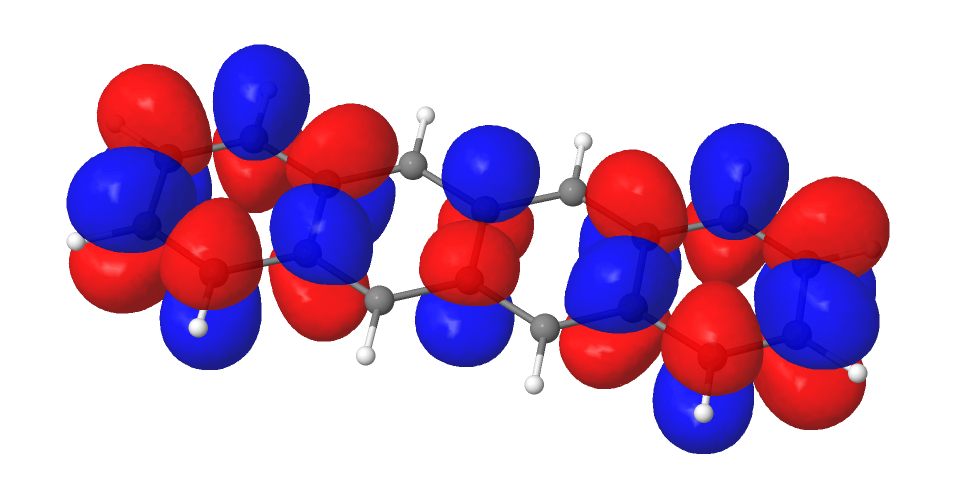}\\
	\end{tabular}
	
        \caption{CASSCF optimized state-average orbitals and their corresponding symmetry (irreducible representation of point group D$_\text{2h}$) of the A$_\text{g}$ symmetry calculation. In the state-averaged calculations, 15 roots with 16 active orbitals and 16 active electrons were employed.}
	\label{fig:omo_sa15-cas1616}
\end{figure}

As described in the main paper, the electronic structure of the neutral tetracene was calculated using a 15 state-averaged CASSCF/CASPT2~\cite{andersson_multiconfigurational_1993,helgaker_molecular_2000} calculation with an active space of 16 active electrons and 16 active orbitals. The calculations were carried out using MOLCAS 8.4\cite{aquilante_span_2016} and OpenMOLCAS~v20.10\cite{fdez_galvan_openmolcas_2019}.
The optimized orbitals of the active space of the A$_\text{g}$ symmetry calculation are shown in figure \ref{fig:omo_sa15-cas1616}.  The orbitals of the B$_\text{1g}$, B$_\text{2u}$, and B$_\text{3u}$ symmetry calculation are qualitatively the same. The orbital transitions and their coefficients of state S$_1$ to S$_6$ are provided in table \ref{tab:orb-trans}.

\begin{table}[h]
	\centering
	\begin{tabular}{>{\centering\arraybackslash}m{1cm}|>{\centering\arraybackslash}m{2cm}>{\centering\arraybackslash}m{1cm}>{\centering\arraybackslash}m{2cm}@{\hskip 1cm}>{\centering\arraybackslash}m{2cm}>{\centering\arraybackslash}m{1cm}>{\centering\arraybackslash}m{2cm}}
		state & \multicolumn{6}{c}{orbital transitions and coefficients} \\
		\hline
		&&&&&&\\[-0.3cm]
		S$_1$  & \includegraphics[width=\linewidth]{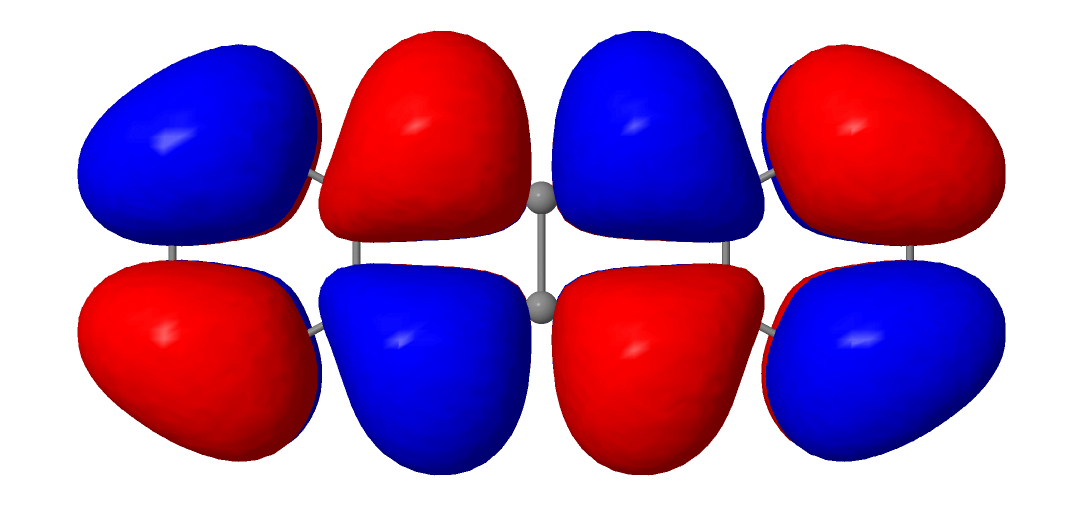} &$\xrightarrow[]{\text{ 0.854}}$& \includegraphics[width=\linewidth]{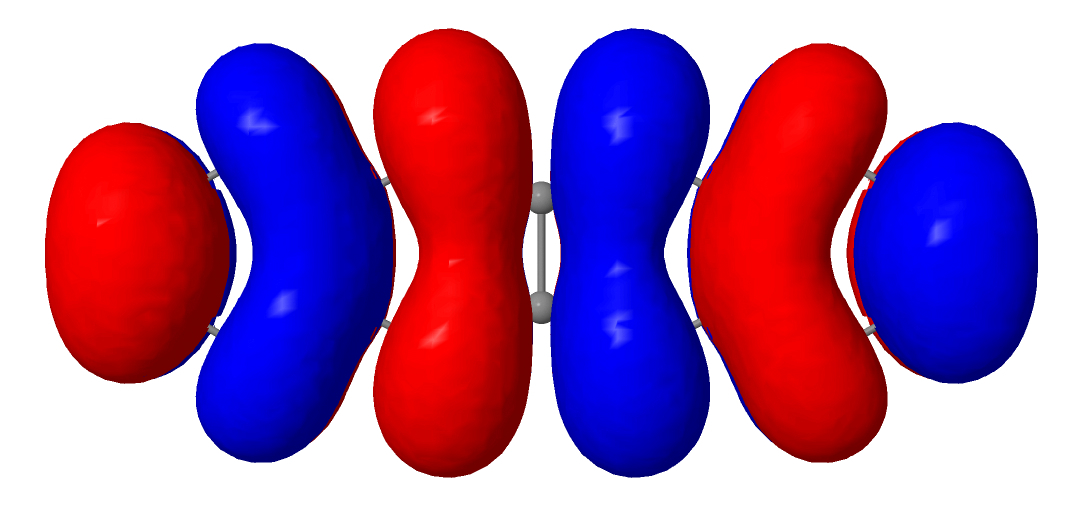} & & & \\
		\hline
		&&&&&&\\[-0.3cm]
		S$_2$ &  \includegraphics[width=\linewidth]{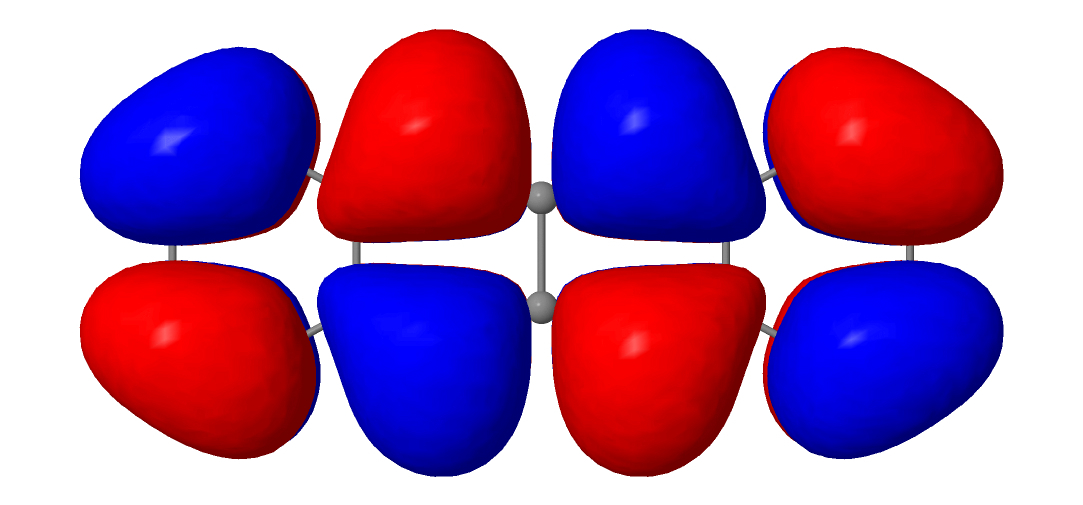}  &$\xrightarrow[]{\text{ 0.539}}$& \includegraphics[width=\linewidth]{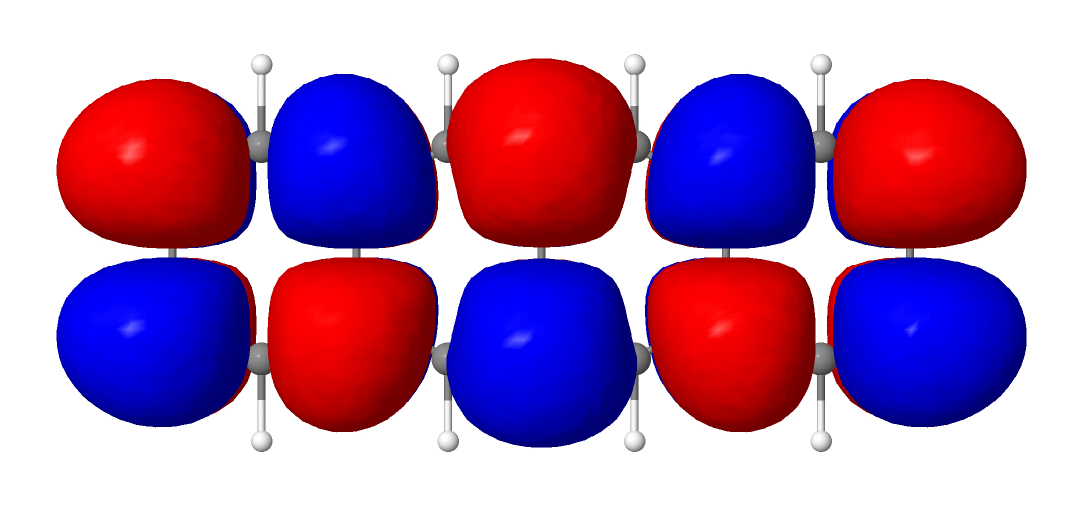} & \includegraphics[width=\linewidth]{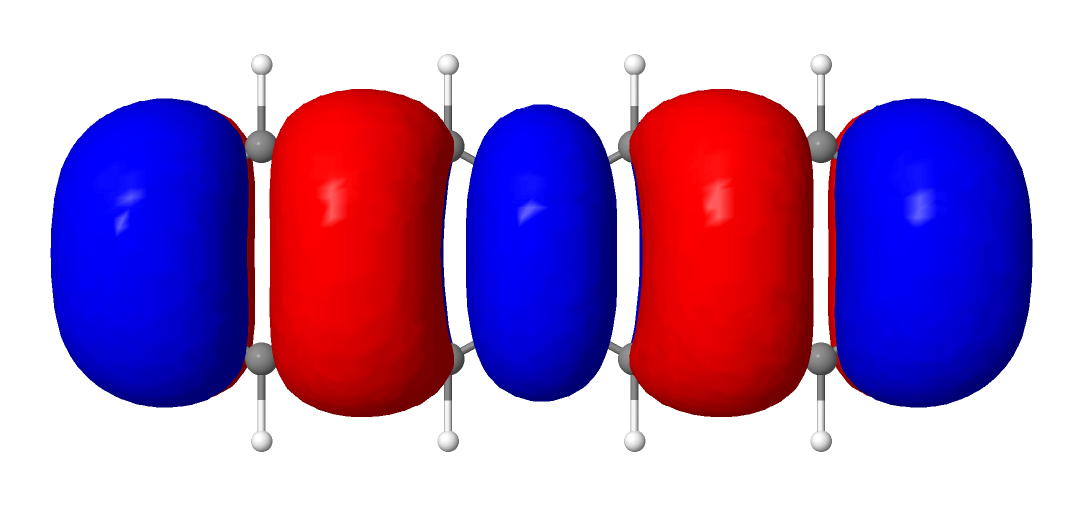} &$\xrightarrow[]{\text{-0.562}}$& \includegraphics[width=\linewidth]{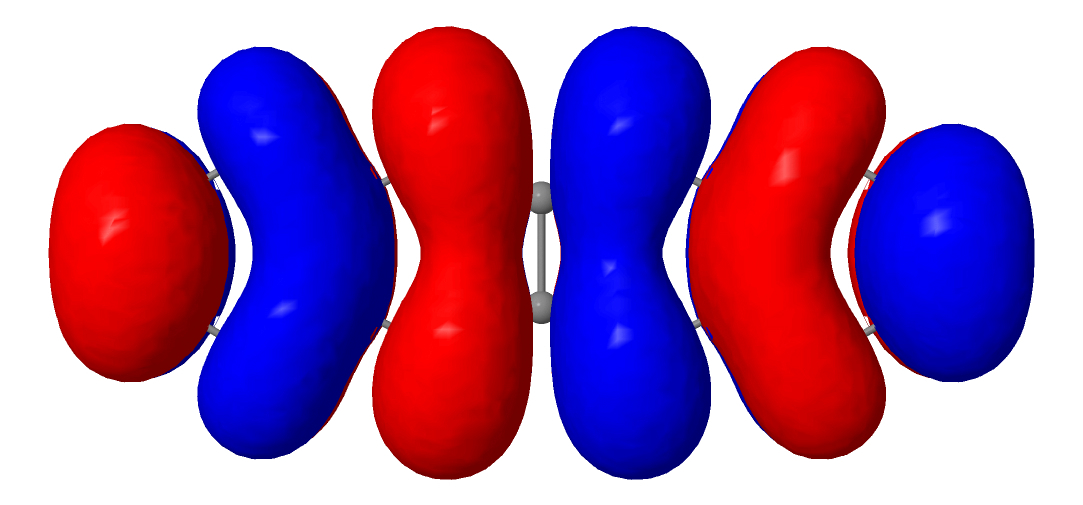} \\
		\hline
		&&&&&&\\[-0.3cm]
		\multirow{6}{*}{S$_3$} &  \includegraphics[width=\linewidth]{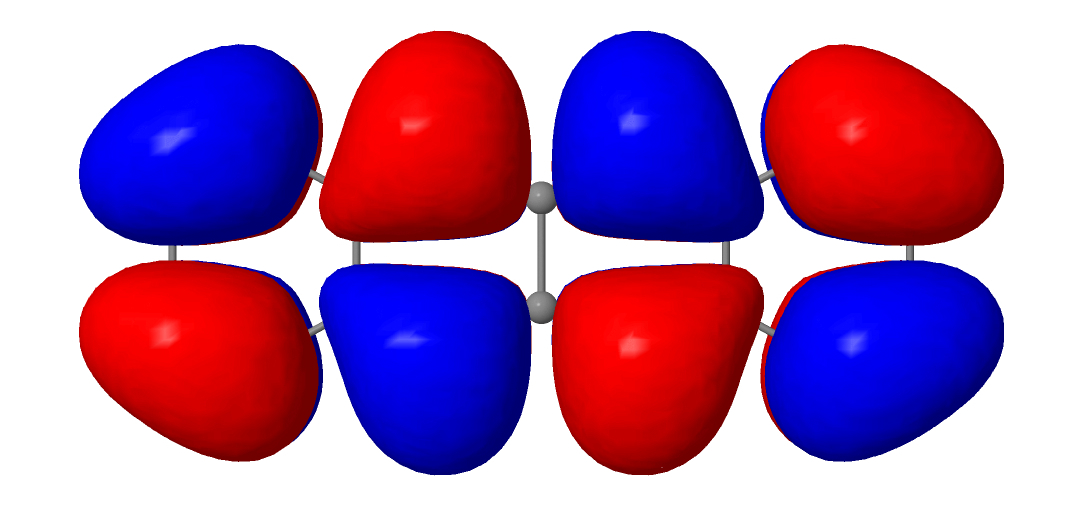}  &$\xRightarrow[]{\text{ 0.607}}$& \includegraphics[width=\linewidth]{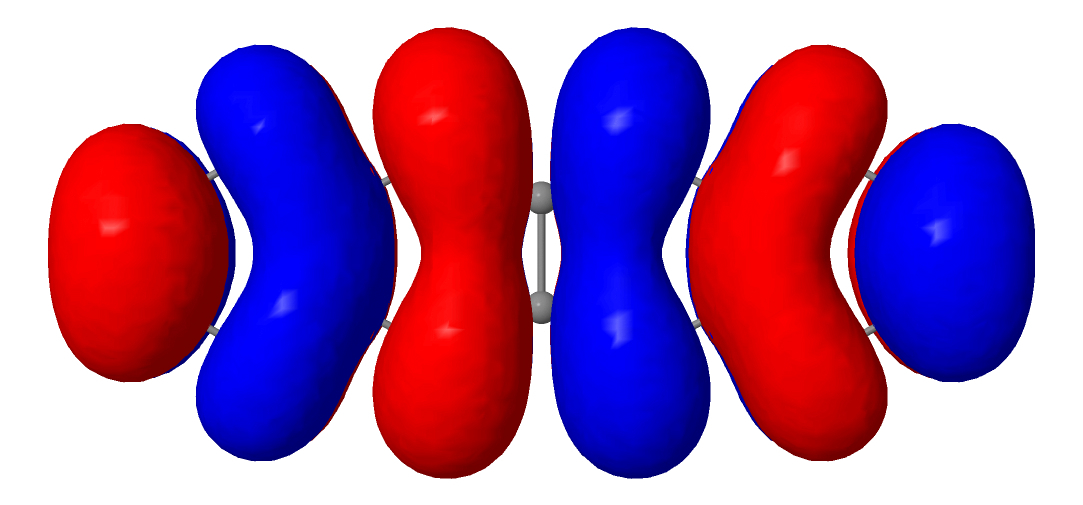} & \includegraphics[width=\linewidth]{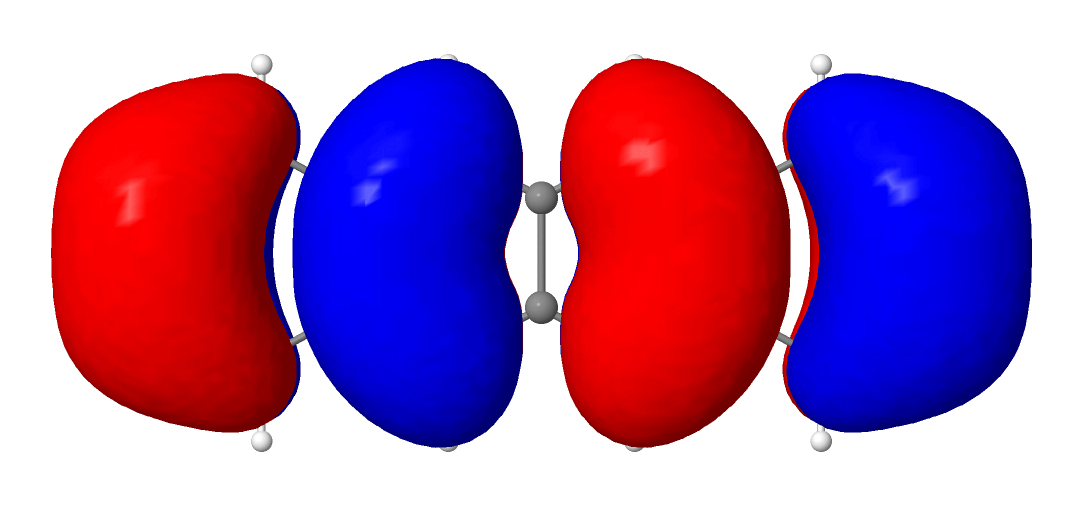} &$\xrightarrow[]{\text{-0.229}}$& \includegraphics[width=\linewidth]{Figures/SI/tetracene_csfs/sym_1/mo_b2g_3_csf} \\[1cm]
		& \includegraphics[width=\linewidth]{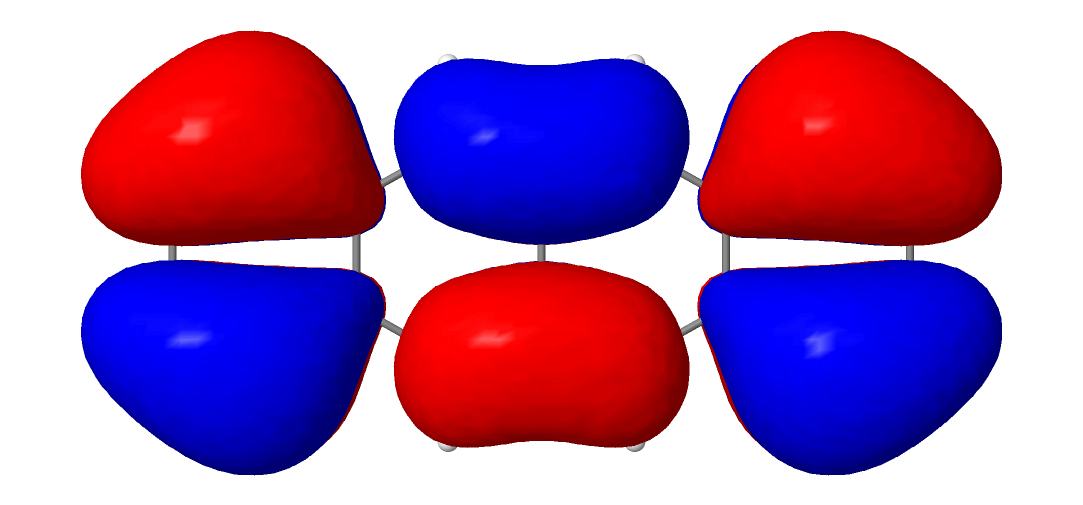}  &$\xrightarrow[]{\text{ 0.206}}$& \includegraphics[width=\linewidth]{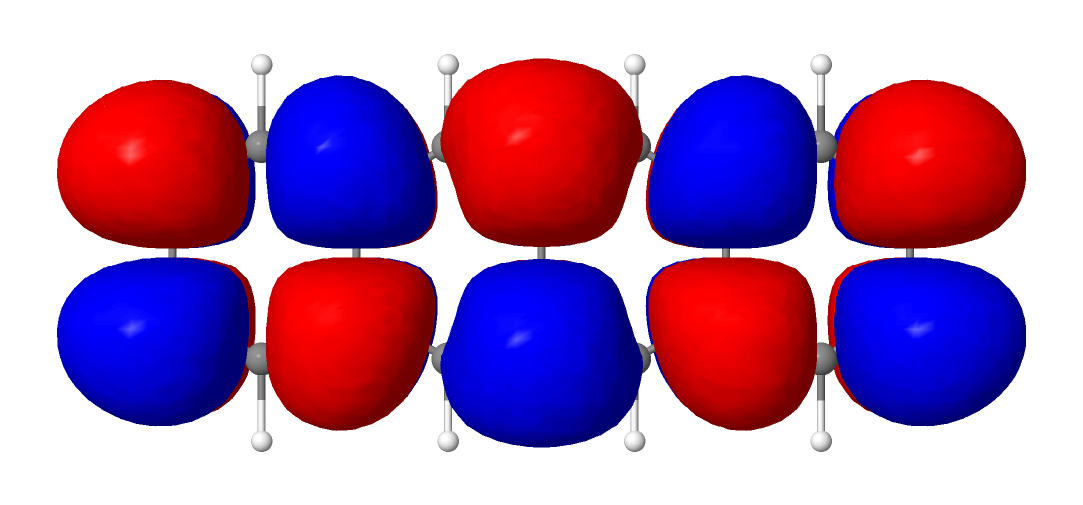} & \includegraphics[width=\linewidth]{Figures/SI/tetracene_csfs/sym_1/mo_au_2_csf}
		\includegraphics[width=\linewidth]{Figures/SI/tetracene_csfs/sym_1/mo_b3g_2_csf} &$\xRightarrow[]{\text{-0.207}}$& \includegraphics[width=\linewidth]{Figures/SI/tetracene_csfs/sym_1/mo_b2g_3_csf} \includegraphics[width=\linewidth]{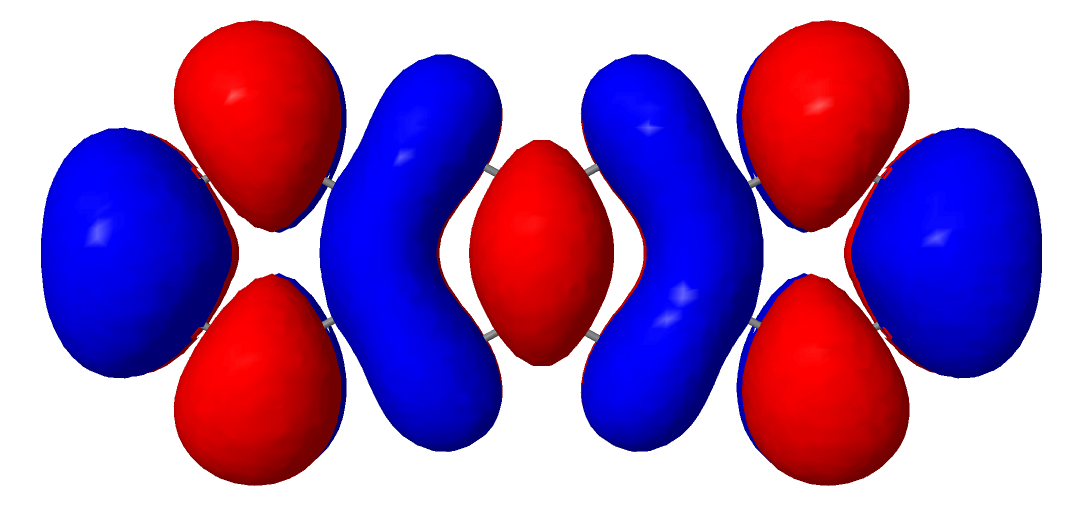} \\
		\hline
		&&&&&&\\[-0.3cm]
		S$_4$ &  \includegraphics[width=\linewidth]{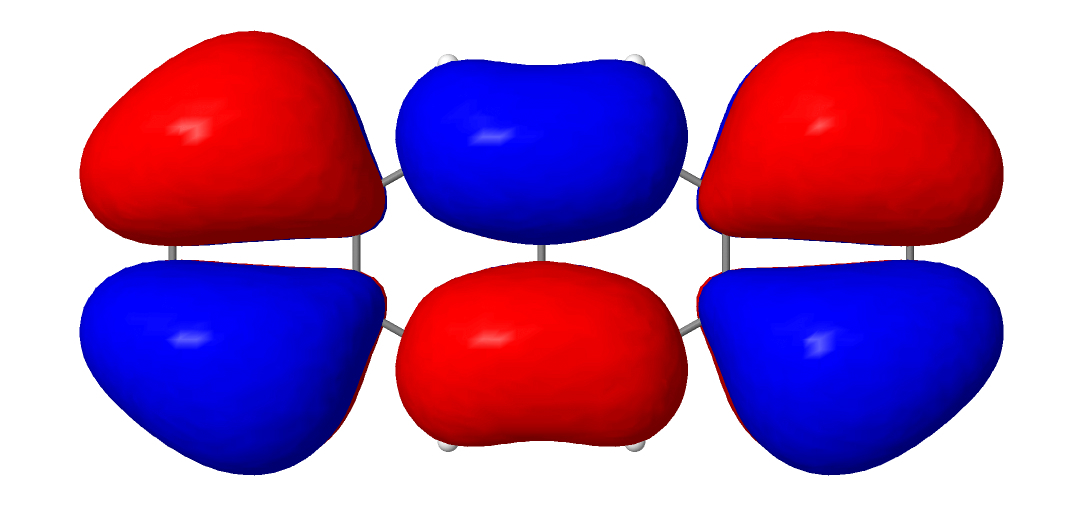}  &$\xrightarrow[]{\text{ 0.535}}$& \includegraphics[width=\linewidth]{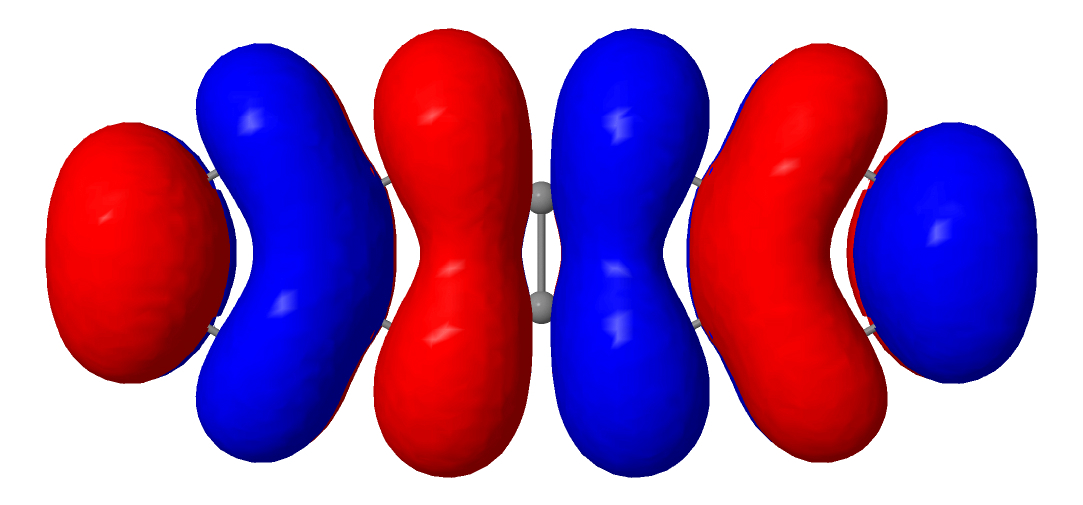} & \includegraphics[width=\linewidth]{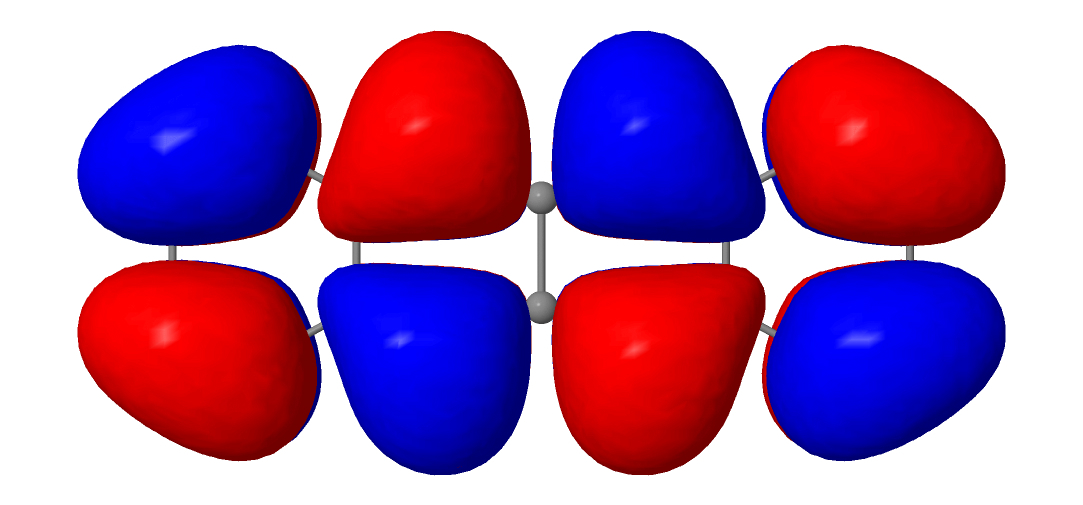} &$\xrightarrow[]{\text{-0.627}}$& \includegraphics[width=\linewidth]{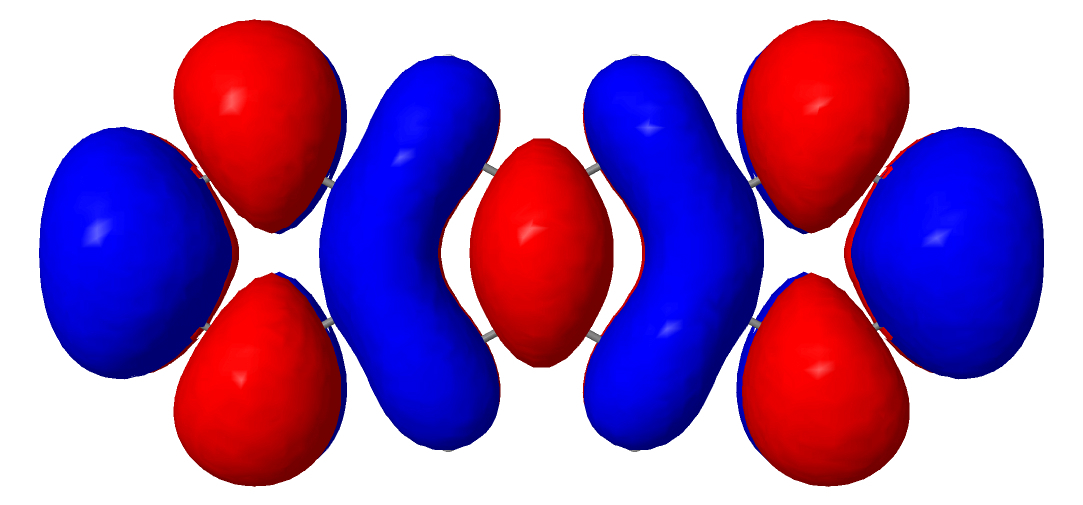} \\
		\hline
		&&&&&&\\[-0.3cm]
		\multirow{6}{*}{S$_5$} & \includegraphics[width=\linewidth]{Figures/SI/tetracene_csfs/sym_3/mo_b3g_2_csf}  &$\xrightarrow[]{\text{ 0.532}}$& \includegraphics[width=\linewidth]{Figures/SI/tetracene_csfs/sym_3/mo_b2g_3_csf} & \includegraphics[width=\linewidth]{Figures/SI/tetracene_csfs/sym_3/mo_au_2_csf} &$\xrightarrow[]{\text{ 0.381}}$& \includegraphics[width=\linewidth]{Figures/SI/tetracene_csfs/sym_3/mo_b1u_4_csf} \\[1cm]
		&  \includegraphics[width=\linewidth]{Figures/SI/tetracene_csfs/sym_3/mo_au_2_csf}  &$\xRightarrow[]{\text{-0.270}}$& \includegraphics[width=\linewidth]{Figures/SI/tetracene_csfs/sym_3/mo_b2g_3_csf} \includegraphics[width=\linewidth]{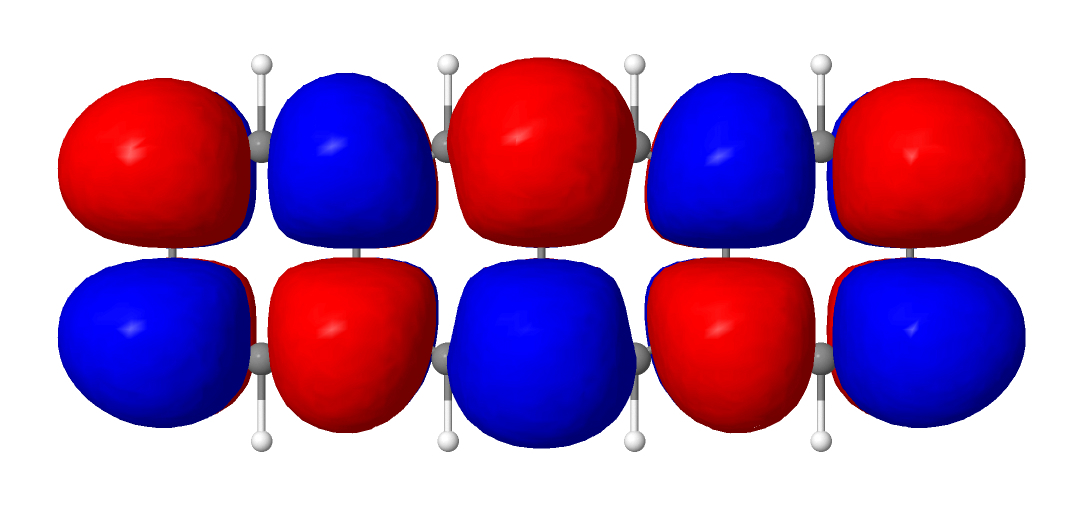} & \includegraphics[width=\linewidth]{Figures/SI/tetracene_csfs/sym_3/mo_au_2_csf} \includegraphics[width=\linewidth]{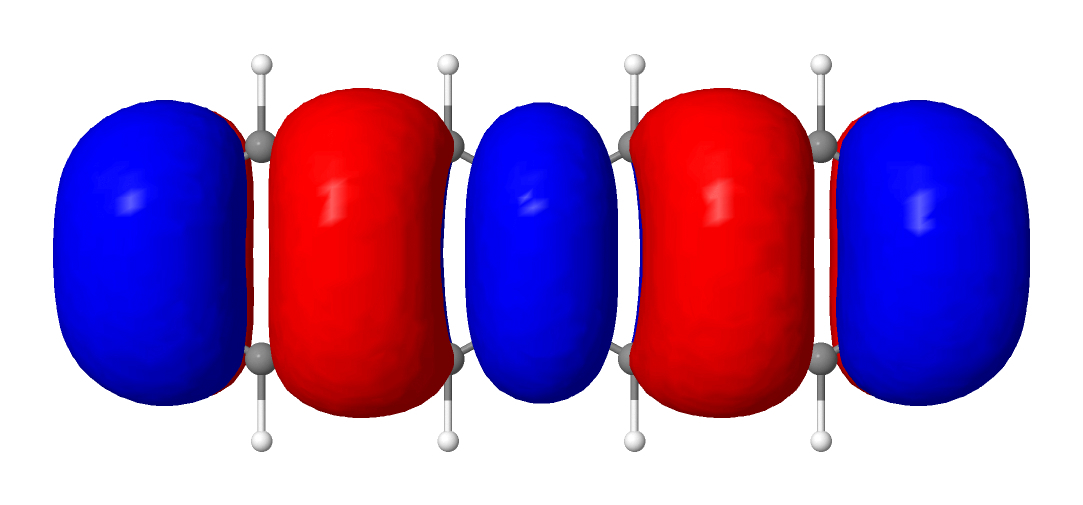} &$\xRightarrow[]{\text{-0.317}}$& \includegraphics[width=\linewidth]{Figures/SI/tetracene_csfs/sym_3/mo_b2g_3_csf} \\
		\hline
		&&&&&&\\[-0.3cm]
		S$_6$ &  \includegraphics[width=\linewidth]{Figures/SI/tetracene_csfs/sym_8/mo_au_2_csf}  &$\xrightarrow[]{\text{ 0.596}}$& \includegraphics[width=\linewidth]{Figures/SI/tetracene_csfs/sym_8/mo_b3g_3_csf} & \includegraphics[width=\linewidth]{Figures/SI/tetracene_csfs/sym_8/mo_b1u_2_csf} &$\xrightarrow[]{\text{ 0.565}}$& \includegraphics[width=\linewidth]{Figures/SI/tetracene_csfs/sym_8/mo_b2g_3_csf} \\
	\end{tabular}
	
	\caption{Orbital transitions and their coefficients (with absolute value $>0.2$) of the states S$_1$ to S$_6$ at CASSCF level of theory with 16 active orbitals and 16 active electrons. The single and double arrows indicate the excitation of one and two electrons, respectively. The number above the arrow describes the coefficient of the specific orbital transition.}
	\label{tab:orb-trans}
\end{table}

\subsection{Ionization potentials and Dyson intensities}

For a complete picture of tetracene as a radical cation, the vertical excitation energies from the ground state equilibrium structure of the neutral tetracene to its cationic states are provided in table \ref{tab:vertical_cation}. The calculations were performed using a 15 state-averaged CASSCF/CASPT2 calculation with an active space of 16 active electrons and 16 active orbitals (similar to the calculation of neutral tetracene). In addition, the Dyson intensities~\cite{tenorio_photoionization_2022} were calculated for transitions from the different neutral states to the respective cationic state. The calculation described above was carried out using MOLCAS 8.4\cite{aquilante_span_2016} and OpenMOLCAS~v20.10\cite{fdez_galvan_openmolcas_2019}. The results are shown in table \ref{tab:vertical_cation}.

\begin{table}[ht]
	\centering
	\begin{tabular}{c|ccc}
		 &     Tc$^{+}$ $^2$A$_\mathrm{u}$& Tc$^{+}$ $^2$B$_\mathrm{1u}$ & Tc$^{+}$ $^2$B$_\mathrm{2g}$  \\\hline
            E$_\mathrm{IP}$ / eV & 6.97 & 8.41 & 8.41 \\
		E$_\mathrm{IP}^\mathrm{theo}$ / eV & 6.938 & 8.397 & 8.843 \\\hline
		D$_\mathrm{S_1}$ & 0.425 & 0.007 & 0.334 \\
		D$_\mathrm{S_2}$ & 0.198 & 0.208 & 0.006 \\
		D$_\mathrm{S_3}$ & 0.041 & 0.033 & 0.561 \\
		D$_\mathrm{S_4}$ & 0.235 & 0.006 & 0.022 \\
		D$_\mathrm{S_5}$ & 0.092 & 0.001 & 0.061 \\
		D$_\mathrm{S_6}$ & 0.245 & 0.231 & 0.005 \\
	\end{tabular}
	\caption{\label{tab:vertical_cation}Comparison of the experimentally observed ionization potentials E$_\mathrm{IP}$\cite{Schmidt1977} to the calculated vertical energies of the radical cation E$_\mathrm{IP}^\mathrm{theo}$ at the ground state equilibrium structure of neutral tetracene, as well as the Dyson intensities $\mathrm{D}$ for transitions from different neutral states to the respective cationic state of tetracene.}
\end{table}

The vertical ionization potentials as shown in table \ref{tab:vertical_cation} for Tc$^{+}$ $^2$A$_\mathrm{u}$ and Tc$^{+}$ $^2$B$_\mathrm{1u}$ agree well with the experimental values and are comparable with the ionization potentials in table 2. The ionization potential of Tc$^{+}$ $^2$B$_\mathrm{2g}$ is significantly larger than the experimental value and not accurately described. Thus, the relaxation of the cation to its ground state equilibrium structure is crucial for a reasonable description of the cationic states. The trends of the Dyson intensities in table 2, however, can also be observed in table \ref{tab:vertical_cation} and the values do not differ significantly. This emphasizes the accuracy of the calculated Dyson intensities.

\subsection{Calibration of the calculations}

\begin{figure}[h]
	\centering
	\begin{tabular}{c|cccc}
		sym. &A$_\text{u}$ & B$_\text{1u}$ & B$_\text{2g}$ & B$_\text{3g}$ \\
		\hline
		&\includegraphics[width=0.2\linewidth]{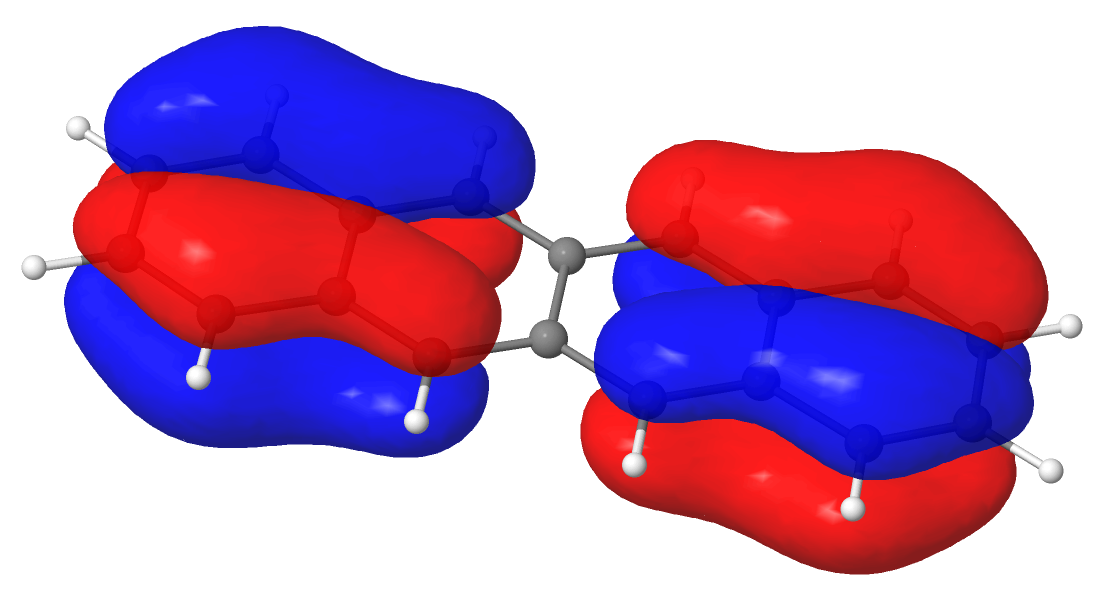} & \includegraphics[width=0.2\linewidth]{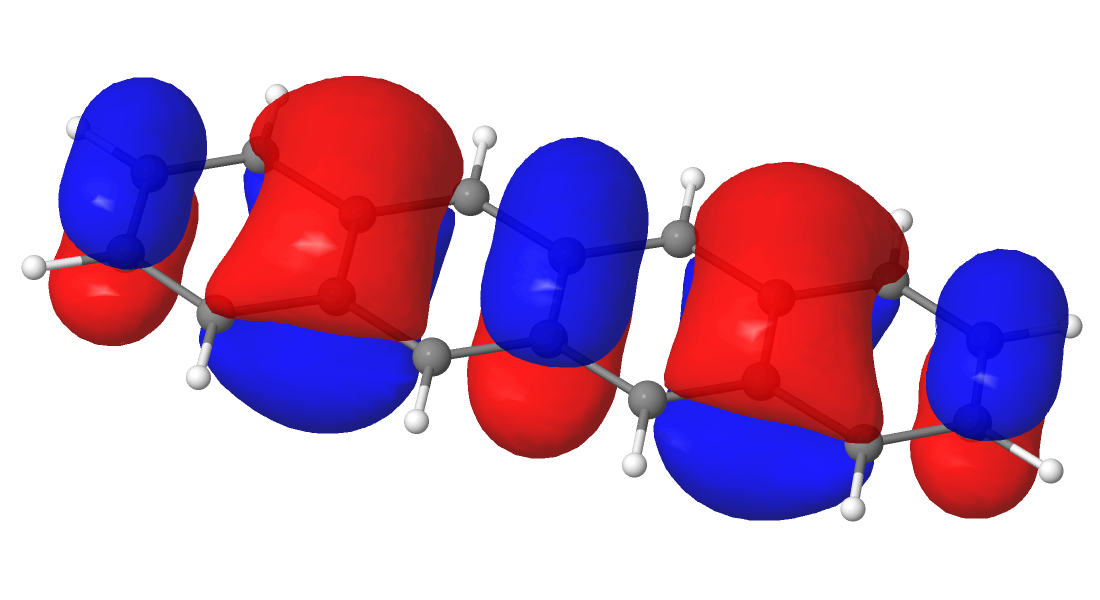} & \includegraphics[width=0.2\linewidth]{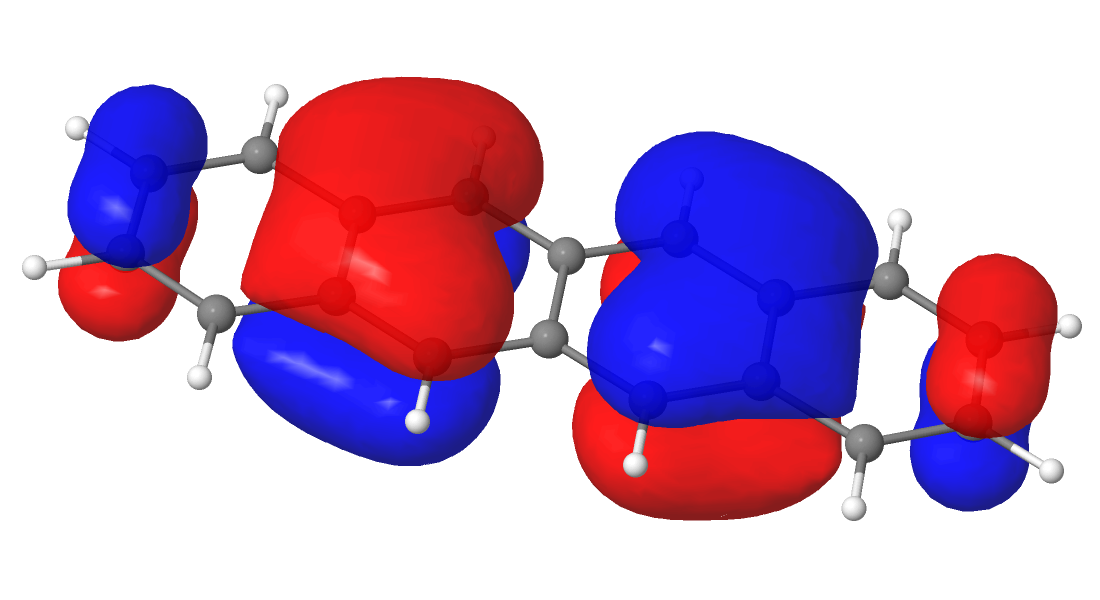} & \includegraphics[width=0.2\linewidth]{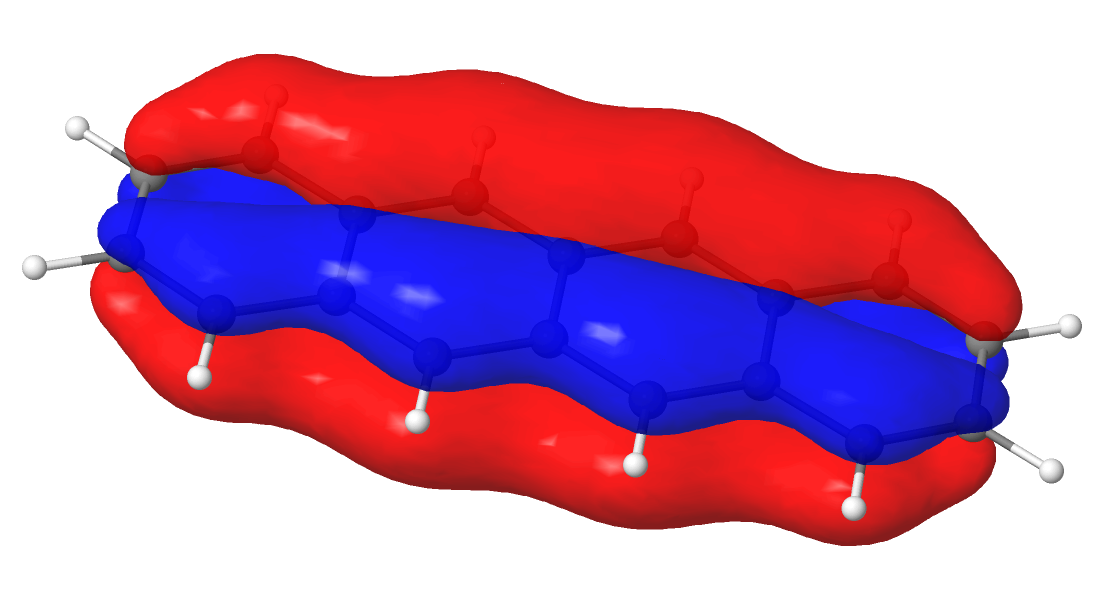} \\
		&\includegraphics[width=0.2\linewidth]{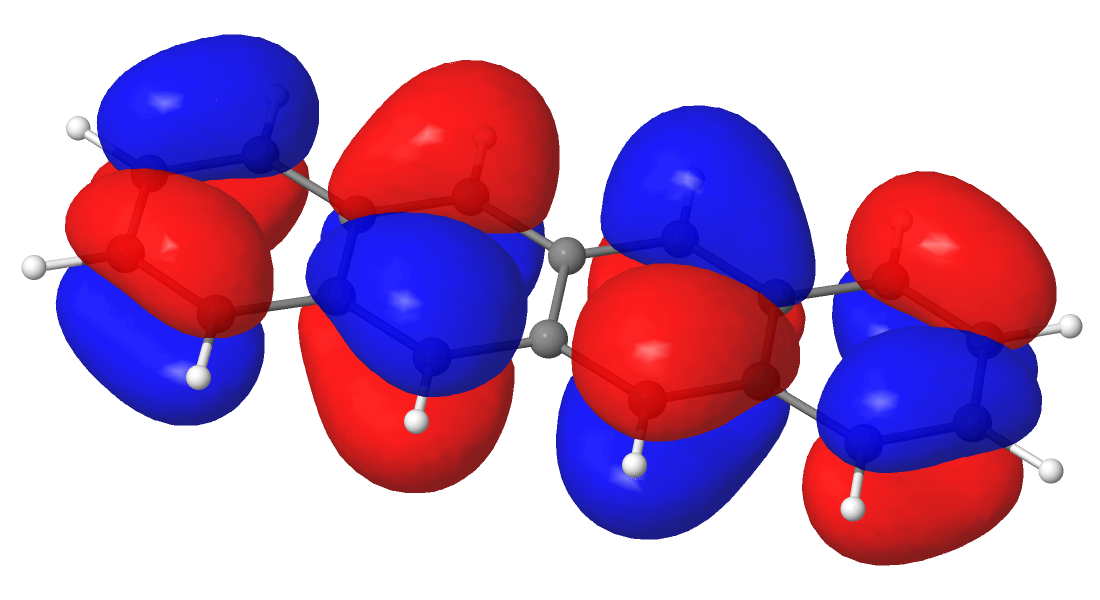} & \includegraphics[width=0.2\linewidth]{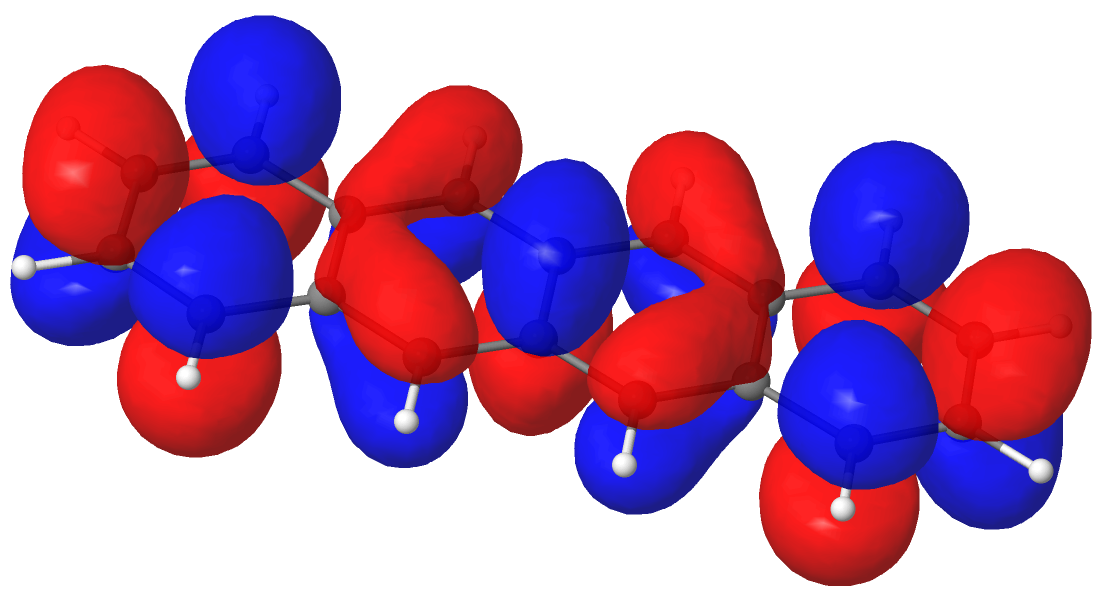} & \includegraphics[width=0.2\linewidth]{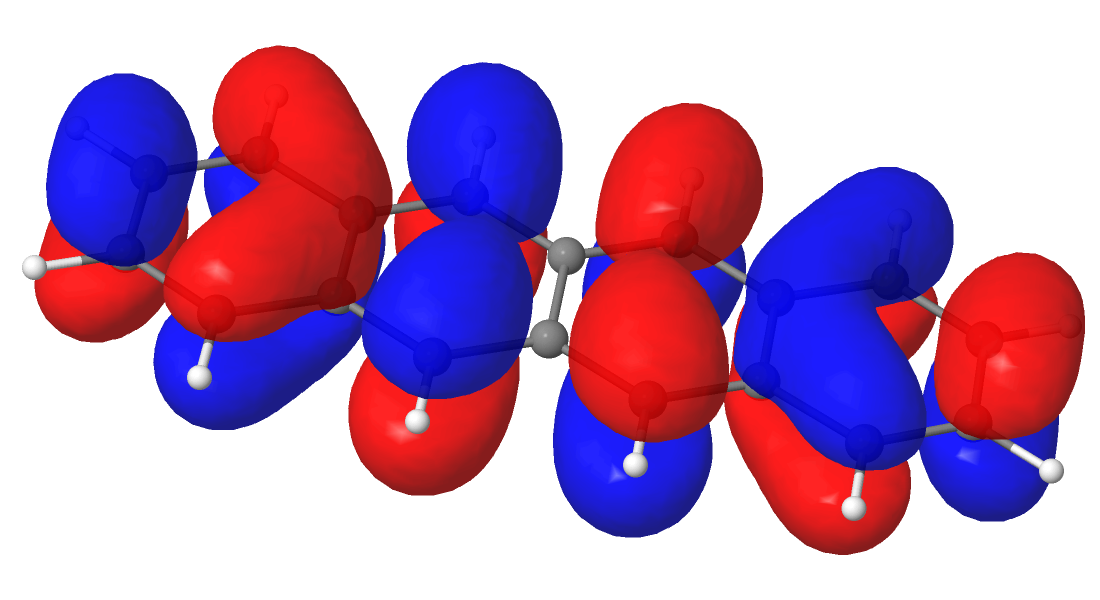} & \includegraphics[width=0.2\linewidth]{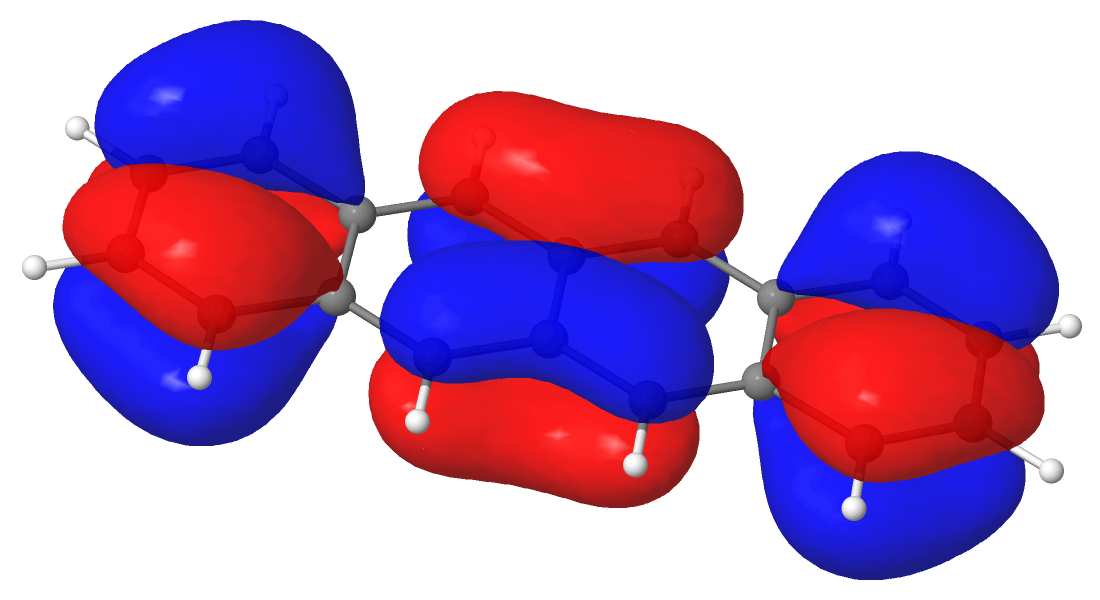} \\
		& \includegraphics[width=0.2\linewidth]{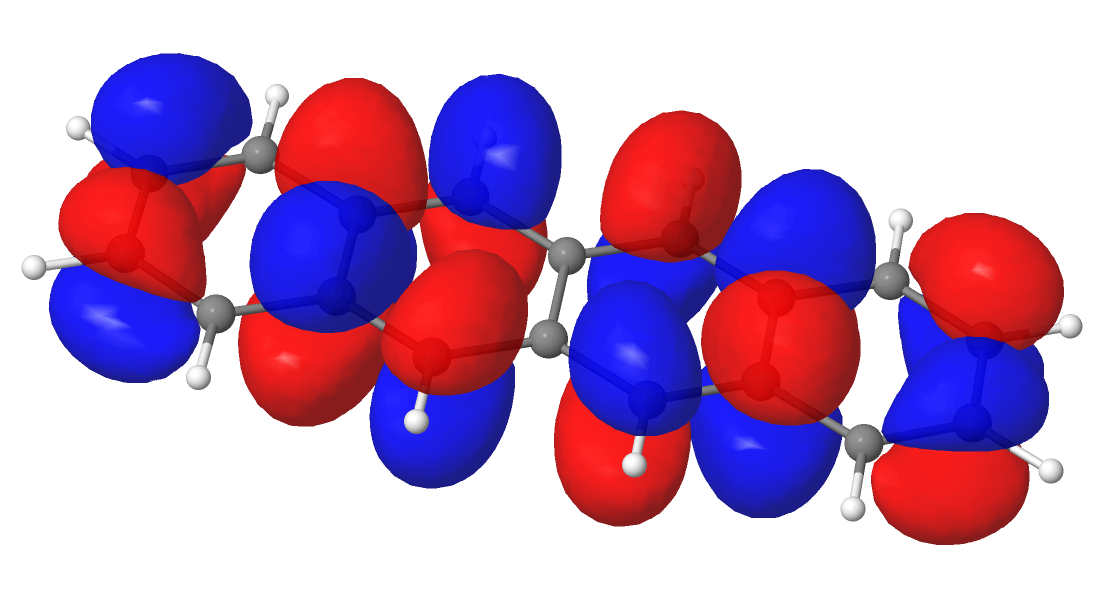} & \includegraphics[width=0.2\linewidth]{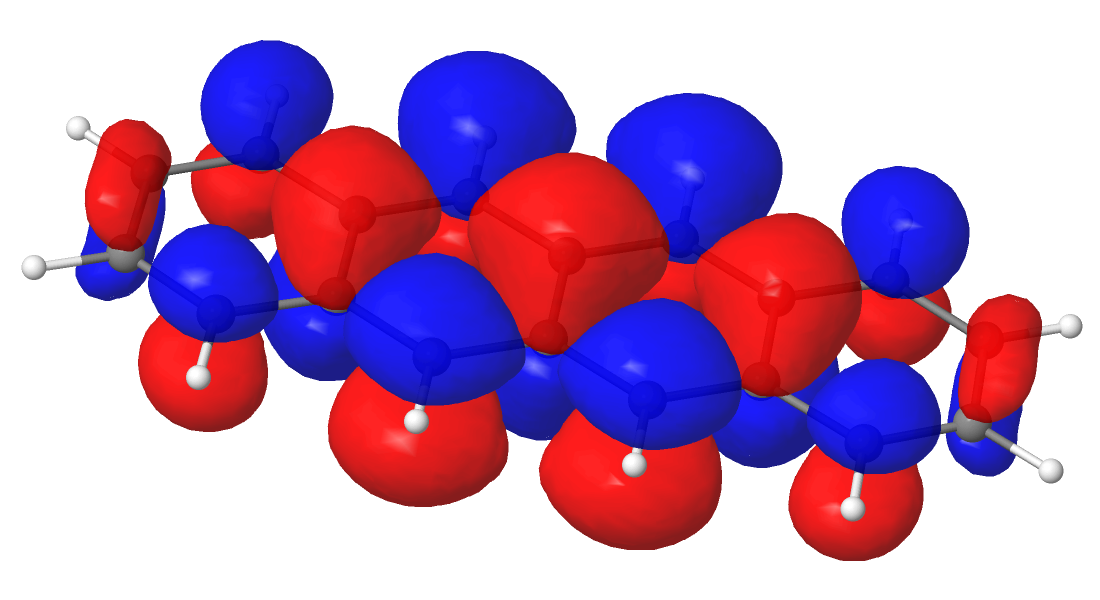} & \includegraphics[width=0.2\linewidth]{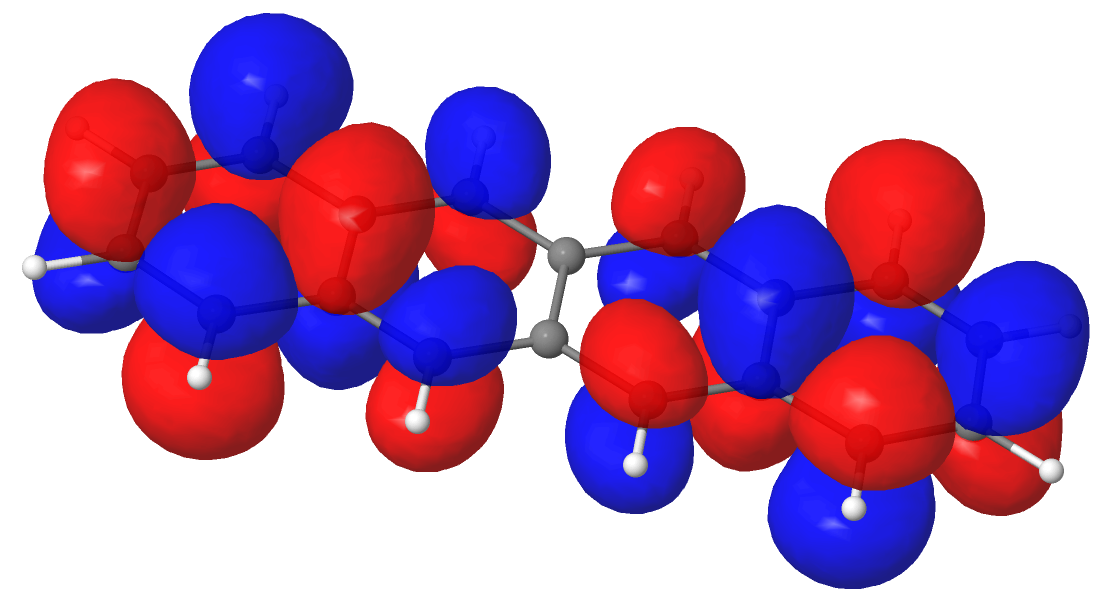} & \includegraphics[width=0.2\linewidth]{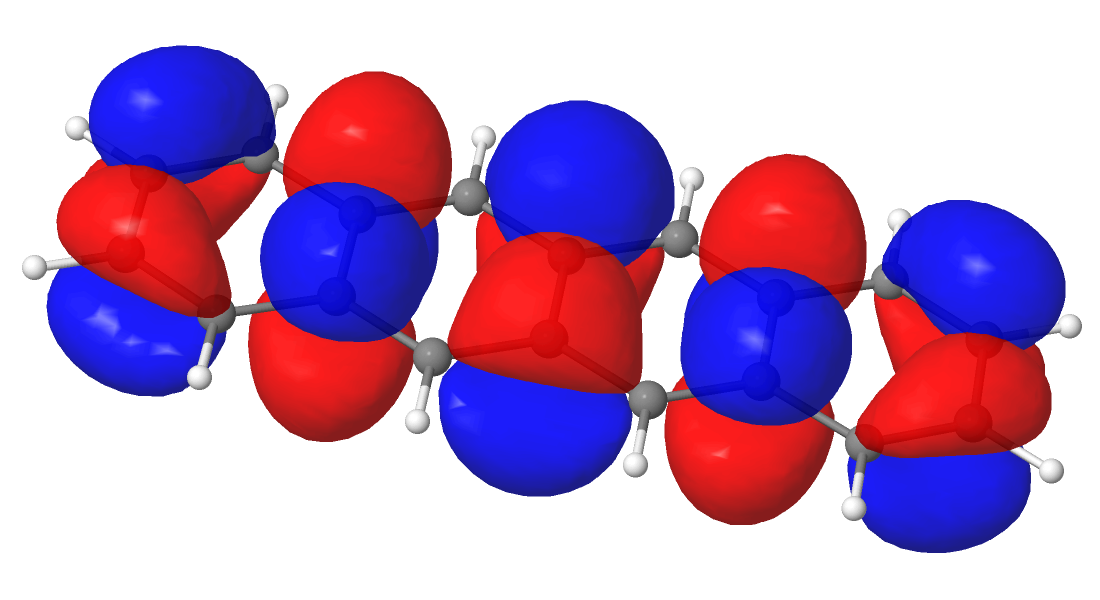} \\
	\end{tabular}
	
        \caption{CASSCF optimized molecular orbitals and their corresponding symmetry of the A$_\text{g}$ symmetry calculation. In the state-average calculations, 10 roots with 12 active orbitals and 12 active electrons were employed.}
	\label{fig:omo_sa10-cas1212}
\end{figure}

\begin{figure}[h]
	\centering
	\begin{tabular}{c|cccc}
		sym. &A$_\text{u}$ & B$_\text{1u}$ & B$_\text{2g}$ & B$_\text{3g}$ \\
		\hline
		&\includegraphics[width=0.2\linewidth]{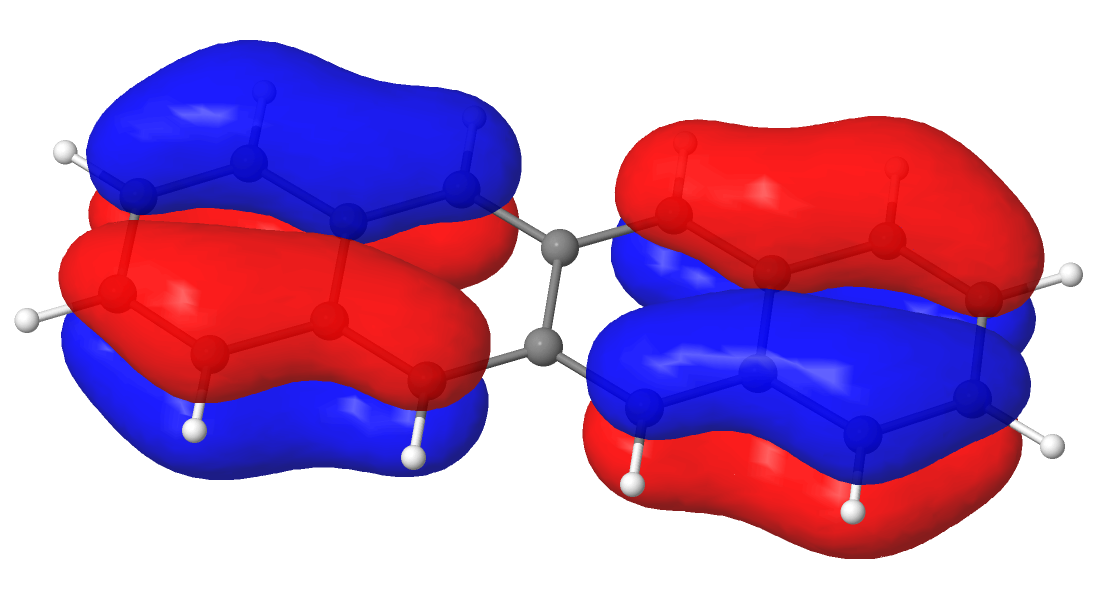} & \includegraphics[width=0.2\linewidth]{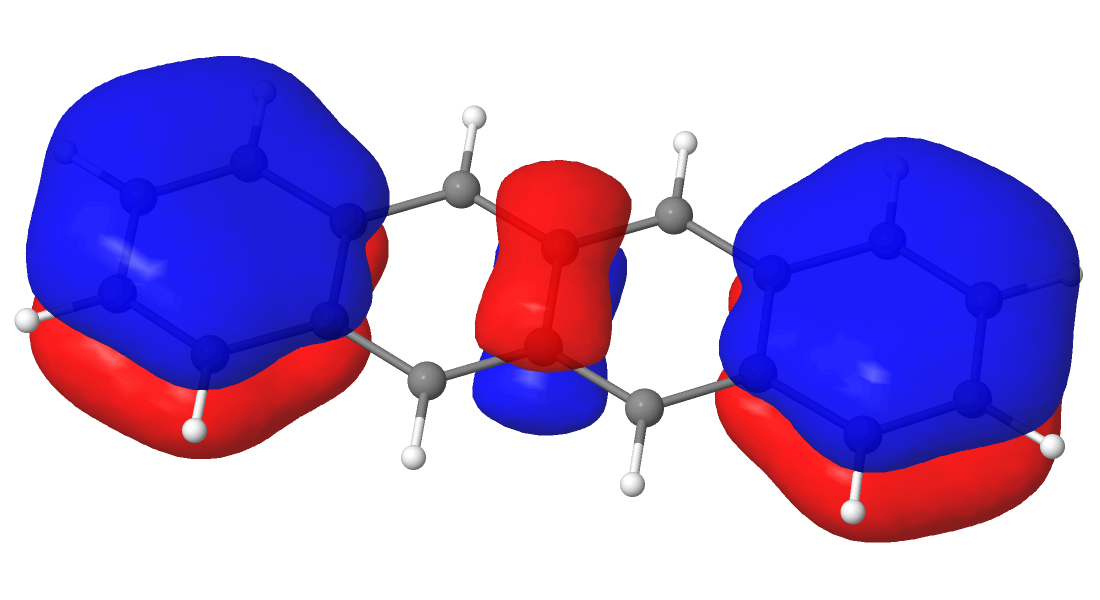} & \includegraphics[width=0.2\linewidth]{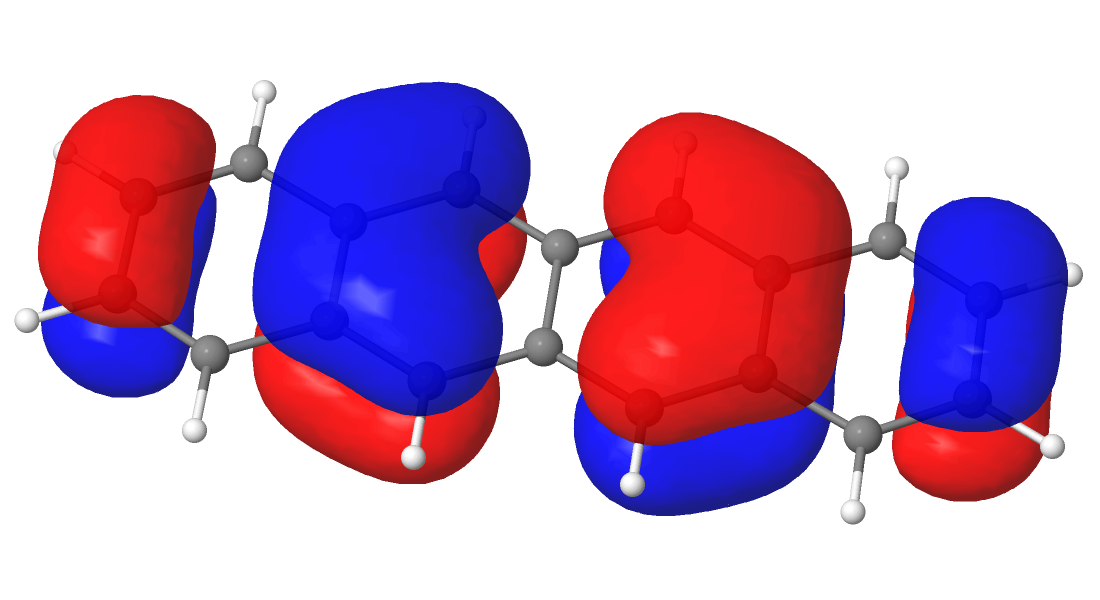} & \includegraphics[width=0.2\linewidth]{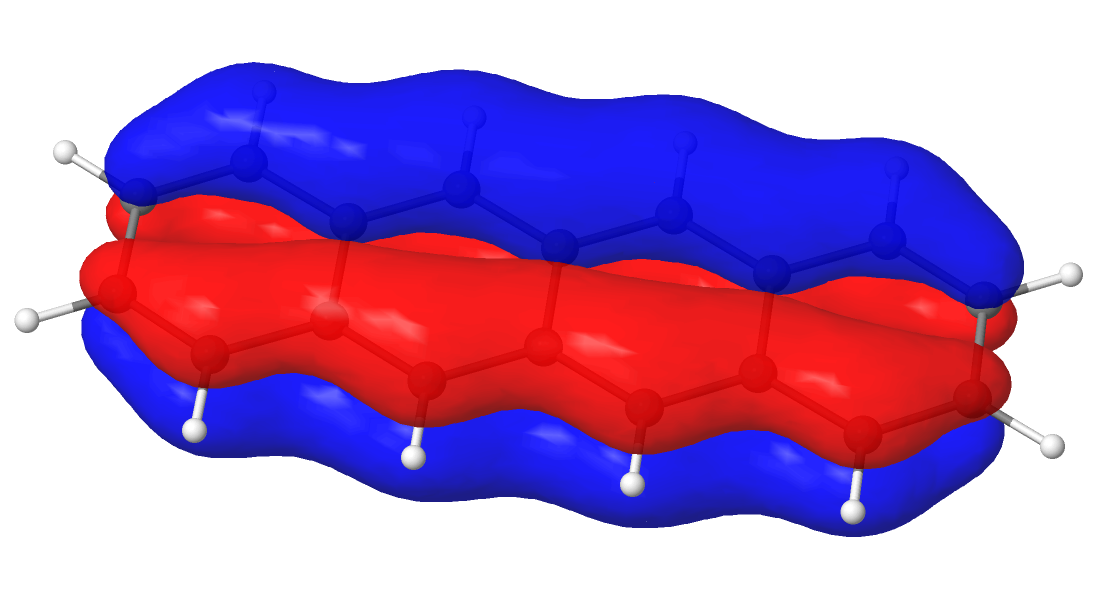} \\
		&\includegraphics[width=0.2\linewidth]{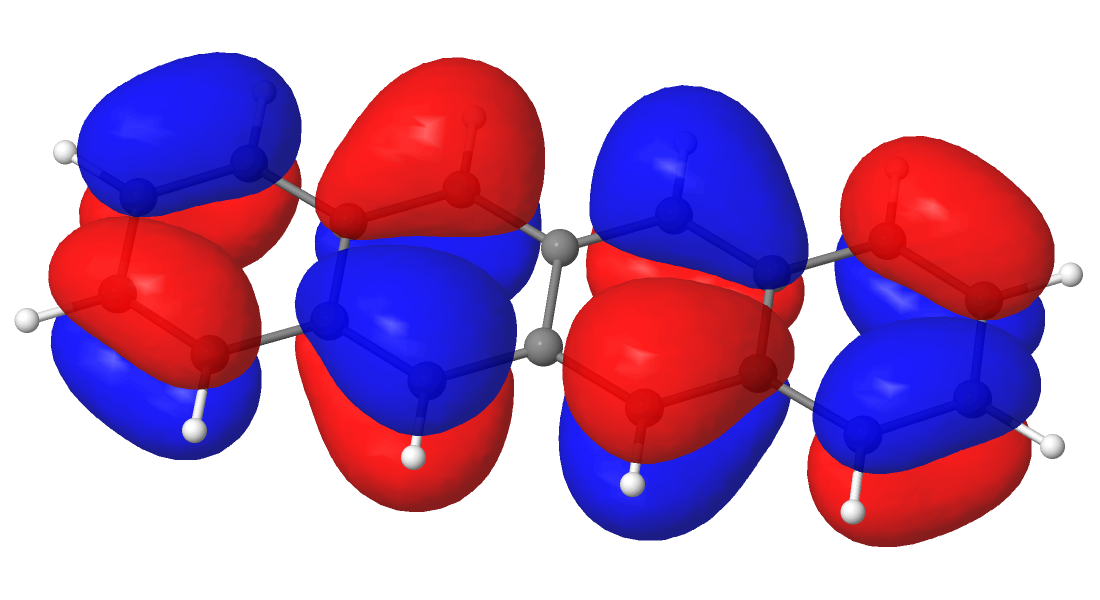} & \includegraphics[width=0.2\linewidth]{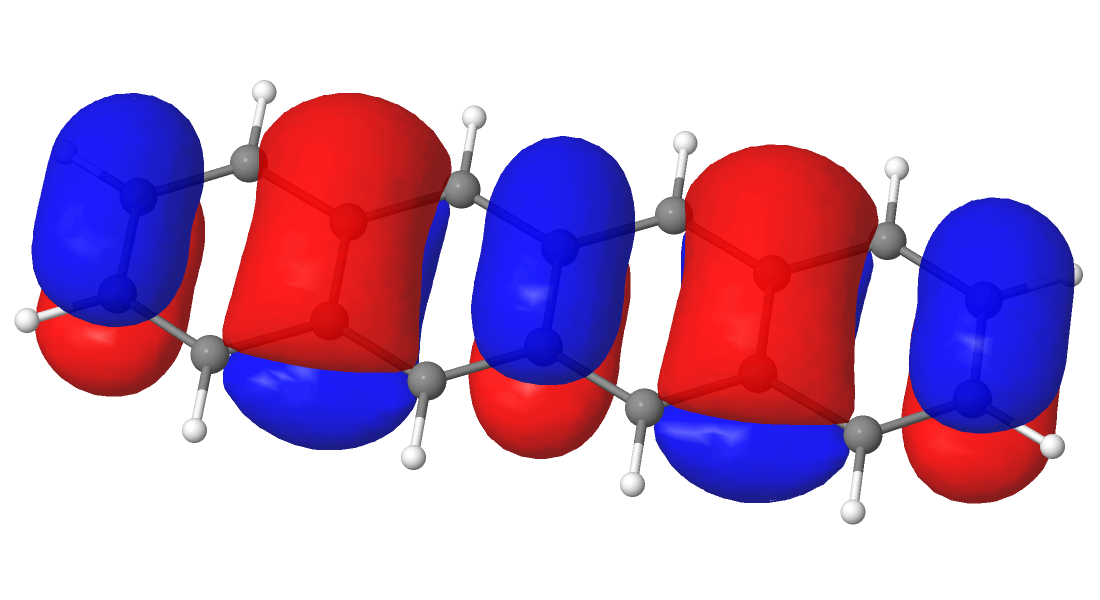} & \includegraphics[width=0.2\linewidth]{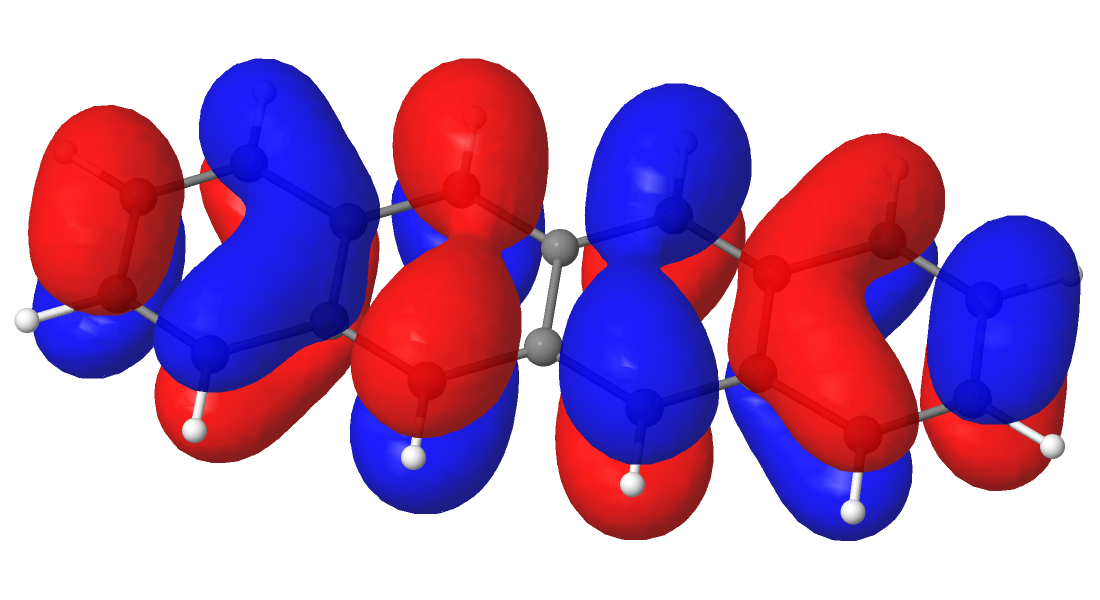} & \includegraphics[width=0.2\linewidth]{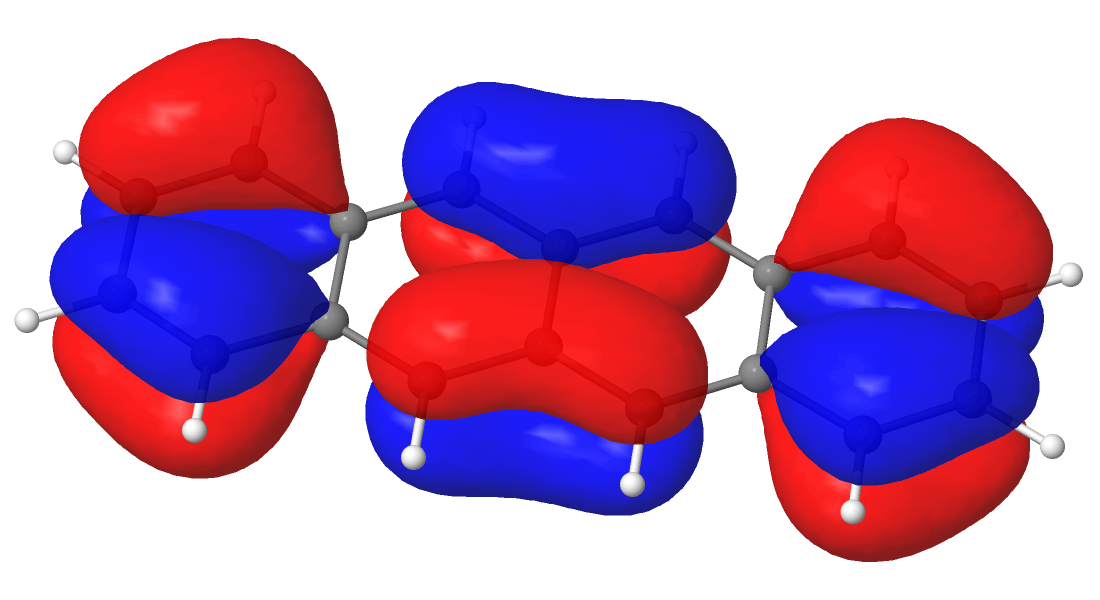} \\
		& \includegraphics[width=0.2\linewidth]{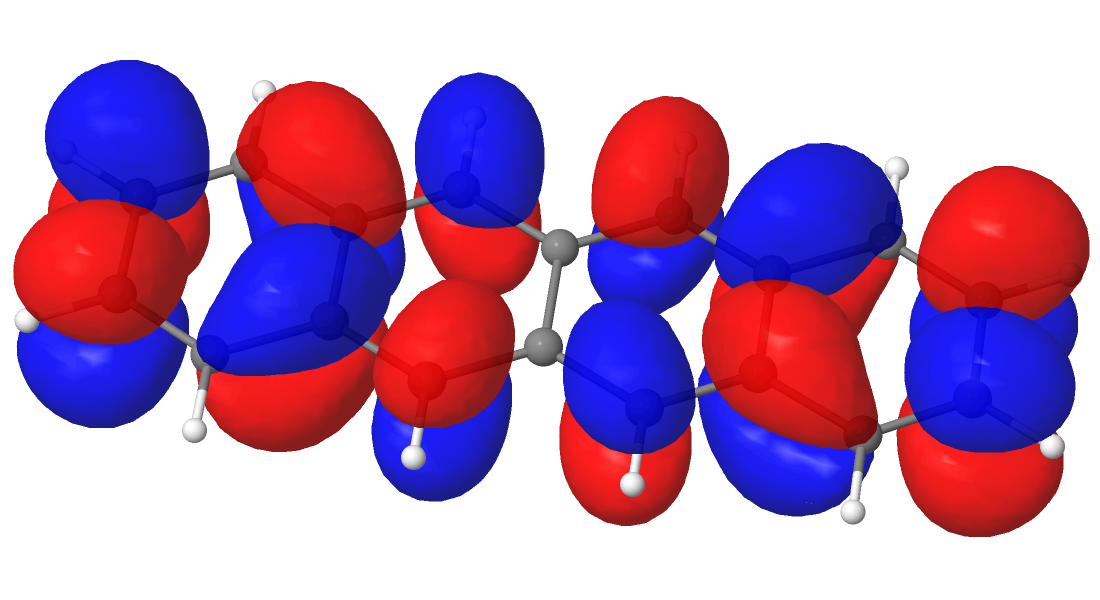} & \includegraphics[width=0.2\linewidth]{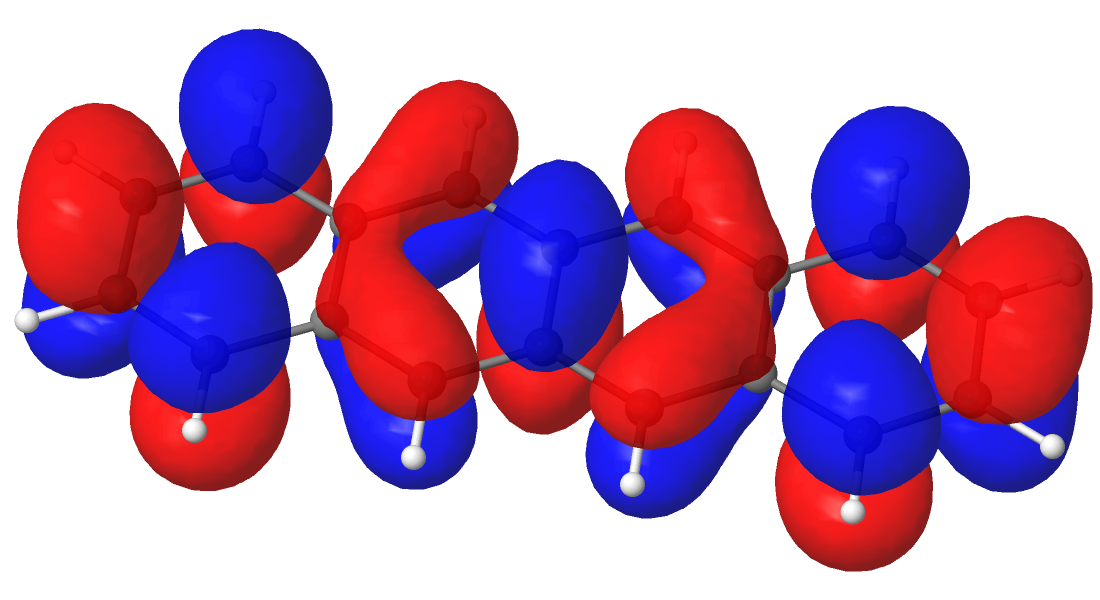} & \includegraphics[width=0.2\linewidth]{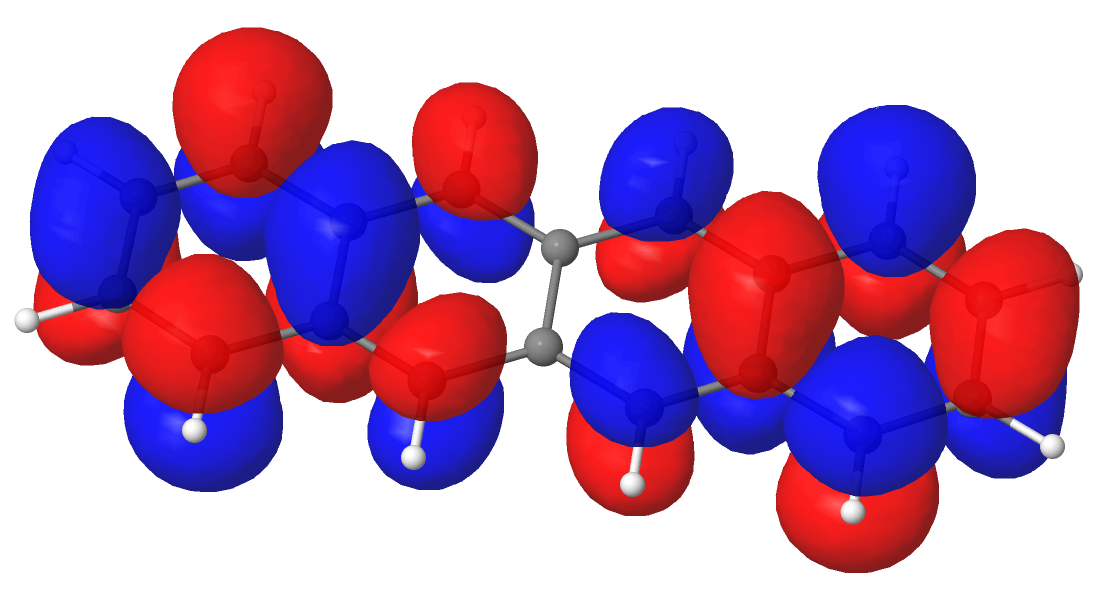} & \includegraphics[width=0.2\linewidth]{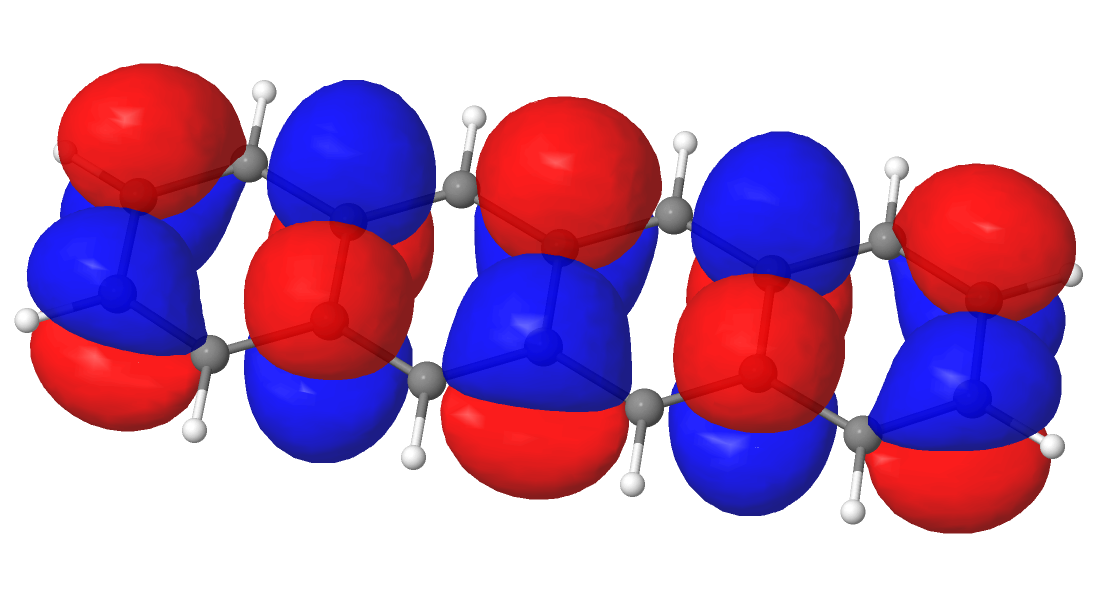} \\
            & \includegraphics[width=0.2\linewidth]{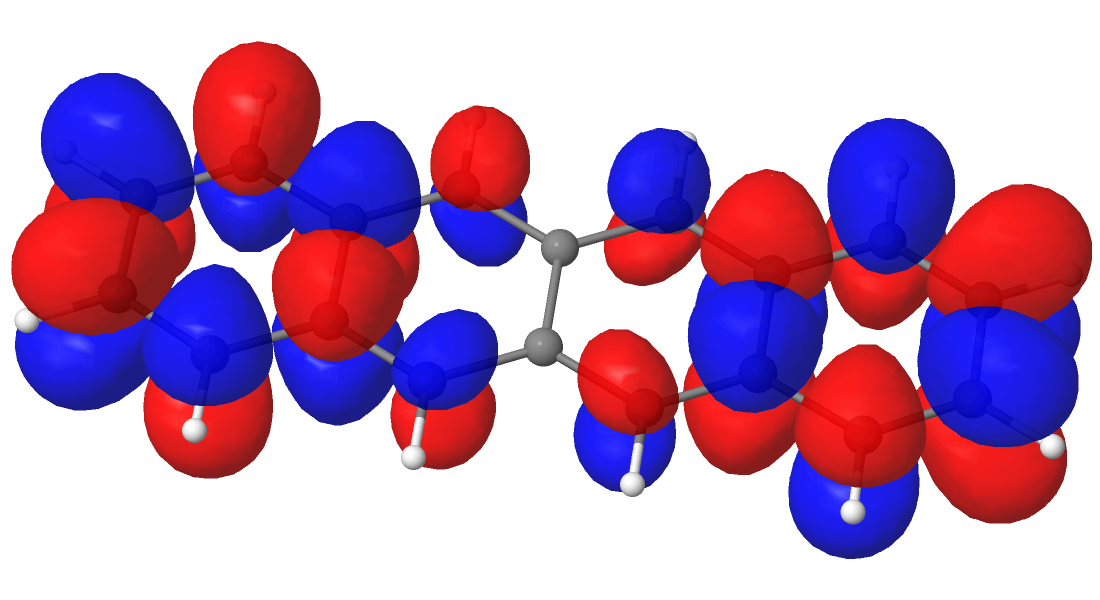} & \includegraphics[width=0.2\linewidth]{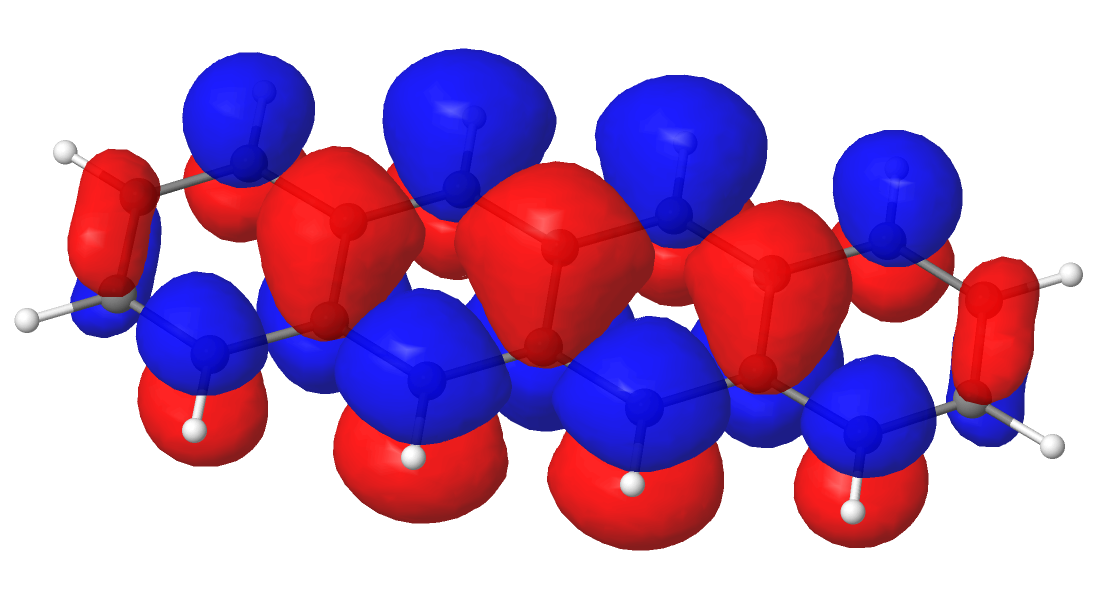} & & \\
	\end{tabular}
	
        \caption{CASSCF optimized molecular orbitals and their corresponding symmetry of the A$_\text{g}$ symmetry calculation. In the state-average calculations, 10 roots with 14 active orbitals and 14 active electrons were employed.}
	\label{fig:omo_sa10-cas1414}
\end{figure}

To verify the selection of our active space, we performed several CASSCF calculations including symmetry. We performed a 10 state-averaged CASSCF calculation with 12 and 14 active electrons and orbitals, respectively, as well as a 15 state-averaged CASSCF calculation with 16 active electrons and orbitals. In addition, an imaginary shift of 0.05~au~\cite{forsberg_multiconfiguration_1997} for the zero order Hamiltonian was included and the calculations were carried out using the ANO-L-VDZP basis set~\cite{widmark_density_1990,widmark_density_1991,pou-amerigo_density_1995,pierloot_density_1995}. The energies $\Delta$E are calculated at CASPT2 level of theory and presented in table \ref{tab:sa10-cas121416_sym}. The corresponding optimized molecular orbitals for the A$_\text{g}$ symmetry calculations with 12 and 14 active electrons and orbitals are shown in figure \ref{fig:omo_sa10-cas1212} and figure \ref{fig:omo_sa10-cas1414}, respectively. The calculations described above were carried out using MOLCAS 8.4\cite{aquilante_span_2016} and OpenMOLCAS~v20.10\cite{fdez_galvan_openmolcas_2019}.

\begin{table}[h]
	\centering
	\begin{tabular}{c|ccc}
		state & $^1\Delta$E / eV (sym.) & $^2\Delta$E / eV (sym.) & $^3\Delta$E / eV (sym.) \\
		\hline
		S$_1$ & 2.85 (B$_\text{2u}$) & 3.04 (B$_\text{2u}$)& 2.83 (B$_\text{2u}$)\\
		S$_2$ & 3.52 (B$_\text{3u}$)& 3.52 (B$_\text{3u}$) & 3.48 (B$_\text{3u}$)\\
		S$_3$ & 3.79 (A$_\text{g}$) & 3.79 (A$_\text{g}$) & 3.88 (A$_\text{g}$)\\
		S$_4$ & 4.10 (B$_\text{1g}$) & 4.19 (B$_\text{1g}$) & 4.23 (B$_\text{1g}$)\\
		S$_5$ & 4.28 (B$_\text{1g}$) & 4.51 (B$_\text{1g}$) & 4.29 (B$_\text{1g}$)\\
            S$_6$ & 4.86 (B$_\text{2u}$) & 4.98 (B$_\text{2u}$) & 4.89 (B$_\text{3u}$)\\
	\end{tabular}
	
        \caption{Vertical excitation energies $\Delta$E (in eV) of the six lowest-lying singlet states, calculated at the ground state equilibrium structure of tetracene. The symmetry of the state is indicated in parenthesis. $^1$The results are obtained from a 10 state-averaged calculation per symmetry with 12 active orbitals and 12 active electrons. $^2$The results are obtained from a 10 state-averaged calculation per symmetry with 14 active orbitals and 14 active electrons. $^3$The results are obtained from a 15 state-averaged calculation per symmetry with 16 active orbitals and 16 active electrons.}
	\label{tab:sa10-cas121416_sym}
\end{table}

The calculated energies of S$_1$ to S$_6$ differ by a maximum of $\sim 0.3$\,eV (for S$_4$) between the different calculations, which shows an accurate selection of the active space. To further increase the accuracy of our calculations, we employed an active space of 16 active electrons and 16 active orbitals with a basis set of triple $\zeta$ quality ANO-L-VTZP~\cite{widmark_density_1990,widmark_density_1991,pou-amerigo_density_1995,pierloot_density_1995}. The energy of S$_1$ decreased with the larger basis set reaching 2.75\,eV, which differs from the experimental value~\cite{Amirav_1979} by only 0.02\,eV. Overall, the relative energies of the singlet states are slightly smaller compared to the calculation with basis set ANO-L-VDZP and are presented in table 1.

\subsection{Geometries}

The geometries were obtained using density functional theory (DFT) with the B3LYP exchange-correlation functional~\cite{vosko_accurate_1980,becke_density-functional_1988,lee_development_1988,becke_density-functional_1992,becke_density-functional_1993} and a def2-TZVP basis set~\cite{weigend_balanced_2005}. For the relaxation, DFT-D3 dispersion interaction corrections~\cite{grimme_consistent_2010} with Becke-Johnson damping~\cite{grimme_effect_2011} were employed. The optimized ground state equilibrium geometry (in Å) of the neutral tetracene and tetracene as a radical cation is provided in table \ref{tab:equ_struc_neutral} and table \ref{tab:equ_struc_cation}, respectively. The geometry optimizations were carried out using the Turbomole V7.5 package~\cite{noauthor_turbomole_nodate,balasubramani_turbomole_2020}.

\begin{table}[h]
    \centering
    \begin{tabular}{cccc}
        atom & x & y & z \\
        \hline
        C   &  2.4392170  &  0.7231002  &  0.0000000 \\
        C   &  1.2300798  &  1.4012461  &  0.0000000 \\ 
        C   &  0.0000000  &  0.7233739  &  0.0000000 \\
        C   &  0.0000000  & -0.7233739  &  0.0000000 \\
        C   &  1.2300798  & -1.4012461  &  0.0000000 \\
        C   &  2.4392170  & -0.7231002  &  0.0000000 \\
        C   &  3.6954713  & -1.4037934  &  0.0000000 \\
        C   &  4.8674639  & -0.7130534  &  0.0000000 \\
        C   &  4.8674639  &  0.7130534  &  0.0000000 \\
        C   &  3.6954713  &  1.4037934  &  0.0000000 \\
        C   & -1.2300798  &  1.4012461  &  0.0000000 \\
        C   & -2.4392170  &  0.7231002  &  0.0000000 \\
        C   & -2.4392170  & -0.7231002  &  0.0000000 \\
        C   & -1.2300798  & -1.4012461  &  0.0000000 \\
        C   & -3.6954713  &  1.4037934  &  0.0000000 \\
        C   & -4.8674639  &  0.7130534  &  0.0000000 \\
        C   & -4.8674639  & -0.7130534  &  0.0000000 \\
        C   & -3.6954713  & -1.4037934  &  0.0000000 \\
        H   &  1.2303876  &  2.4855248  &  0.0000000 \\
        H   &  1.2303876  & -2.4855248  &  0.0000000 \\
        H   &  3.6938585  & -2.4873222  &  0.0000000 \\
        H   &  3.6938585  &  2.4873222  &  0.0000000 \\
        H   & -1.2303876  &  2.4855248  &  0.0000000 \\
        H   & -1.2303876  & -2.4855248  &  0.0000000 \\
        H   &  5.8116057  & -1.2428699  &  0.0000000 \\
        H   &  5.8116057  &  1.2428699  &  0.0000000 \\
        H   & -3.6938585  &  2.4873222  &  0.0000000 \\
        H   & -5.8116057  &  1.2428699  &  0.0000000 \\
        H   & -5.8116057  & -1.2428699  &  0.0000000 \\
        H   & -3.6938585  & -2.4873222  &  0.0000000 \\
    \end{tabular}
    \caption{Cartesian coordinates of the equilibrium structure of the neutral tetracene (in Å).}
    \label{tab:equ_struc_neutral}
\end{table}

\begin{table}[h]
    \centering
    \begin{tabular}{cccc}
        atom & x & y & z \\
        \hline
        C  &   2.4462484  &  0.7195624  &  0.0000000 \\ 
        C  &   1.2234980  &  1.4030610  &  0.0000000 \\ 
        C  &   0.0000000  &  0.7228147  &  0.0000000 \\ 
        C  &   0.0000000  & -0.7228147  &  0.0000000 \\ 
        C  &   1.2234980  & -1.4030610  &  0.0000000 \\ 
        C  &   2.4462484  & -0.7195624  &  0.0000000 \\ 
        C  &   3.6824827  & -1.4037387  &  0.0000000 \\ 
        C  &   4.8670831  & -0.7042880  &  0.0000000 \\ 
        C  &   4.8670831  &  0.7042880  &  0.0000000 \\ 
        C  &   3.6824827  &  1.4037387  &  0.0000000 \\ 
        C  &  -1.2234980  &  1.4030610  &  0.0000000 \\ 
        C  &  -2.4462484  &  0.7195624  &  0.0000000 \\ 
        C  &  -2.4462484  & -0.7195624  &  0.0000000 \\ 
        C  &  -1.2234980  & -1.4030610  &  0.0000000 \\ 
        C  &  -3.6824827  &  1.4037387  &  0.0000000 \\ 
        C  &  -4.8670831  &  0.7042880  &  0.0000000 \\ 
        C  &  -4.8670831  & -0.7042880  &  0.0000000 \\ 
        C  &  -3.6824827  & -1.4037387  &  0.0000000 \\ 
        H  &   1.2254444  &  2.4867158  &  0.0000000 \\ 
        H  &   1.2254444  & -2.4867158  &  0.0000000 \\ 
        H  &   3.6853418  & -2.4862714  &  0.0000000 \\ 
        H  &   3.6853418  &  2.4862714  &  0.0000000 \\ 
        H  &  -1.2254444  &  2.4867158  &  0.0000000 \\ 
        H  &  -1.2254444  & -2.4867158  &  0.0000000 \\ 
        H  &   5.8085655  & -1.2371120  &  0.0000000 \\ 
        H  &   5.8085655  &  1.2371120  &  0.0000000 \\ 
        H  &  -3.6853418  &  2.4862714  &  0.0000000 \\ 
        H  &  -5.8085655  &  1.2371120  &  0.0000000 \\ 
        H  &  -5.8085655  & -1.2371120  &  0.0000000 \\ 
        H  &  -3.6853418  & -2.4862714  &  0.0000000 \\
    \end{tabular}
    \caption{Cartesian coordinates of the equilibrium structure of tetracene as a radical cation~(in~Å).}
    \label{tab:equ_struc_cation}
\end{table}

\clearpage

\bibliography{SI} 
\bibliographystyle{SI} 